\documentclass[12pt]{article}
\pdfoutput=1

\newlength{\abstractwidth}
\usepackage{amsfonts}

\usepackage{epsf}
\usepackage{color}
\usepackage{graphicx}
\usepackage{tikz}
\usepackage{dsfont}
\usepackage{subfigure}

\usepackage{hyperref}
\definecolor{darkred}{rgb}{0.8,0.1,0.1}
\hypersetup{colorlinks=true, linkcolor=darkred, citecolor=blue, linktoc=page}

\usetikzlibrary{arrows,shapes,positioning}
\usetikzlibrary{decorations.markings}
\usepackage[rightcaption]{sidecap}
\tikzstyle arrowstyle=[scale=1]
\tikzstyle directed=[postaction={decorate,decoration={markings,
    mark=at position .65 with {\arrow[arrowstyle]{stealth}}}}]
\tikzstyle reverse directed=[postaction={decorate,decoration={markings,
    mark=at position .65 with {\arrowreversed[arrowstyle]{stealth};}}}]
\usetikzlibrary{positioning}

\usepackage{amssymb}
\usepackage{latexsym}

\renewcommand{\thefootnote}{\fnsymbol{footnote}}
\renewcommand{\thanks}[1]{\footnote{#1}}
\newcommand{\starttext}{
\setcounter{footnote}{0}
\renewcommand{\thefootnote}{\arabic{footnote}}}

\newcommand{\bea}{\begin{eqnarray}}
\newcommand{\eea}{\end{eqnarray}}
\newcommand{\be}{\begin{eqnarray}}
\newcommand{\ee}{\end{eqnarray}}

\def\cA{{\cal A}}
\def\cB{{\cal B}}
\def\cC{{\cal C}}

\def\cG{{\cal G}}

\def\cN{{\cal N}}
\def\cO{{\cal O}}

\def\cS{{\cal S}}

\def\mG{\mathfrak{G}}

\def\mI{\mathfrak{I}}

\def\ZZ{{\mathbb Z}}
\def\RR{{\mathbb R}}

\def\CC{{\mathbb C}}
\def\HH{{\mathbb H}}

\def\Re{{\rm Re \,}}
\def\Im{{\rm Im \,}}

\def\half{{1\over 2}}
\def\p{\partial}

\def\f{\varphi}
\def\tet{\vartheta}
\def\ep{\varepsilon}
\def\om{\omega}

\def\hI{{\hat I}}
\def\hJ{{\hat J}}

\def\no{\nonumber}
\def\sm{\smallskip}

\makeatletter\def\l@subsubsection#1#2{}%
\makeatother

\begin{document}
\starttext
\setcounter{footnote}{0}

\begin{flushright}
2017 March 21st 
\end{flushright}

\vskip 0.3in

\begin{center}

{\Large \bf Warped $AdS_6\times S^2$ in Type IIB supergravity II}

\vskip 0.1in

{\large \bf Global solutions and  five-brane webs}

\vskip 0.4in

{\large   Eric D'Hoker,  Michael Gutperle and Christoph F.~Uhlemann} 

\vskip 0.15in

{ \sl 
Mani L. Bhaumik Institute for Theoretical Physics}\\
{\sl  Department of Physics and Astronomy}\\
{\sl University of California, Los Angeles, CA 90095, USA}

\vskip 0.05in

{\tt \small dhoker@physics.ucla.edu; gutperle@ucla.edu; uhlemann@physics.ucla.edu}

\vskip 0.5in

\begin{abstract}

Motivated by the construction of holographic duals to five-dimensional superconformal quantum field theories, we obtain global solutions to Type IIB supergravity invariant under the superalgebra $F(4)$ on a space-time of the form $AdS_6 \times S^2$ warped over a two-dimensional Riemann surface $\Sigma$. In earlier work, the general local solutions were expressed in terms of two locally holomorphic functions $\cA_\pm$ on $\Sigma$ and global solutions were   sketched when $\Sigma$ is a disk.  In the present paper, the physical  regularity conditions on the supergravity fields required for global solutions are implemented on $\cA_\pm$ for arbitrary $\Sigma$. Global solutions exist only when $\Sigma$ has a non-empty boundary $\p \Sigma$. The differentials $\p \cA_\pm$ are allowed to have poles only on $\p \Sigma$ and each pole corresponds to a semi-infinite $(p,q)$ five-brane.  
The construction for the disk is  carried out in detail and the conditions for the existence of global solutions are articulated for surfaces with more than one boundary and higher genus.

\end{abstract}
\end{center}

\baselineskip=15pt
\setcounter{equation}{0}
\setcounter{footnote}{0}

\newpage
\tableofcontents
\newpage

\section{Introduction}
\setcounter{equation}{0}
\label{sec:1}

In a recent paper \cite{D'Hoker:2016rdq}, we solved the BPS equations of Type IIB supergravity  for configurations with 16 residual supersymmetries on a space-time of the form $AdS_6 \times S^2$ warped over a Riemann surface $\Sigma$. The isometry algebra $SO(2,5) \times SO(3)$ of this space-time is uniquely completed into the exceptional Lie superalgebra $F(4)$ with 16 fermionic generators \cite{Kac:1977em,Shnider:1988wh}. In the present paper we shall impose on the local solutions of \cite{D'Hoker:2016rdq} the positivity and regularity conditions needed to obtain physically sensible supergravity fields and we shall construct the corresponding general global solutions. The first part of the present paper provides detailed derivations and discussions of  results which were described in a short note  \cite{DHoker:2016ysh}.

\sm

The motivation for this work stems from the study of five-dimensional supersymmetric quantum field theories with the help of holographic methods. Minimal Poincar\'e supersymmetry in five dimensions has eight supersymmetry generators and $SU(2)$ R-symmetry. In their conformal phase, five-dimensional supersymmetric field theories  will have $SO(2,5)$ conformal symmetry, $SU(2)$ R-symmetry and sixteen supersymmetries and therefore be invariant under the $F(4)$ superconformal algebra. Unlike in space-time dimensions three, four, and six, where the maximal superconformal algebra has thirty two supersymmetries, in five dimensions the superconformal algebra $F(4)$ is unique and has sixteen supersymmetries.

\sm

The existence of physically consistent interacting five-dimensional gauge theories is far from obvious since standard perturbative methods show them to be non-renormalizable. Arguments for the existence of non-Abelian gauge theories were advanced in \cite{Seiberg:1996bd} and in \cite{Intriligator:1997pq} and drew heavily on the structure imposed by five-dimensional Poincar\'e supersymmetry  and the dynamics on the Coulomb branch.  It was further argued that the theory in the non-conformal Coulomb phase can flow to a strongly coupled fixed point with superconformal symmetry.  The five-dimensional gauge theories  of \cite{Seiberg:1996bd,Intriligator:1997pq} can have gauge groups  $SU(N)$ or $USp(N)$ for arbitrary $N$, as well as exceptional gauge groups.  Consequently, one should expect that, in the large $N$  limit, the theories with gauge groups $SU(N)$ and $USp(N)$ will admit weakly coupled $AdS_6$ gravity duals.   The explicit construction of such gravity duals is the goal of  \cite{D'Hoker:2016rdq} as 
well as of the present paper.

\sm

Several brane realizations of five-dimensional gauge theories with eight Poincar\'e supersymmetries have appeared in the string theory literature. In  \cite{Seiberg:1996bd, Brandhuber:1999np,Bergman:2012kr}  Type IIA supergravity solutions were constructed using $N$ D4-branes near a stack of D8/O8 branes.  These solutions suffer from singularities associated with the presence of the D8/O8 branes. 

\sm

In \cite{Aharony:1997ju,Aharony:1997bh,DeWolfe:1999hj}   a large class of five-dimensional theories was constructed in Type IIB string theory using brane webs. The five-dimensional field theories live on D5 branes which are of finite extent in one dimension and are embedded in a web of semi-infinite $(p,q)$ five-branes as well as optional seven-branes. The brane webs realize the moduli space and the matter content  of such theories geometrically, generalizing the construction of three-dimensional theories \cite{Hanany:1996ie}  to five dimensions.

\sm

The solutions constructed in \cite{D'Hoker:2016rdq} are local in the sense that all BPS conditions, Bianchi identities and field equations are satisfied point-wise by the choice of two locally holomorphic functions ${\cal A}_\pm$ on a two-dimensional Riemann surface $\Sigma$ whose global structure is as yet undetermined.\footnote{Previous attempts to construct warped $AdS_6$ solutions have appeared in \cite{Apruzzi:2014qva,Kim:2015hya,Kim:2016rhs}, but the partial differential equations derived from the BPS conditions have remained unsolved in these papers.} In  \cite{D'Hoker:2016rdq}, we followed the strategy developed some time ago  in  \cite{D'Hoker:2007xy,D'Hoker:2007fq,D'Hoker:2008wc}, and utilized the integrable structure of the reduced BPS equations \cite{DHoker:2014bft} along with the holomorphic structure of the Riemann surface $\Sigma$  to its fullest. These techniques have generated the complete local solution to the BPS equations. 

\sm

The local solutions of \cite{D'Hoker:2016rdq} need to be supplemented by positivity and regularity conditions in order to produce genuine global supergravity solutions with physically sensible behavior of all supergravity fields. These positivity and regularity conditions are as follows.  First, all supergravity fields must satisfy the reality conditions appropriate for Type IIB, such as for example the space-time metric and the dilaton field must be real. Second, the space-time metric must have the correct ten-dimensional Minkowski signature. Third, the space-time metric must be geodesically complete. Fourth, any singularities that might occur in the solutions, such as curvature singularities or  a divergent  string coupling, must be associated with physically sensible sources, such as BPS branes already present   in Type IIB.

\sm

The main new result of the present paper is the complete construction of  global solutions which satisfy all the positivity and physical regularity conditions enumerated above.  Our solutions evade a recently obtained theorem \cite{Gutowski:2017edr} stating that no supersymmetric warped $AdS_6$ solutions exist in Type IIB supergravity, since our solutions are warped over a non-compact  Riemann surface which  violates one of the assumptions of the no-go theorem of \cite{Gutowski:2017edr}. We shall show that the resulting global solutions provide viable candidates for holographic duals to five-dimensional superconformal field theories in the large-$N$ limit. 

\sm

The organization of this paper is as follows. In section \ref{sec:2}, we present a brief review of the local supergravity solutions in terms of holomorphic functions obtained in \cite{D'Hoker:2016rdq}. We articulate the positivity and regularity conditions in quantitative terms, formulate our strategy for the construction of global solutions, and show that no physically sensible solutions exist unless $\Sigma$ has a non-empty boundary. In section \ref{sec:3}, we present a new class of  global solutions for which the Riemann surface $\Sigma$ is the upper half plane and which are parameterized by holomorphic functions with simple poles on the boundary of $\Sigma$. For these  general solutions we provide a detailed counting of the number of free moduli of the space of solutions. In section \ref{sec:examples} we present explicit examples with three and four poles and discuss natural connections to $(p,q)$ 5-brane webs.  In section \ref{sec:annulus} we extend the general construction to the case where $\Sigma$ is 
 an annulus, and in section \ref{sec:generalRiemann} to the case where $\Sigma$ has arbitrary genus and an arbitrary non-zero number of boundary components.  In the case of more than one boundary component some global regularity conditions remain to be imposed on the solutions. Finally, in section \ref{sec:6}, we conclude with a  discussion of our results and present some directions for  future  research.

\section{ Local solutions and strategy for global solutions }
\setcounter{equation}{0}
\label{sec:2}

In this section, we begin by summarizing from \cite{D'Hoker:2016rdq} the structure of the local solutions to the BPS equations in Type IIB supergravity with 16 supersymmetries and $SO(2,5) \times SO(3)$ isometry on a space-time of the form $AdS_6 \times S^2$ warped over a Riemann surface $\Sigma$. We then review the reality and positivity conditions required on the physical supergravity fields, and impose mild regularity conditions. Finally, we discuss our strategy for the construction of global solutions starting from an analogy with two-dimensional electrostatics.

\subsection{Supergravity fields in terms of holomorphic functions}

The $SO(2,5) \times SO(3)$ isometry dictates the Ansatz for the bosonic supergravity fields, while the fermionic fields vanish. The five-form field strength vanishes by symmetry, while the metric and the complex three-form flux fields are as follows, 
\bea
\label{2a1}
ds^2 & =  & f_6^2 \, ds^2 _{AdS_6} + f_2 ^2 \, ds^2 _{S^2} + 4\rho^2 | dw |^2
\no \\
  F_{(3)} & = & d \cC  \wedge {\rm vol}_{S^2}
\eea
Here,  $f_6, f_2, \rho, \, \cC$, and the dilaton-axion scalar field $B$ are functions of $\Sigma$ only. The metrics
$ds^2 _{AdS_6}$ and $ds^2 _{S^2}$ are the $SO(2,5)$ and $ SO(3)$ invariant metrics with unit radius respectively on $AdS_6$ and $S^2$, and  ${\rm vol} _{S^2}$ is the corresponding volume form on $S^2$. Finally, we have introduced local complex coordinates $w, \bar w$ in terms of which the metric on $\Sigma$ is conformally flat.

\sm

The supergravity fields for the local solutions were obtained in  \cite{D'Hoker:2016rdq}, and expressed in terms of locally holomorphic functions $\cA_\pm$ on the Riemann surface $\Sigma$. Of fundamental importance to the structure of the local solutions and, as we shall see shortly also to the structure of the global solutions, are the following real functions, (more properly, one should think of $\kappa ^2$ as a $(1,1)$ differential on $\Sigma$), 
\bea
\label{2a2}
\kappa ^2 & = & - |\p_w \cA_+|^2 + |\p_w \cA_-|^2
\no \\
\cG & = & |\cA_+|^2 - |\cA_-|^2 + \cB + \bar \cB
\eea
where the locally holomorphic function $\cB$ is defined up to an additive constant by the relation, 
\bea
\label{2a3}
\p_w \cB = \cA_+ \p_w \cA_- - \cA_- \p_w \cA_+
\eea
In Einstein frame, the metric functions $f_6^2$, $f_2^2$ and $\rho^2$ are given by,
\bea
\label{2a4}
 f_6^2=\frac{c_6^2 \, \kappa^2 (1+R) }{\rho^2 \, (1-R)}
\hskip 0.6in
 f_2^2 =\frac{c_6^2 \, \kappa^2 (1-R)}{9 \, \rho^2 (1+R)}
\hskip 0.6in
 \rho^2= {  c_6(R+R^2)^\half \, (\kappa ^2)^{{ 3 \over 2}}  \over |\partial_w \cG| \, (1-R)^{3 \over 2} }
\eea
where the real positive function $R$ is defined by the following quadratic equation,
\bea
\label{2a5}
R+\frac{1}{R} =  2+ { 6 \, \kappa^2 \, \cG \over |\partial_w\cG|^2}
\eea
The axion-dilaton field $B$  is given by,
\bea 
\label{2a6}
B ={\p_w \cA_+ \,  \partial_{\bar w} \cG - R \, \p_{\bar w} \bar \cA_-   \partial_w \cG \over 
R \, \p_{\bar w}  \bar \cA_+ \partial_w \cG - \p_w \cA_- \partial_{\bar w}  \cG}
\eea
while the flux potential function $\cC$ for the three-form field strength $F_{(3)}$  takes the form, 
\bea
\label{2a7}
 \cC = \frac{4 i c_6}{9}\left (  
{\p_{\bar w} \bar \cA_- \, \p_w \cG \over \kappa ^2} 
- 2 R \, {  \partial_w \cG \, \p_{\bar w} \bar \cA_- +  \partial_{\bar w}  \cG \, \p_w \cA_+ \over (R+1)^2 \, \kappa^2 }  
 - \bar  \cA_- - 2 \cA_+ \right )
\eea
Residual gauge transformations, which act on $\cC$ by a constant shift, leave $F_{(3)}$ invariant. 
The constant $c_6$ can be absorbed into a rescaling of $\cA_\pm$ and we will therefore set it to one in the following.

\subsection{\texorpdfstring{$SU(1,1)$}{SU(1,1)} symmetry and monodromy}
\label{sec:5-3}

Under the $SU(1,1)$ symmetry of Type IIB supergravity, the metric in the Einstein frame and the five-form field strength are invariant while the axion-dilaton field $B$ and the three-form field strength transform non-trivially. Upon reduction to the $AdS_6 \times S^2 \times \Sigma$ Ansatz, the metric factors $f_6^2, f_2^2, \rho^2$ are invariant under $SU(1,1)$, the five-form field strength is zero which is $SU(1,1)$-invariant, while $B$ and $\cC$ transform as follows,
\bea
 B & \rightarrow & B'=\frac{uB+v}{\bar v B+\bar u}
 \hskip 1.9in u,v \in \CC
 \no \\
 \cC & \rightarrow & \cC ' = u \, \cC +v\, \bar \cC + \hbox{constant}
  \hskip 1in |u|^2 - |v|^2=1
\eea
where the additive constant in $\cC'$ allows for a gauge transformation.

\sm

Correspondingly, the building blocks $\kappa ^2$ and $\cG$ of the metric factors must be invariant under $SU(1,1)$. The invariance of $\kappa ^2$ combined with the transformation rule of $B$ dictate the transformation rule for the differentials $\p_w \cA_\pm$, 
\bea
\p_w \cA_+ & \to & + u \, \p_w \cA_+ - v \, \p_w \cA_- 
\no \\
\p_w \cA_- & \to & - \bar v \, \p_w \cA_+ + \bar u \, \p_w \cA_- 
\eea
Integration of the differentials $\p_w \cA_\pm$ and $\p_w \cB$  gives the transformation rule of $\cA_\pm$ and $\cB$, 
\bea
\label{Atrans}
\cA_+ & \to & \cA_+'=  + u \cA_+ - v \cA_- + a_+ 
\no \\
\cA_- & \to & \cA_-' =   - \bar v \cA_+ + \bar u \cA_- + a_-
\no \\
\cB ~ \, & \to & \, \cB ' ~ = ~ \cB + a_+ \cA_- - a_- \cA_+ + b_0
\eea
where $a_\pm$ and $b_0$ are complex constants. The function   $\cG$ is invariant provided we impose, 
\bea
\label{moncon}
a_+ - \bar a_- = b_0 + \bar b_0 =0
\eea
The combined transformations on $\cA_\pm$ form a group $\mG = SU(1,1) \times \CC$, whose action on the differentials $\p_w \cA_\pm$ reduces to the action of the subgroup $SU(1,1)$.

\sm

When $\Sigma$ has a non-trivial fundamental group  $\pi_1(\Sigma)$ the functions $\cA_\pm$ are allowed to have $\mG$-valued monodromy around non-trivial closed loops on $\Sigma$.  The monodromy on the differentials $\p_w \cA_\pm$ is simplest and given by a representation of $\pi _1 (\Sigma) \to SU(1,1)$. Mathematically, this representation could be arbitrary. Physically, for a supergravity embedded into Type IIB string theory, however, the monodromies generated by parabolic elements acting on the axion field by a shift correspond to the presence of D7 branes, and have a well-known interpretation. It is also possible for the differentials $\p_w \cA_\pm$ to have a branch point in the interior of $\Sigma$ around which axion monodromy is allowed. 

\sm

In the present paper, we shall consider solutions only for which the dilaton/axion field is single-valued and no D7 branes are present. We plan to investigate solutions with nontrivial monodromies in the future.

\subsection{Positivity, regularity  and boundary conditions}
\label{sec:22}

Reality and positivity of the function $R$ in (\ref{2a5}) require $\kappa^2$ and $\cG$ to have the same sign. Reality of $\rho^2$ and positivity of the metric factors require $\kappa^2$  to have the same sign as $1-R$. Without loss of generality, we may choose the branch $0\leq R \leq 1$ in (\ref{2a5}) resulting in the following positivity conditions in the interior of the Riemann surface $\Sigma$,
\bea
\label{2b1}
\kappa ^2  >  0
\hskip 1in
\cG  >  0
\eea
These necessary conditions are sufficient to guarantee that $f_6^2, f_2^2 , \rho^2$ are real and positive.

\sm

The topology of  $\Sigma$, namely its genus and number of its boundary components, remains to be determined by global considerations.  The boundary $\p \Sigma$ of the surface $\Sigma$ does not correspond to a boundary of the physical space-time manifold of the supergravity solution. Instead,  $\p \Sigma$ corresponds to the shrinking of  $S^2$ as part of the slicing of a regular $S^3$ cycle, thereby ensuring the geodesic completeness of the full space-time geometry. As such, we may characterize $\p \Sigma$ by $f_2^2=0$ and $f_6^2 \not= 0$ or, equivalently, by the following conditions,  
\bea
\label{2b2}
\kappa ^2 \Big |_{ \p \Sigma} =0 \hskip 1in  \cG \Big |_{\p \Sigma} =0
\eea
We shall now prove the following  {\sl Proposition},
\begin{enumerate}
\itemsep =0in
\item {\sl There are no regular solutions satisfying (\ref{2b1})  when $\Sigma$ is compact without boundary.}
\item {\sl For an arbitrary Riemann surface $\Sigma$, with non-empty boundary $\p \Sigma$, the condition $\cG | _{\p \Sigma } =0$ implies $ \cG >0$ everywhere in the interior of $\Sigma$ provided $\kappa ^2 >0$. }
\end{enumerate}
A practical implication of this proposition is that it will suffice to establish the property $\cG=0$ on $\p \Sigma$ in order to guarantee the condition $\cG>0$ of (\ref{2b1}) in the interior of $\Sigma$.

\sm

To prove the {\sl Proposition} we begin by evaluating the mixed derivative of $\cG$ in the second line of (\ref{2a2}) in terms of $\kappa ^2$ with the help of the first line of (\ref{2a2}),
\bea
\label{2b3}
\p_{\bar w} \p_w \cG = - \kappa ^2
\eea
To prove item 1 of the {\sl Proposition}, we assume that $\Sigma$ is  compact without boundary and that $\kappa ^2$ and $\cG$ are sufficiently regular.  Integrating both sides of  (\ref{2b3})  over $\Sigma$, the left side vanishes since $\Sigma$ has no boundary, yet the right side must be strictly negative since $\kappa^2 > 0$ in the interior of $\Sigma$.  Therefore, (\ref{2b3})  has no solutions consistent with the requirements of (\ref{2b1}). This statement is in agreement with results found in a recent paper  \cite{Gutowski:2017edr}   stating that there are no supersymmetric $AdS_6$ solutions which are warped over a compact space.

\sm

To prove item 2 of the {\sl Proposition}, we solve equation (\ref{2b3}) along with the boundary condition on $\cG$ in (\ref{2b2}) to obtain the following integral equation,\footnote{Throughout, we shall use the following normalizations $d^2 z = { i \over 2} dz \wedge d\bar z$ and $\int _\Sigma d^2 z \, \delta (z,w)=1$.} 
\bea
\label{2b4}
\cG (w) = { 1 \over \pi} \int _\Sigma d^2 z \, G(w,z) \, \kappa ^2 (z)
\eea
Here,  $G(w,z)$ is the scalar Green function on $\Sigma$, which is symmetric $G(z,w)=G(w,z)$ and vanishes on the boundary $\p \Sigma$,
\bea
\label{2b5}
\p_{\bar w} \p_w \, G(w,z) & = & -\pi \delta (w,z) 
\no \\
G(w,z) \Big | _{w \in \p \Sigma} & = & 0 
\eea
For any two points $w,z$ in the interior of $\Sigma$, the function $G(w,z)$ is strictly positive, as we shall review below, and therefore $\cG>0$  in view of $\kappa ^2 >0$ in the interior of $\Sigma$. 

\sm

To prove positivity of the Green function $G(w,z)$ for arbitrary points $w,z$ in the interior of $\Sigma$, we fix $z$ and define $\Sigma _\ep$ to be the Riemann surface $\Sigma$ minus the disk of radius $\ep>0$ centered at the point $z$. In the interior of $ \Sigma_\ep$ the function $G(w,z)$ is real harmonic in the variable $w$. By the minimum-maximum principle for harmonic functions, $G(w,z)$ takes its minimum and maximum values on the boundary of $\Sigma _\ep$. Since $G$ vanishes on $\p \Sigma$ by (\ref{2b2}) and, for sufficiently small $\ep$ is positive on the circle of radius $\ep$ centered at $z$, it follows that $G$ must be strictly positive  in the interior of $\Sigma _\ep$ and thus in the interior of $\Sigma$ as we take the limit $\ep \to 0$. 

\newpage

\subsection{Strategy for global solutions without monodromy}
\label{sec:23}

The construction of global supergravity solutions has now been reduced to solving for two locally holomorphic functions $\cA_\pm$ on a Riemann surface $\Sigma$ with non-empty boundary $\p \Sigma$, subject to the positivity condition $\kappa ^2 >0$ in the interior of $\Sigma$ of (\ref{2b1}) and the boundary conditions $\kappa ^2 = \cG=0$ on $\p \Sigma$. The positivity condition $\cG>0$ in the interior of $\Sigma$ in (\ref{2b1}) is then automatic by the {\sl Proposition}  of the preceding subsection.   In the present subsection, we shall present our general strategy for solving these conditions, with the assumption that the differentials $\p_w \cA_\pm$ are meromorphic on $\Sigma$, and in the absence of $SU(1,1)$ monodromy. 

\sm

We fix a Riemann surface $\Sigma$ with non-empty boundary $\p \Sigma$. We shall use an analogy with two-dimensional electrostatics to solve for the meromorphic differentials $\p_w \cA_\pm$. We begin by introducing the meromorphic function $\lambda$ on $\Sigma$, 
\bea
\label{2c1}
\lambda (w) = { \p _w \cA_+ (w) \over \p_w \cA_- (w)} 
\eea
Our strategy will consist of the following steps,
\begin{enumerate}
\itemsep=-0.05in
\item  solving for the  function $\lambda$ using the properties of $\kappa ^2$ and an electrostatics analogy;
\item constructing  the  meromorphic differentials $\p_w \cA_\pm$ and their integrals $\cA_\pm$;
\item enforcing the condition $\cG=0$ on the boundary of $ \Sigma$.
\end{enumerate}

\subsubsection{Constructing the function \texorpdfstring{$\lambda$}{lambda} via an electrostatics analogy}

We begin by constructing the meromorphic function $\lambda$. The further  properties of $\lambda$ follow from its definition (\ref{2c1}) and the properties of $\kappa ^2$ in (\ref{2b1}) and (\ref{2b2}), and we have,
\begin{itemize}
\itemsep=-0.05in
\item  $|\lambda |<1$ in the interior of $\Sigma$ in view of $\kappa ^2 >0$;
\item $|\lambda |=1$ on the boundary of $\Sigma$ in view of $\kappa ^2=0$ on $\p \Sigma$;
\item  $\lambda$ is holomorphic in $\Sigma$, since it is meromorphic and bounded in $\Sigma$.
\end{itemize}
To solve for $\lambda$ subject to these properties, we use an electrostatics analogy. We view the function  
$ - \ln |\lambda|^2$  as an electrostatic potential for an array of $N$ point-like charges.  This potential is real, and the properties of $\lambda$ imply that $ - \ln |\lambda|^2$ is locally harmonic, strictly positive in the interior of $\Sigma$, and zero on the boundary $\p \Sigma$. The potential $ - \ln |\lambda|^2$ must have singularities, since otherwise it would  vanish throughout. Its  singularities are located at the zeros of $\lambda$ in the interior of $\Sigma$ and are logarithmic. Combining these results,  we find, 
 \bea
 \label{2c2}
- \ln |\lambda (w) |^2 = \sum _{n=1}^N q_n \, G(w, s_n)
\eea
where $G$ is the scalar Green function defined in  (\ref{2b5}).  The charges $q_n$  are real and strictly positive for all $n =1, \cdots, N$. Since we have $|\lambda | =1$ on $\p \Sigma$, the points $s_n$  must lie in the interior of $\Sigma$, away from the boundary.  

\sm

Conversely, when the charges $q_n$ are all positive, the potential  $ - \ln |\lambda |^2$ constructed in (\ref{2c2})  is positive throughout the interior of $\Sigma$. This result may be established by using the positivity  of the scalar Green function $G(w,z)$, already proven at the end of subsection \ref{sec:22}. To construct a single-valued $\lambda (w)$ from (\ref{2c2}) further requires the charges $q_n$ to be integers. The holomorphic function $\lambda (w) $ may be obtained  by holomorphically splitting (\ref{2c2}), and the result is unique up to an arbitrary phase factor. The precise expressions will be given in the sequel when $\Sigma$ is the upper half plane, an annulus, or a Riemann surface of higher genus.

\subsubsection{Constructing the differentials \texorpdfstring{$\p_w \cA_\pm$}{dA}}
\label{sec:232}

To construct the differentials $\p_w \cA_\pm$ we decompose  $\lambda$ into a ratio of holomorphic  functions (or  holomorphic differentials of equal weight) $\lambda _\pm$,
\bea
\label{2c3}
\lambda (w) = { \lambda _+ (w) \over \lambda _- (w)}
\eea
The functions $\lambda _\pm$ are clearly not unique, but we may impose the irreducibility condition that  $\lambda _-$ has no zeros in the interior of $\Sigma$ or on its boundary. In that case, the zeros of $\lambda _+$ coincide with the zeros of $\lambda$ and are given by the points $s_n$ of (\ref{2c2}).  Choosing local complex coordinates $w, \bar w$ near a given boundary component in which the boundary is represented by a segment of the real line, we require $\lambda _\pm $ to obey the following conjugation property,
\bea
\label{2c4}
 \overline{\lambda _\pm (\bar w)} = \lambda _ \mp (w) 
\eea
which guarantees the relation $|\lambda (w) |=1$ when $w$ is real and on the boundary of $\Sigma$. The meromorphic differentials $\p_w \cA_\pm$ are then given as follows, 
\bea
\label{2c5}
\p_w \cA_\pm (w) = \lambda _\pm (w) \, \varphi (w)
\eea
where $\varphi (w)$ is a meromorphic function or differential on $\Sigma$.

\sm

The properties of $\p_w \cA_\pm$, $\kappa ^2$ and $\cG$ narrow down the  properties of $\varphi $.  The combinations $\lambda _\pm \varphi$ must be single-valued meromorphic 1-forms on $\Sigma$.
The form $\varphi$ can have neither zeros nor poles in the interior of $\Sigma$, since a zero would make  $\kappa ^2$  vanish  in contradiction with (\ref{2b1}), and a pole would  produce local monodromy in the supergravity fields. Thus, the zeros and poles of $\varphi$ must all lie on $ \p \Sigma$, so that  $\varphi$ is an imaginary function satisfying $\overline{\varphi (\bar w) } = - \varphi (w)$ in local complex coordinates where the local boundary corresponds to real $w$. Therefore we have the following conjugation property for the meromorphic differentials $\p _w \cA_\pm$, 
\bea
\label {2c6}
 \overline{ \p_{\bar w} \cA_\pm (\bar w) }  =  - \, \p_w \cA_\mp (w) 
\eea 
On an arbitrary Riemann surface $\Sigma$, with non-empty boundary $\p \Sigma$, the conjugation condition may be formulated globally in terms of the {\sl double surface} $\hat \Sigma$, equipped with an anti-conformal involution $\mI$, such that $\Sigma = \hat \Sigma / \mI$, and each point on the boundary $ \p \Sigma$ is mapped to itself under $\mI$. This formalism will be developed and used to construct $\p_w \cA_\pm$ in section \ref{sec:generalRiemann}.

\subsubsection{Enforcing the condition 
\texorpdfstring{$\cG=0$ on $\p \Sigma$}{G=0 on dSigma}}

To analyze the vanishing of $\cG$ on $\p \Sigma$, we locally integrate the conditions (\ref{2c6}). By choosing the integration constants, we may  impose the following conjugation property, 
\bea
\label {2c7}
 \overline{ \cA_\pm (\bar w ) }  =  - \, \cA_\mp (w) 
\eea 
While the differentials $\p_w \cA_\pm$ are meromorphic and have poles on $\p \Sigma$, their integrals $\cA_\pm$ will have logarithmic branch points on $\p \Sigma$ whose branch cuts must be chosen  consistently with the conjugation condition (\ref{2c7}).  

\sm

The condition (\ref{2c6}) suffices to guarantee that $\cG$ is constant along any line segment of an arbitrary  boundary component which is free of poles in $\varphi$.  To prove this statement, we consider the derivative tangent to the boundary component of the real function $\cG$. Since here we have parametrized the boundary component locally by the real part of $w$ we find, using (\ref{2a2}) and (\ref{2c6}),  
\bea
\p_w \cG + \p_{\bar w} \cG & = & \Big ( \cA_+ (w) - \overline{\cA_- (w)} \Big )
\Big (  \p_w \cA_- (w) - \p_{\bar w} \cA_-   (\bar w) \Big ) 
\no \\ && 
- \Big ( \cA_- (w) - \overline{\cA_+ (w)} \Big )
\Big (  \p_w \cA_+ (w) - \p_{\bar w} \cA_+   (\bar w) \Big ) 
\eea
Along a boundary component away from poles in $\p_w \cA_\pm$, the variable  $w$ is real and  the differences of the derivative terms vanish identically. Therefore $\p_w \cG+ \p_{\bar w} \cG$ vanishes and $\cG$ is constant on any such segment of an arbitrary boundary component.  To promote the piecewise constancy of $\cG$  to the vanishing of $\cG$ throughout the boundary will require restrictions  on the free parameters of the functions $\cA_\pm$. We shall establish later that there will be one real condition for each pole. When $\Sigma$ has several disconnected boundary components, there will be further conditions to ensure that $\cG=0$ along each boundary component separately.

\newpage

\section{Global solutions  for the upper half plane}
\setcounter{equation}{0}
\label{sec:3}

In this section, we shall construct explicitly the global solutions for which $\Sigma$ has genus zero and a single boundary component. In this case $\Sigma$ may be mapped to the unit disk whose boundary is the unit circle,  or to the upper half complex plane whose boundary is the real axis. Our construction will follow closely the strategy outlined in section \ref{sec:23}, and it will be convenient to identify $\Sigma$ with the upper half plane $\Sigma = \HH$ and $\p \Sigma = \RR$.

\subsection{The function \texorpdfstring{$\lambda$}{lambda}}

\begin{figure}
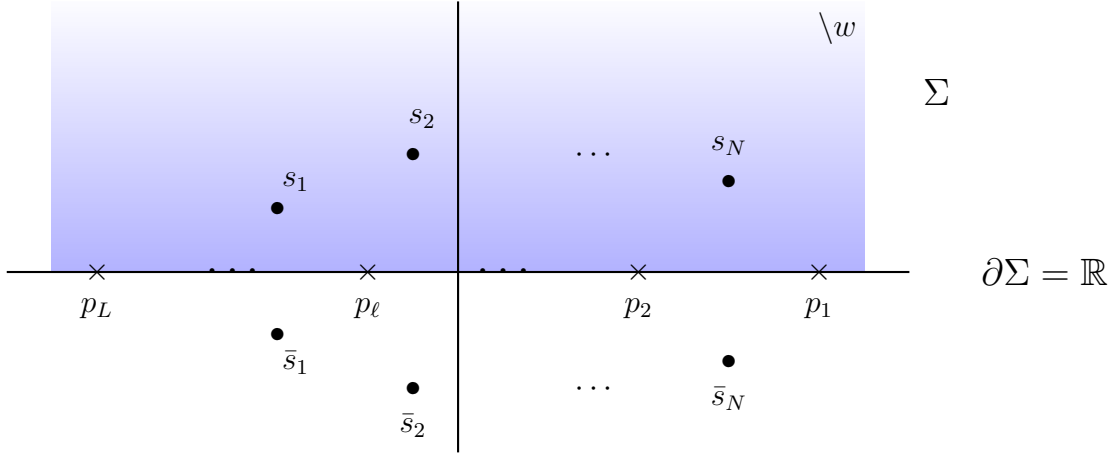

\begin{center}
\tikzpicture[scale=1.2]
\scope[xshift=-5cm,yshift=0cm]
\shade [ top color=blue! 1, bottom color=blue! 30] (0.5,0)  rectangle (9.5,3);
\draw [thick] (0,0) -- (10, 0) ;
\draw [thick] (5,-2) -- (5,3) ;
\draw (10.3,2) node{{\large $\Sigma$}};
\draw (11.5,0) node{{\large $\p \Sigma = \RR$}};
\draw (3,0.7) node{$\bullet$};
\draw (4.5,1.3) node{$\bullet$};
\draw (6.5,1.3) node{$\cdots$};
\draw (8,1) node{$\bullet$};
\draw (3,-0.7) node{$\bullet$};
\draw (4.5,-1.3) node{$\bullet$};
\draw (6.5,-1.3) node{$\cdots$};
\draw (8,-1) node{$\bullet$};
\draw (3.2,1) node{$s_1$};
\draw (4.6,1.7) node{$s_2$};
\draw (8,1.4) node{$s_N$};
\draw (3.2,-1) node{$\bar s_1$};
\draw (4.5,-1.7) node{$\bar s_2$};
\draw (8,-1.4) node{$\bar s_N$};
\draw (9.2, 2.7) node{$\backslash w$};
\draw (1,0) node{$\times$};
\draw (1,-0.4) node{$p_L$};
\draw (4,0) node{$\times$};
\draw (4,-0.4) node{$p_\ell$};
\draw (7,0) node{$\times$};
\draw (7,-0.4) node{$p_2$};
\draw (9,0) node{$\times$};
\draw (9,-0.4) node{$p_1$};
\draw (2.5,0) node{\Large $\cdots$};
\draw (5.5,0) node{\Large $\cdots$};
\endscope
\endtikzpicture
\caption{The upper-half plane $\Sigma= \HH$; its boundary $\p \Sigma = \RR$; an array of zeros of the function $\lambda$  at the points $s_n$ and poles at $\bar s_n$, with $n=1,\cdots, N$ and $\Im (s_n) >0$;  and an array of poles of $\p_w \cA_\pm$ at the points $p_\ell \in \RR$ with $\ell =1, \cdots, L$.\label{fig1}}
\end{center}
\end{figure}

The starting point is the general formula for $- \ln |\lambda|^2$ given in (\ref{2c2}), for an array of $N$ positive integer charges $q_n$ located at points $s_n$ in the upper half plane $\HH$ with $n=1, \cdots, N$. Without loss of generality, we  set $q_n=1$ for all $n$ and obtain higher integer charges by coalescing charges at different points $s_n$ if need be.  The set-up is represented schematically in Figure \ref{fig1}.  The  scalar Green function $G$ for $\HH$  is given by,
\bea
\label{3a1}
G(w,z) = - \ln \left | { w-z \over w- \bar z} \right |^2
\eea
The resulting formula for $ \lambda$ is,
\bea
\label{3a2}
\lambda (w)  = \lambda _0^2 \, \prod _{n=1}^N   { w-s_n \over w- \bar s_n} 
\hskip 1in \Im (s_n) >0
\eea
Here,  $\lambda_0$ is an arbitrary phase factor, which may be viewed as effecting an $SU(1,1)$ transformation $\cA_+ \to \lambda _0 \cA_+$ and $\cA_- \to \bar \lambda _0 \cA_-$ on the solution with $\lambda _0=1$. 
Following the process leading to (\ref{2c3}), we express the  function $\lambda$, which is holomorphic on $\HH$ and meromorphic in $\CC$, as the ratio of  functions $\lambda _\pm$ which are both holomorphic in $\CC$, and which are complex conjugate functions of one another, 
\bea
\label{3a3}
\lambda (w) = { \lambda _+ (w) \over \lambda _- (w) } 
\hskip 1in
 \overline{ \lambda _\pm (\bar w) } = \lambda _\mp (w)
\eea
The irreducible solution for $\lambda _\pm$, in the sense of subsection \ref{sec:232},  is unique and given by,
\bea
\label{3a4}
\lambda _+ (w) = \lambda _0 \prod _{n=1}^N (w-s_n)
\hskip 0.6in
\lambda _- (w) = \bar \lambda _0 \prod _{n=1}^N (w-\bar s_n)
\eea
We note that  $\lambda $ has a smooth limit as any one of its complex zeros approaches the real line, $\Im (s_n) \to 0$, so that the condition $\Im (s_n) >0$ could actually be smoothly relaxed to $\Im (s_n)  \geq 0$, with the understanding that at least one of the zeros $s_n$ must satisfy $\Im (s_n) >0$.

\subsection{The differentials \texorpdfstring{$\p_w \cA_\pm$}{dA}}

The holomorphic functions $\lambda _\pm$ and the meromorphic differentials $\p_w \cA_\pm$ satisfy the relation, 
\bea
\label{3b1}
\lambda _+ (w) \, \p_w \cA_-(w)  =  \lambda _-(w) \, \p_w \cA_+ (w) 
\eea
Eliminating $\lambda _\pm$ between this relation and its complex conjugate  relation obtained from (\ref{3a3}),  requires  the doublet of  differentials $\p_w \overline{  \cA_\pm (\bar w) } $ to be proportional to the doublet of differentials $  \p_w \cA_\mp (w)  $ by a meromorphic  function we shall denote $\gamma (w)$. Involution of the complex conjugation operation then imposes a conjugation condition on $\gamma$, given as follows, 
\bea
\label{3b2}
\p_w \overline{  \cA_\pm (\bar w) } = \gamma (w) \, \p_w \cA_\mp (w)  
\hskip 1in 
\gamma (w) \, \overline{ \gamma (\bar w) } =1
\eea
To construct the differentials $\p_w \cA_\pm$ we analyze the zeros on both sides of (\ref{3b1}). 
Since $\lambda_-$ has no zeros in the upper half plane, the set of zeros of $\p_w \cA_+$ is the union of the set of zeros of $\lambda _+$ and the set of zeros of $\p_w \cA_-$ (counted with multiplicities). But if the differentials $\p_w \cA_\pm$ have a common zero, then $\kappa ^2=0$ must vanish there. Strict positivity of $\kappa ^2$ in the interior of $\Sigma$ prohibits such zeros. Thus, any zero common to  $\p_w \cA_\pm$ must be on the real line. 

\sm

All poles  must be common to $\p_w \cA_\pm$ since their ratio $\lambda$ is bounded in the upper half plane. A simple pole in the interior of the upper half plane will produce a logarithmic  branch cut in $\cA_\pm$. In appendix \ref{appenda} we give an argument that  no such poles occur and, by complex conjugation,  that no poles occur in the lower half plane as well. Thus, all poles in $\p_w \cA_\pm$ must be common and on the real axis. We shall denote the positions of the poles by $p_\ell \in \RR$ with $\ell=1, \cdots, L$, for an as yet undetermined number $L$. 

\sm

Putting all together, we find the following general expression  for the differentials $\p_w \cA_\pm$,
\bea
\label{3b3}
\p_w \cA_+ (w) & = & \omega _0 \, \lambda _0 \, 
\prod _{n=1}^N (w-s_n)  \prod _{\ell=1}^L { 1 \over (w-p_\ell)}
\no \\
\p_w \cA_- (w) & = & \omega _0 \, \bar \lambda _0 \,
\prod _{n=1}^N (w- \bar s_n)  \prod _{\ell=1}^L { 1 \over (w-p_\ell)}
\eea
The zeros $s_n$  satisfy $\Im (s_n) \geq 0$ for all $n=1,\cdots, N$ with at least one zero having strictly positive imaginary part;  the poles $p_\ell$  are real for all $\ell=1,\cdots, L$; and the constant $\omega _0$ is  complex.  
Regularity of the differentials $\p_w \cA_\pm$  at $w=\infty$ imposes the condition 
$\p_w \cA_\pm \sim w^{-2}$ as $w \to \infty$, which provides us with a relation between $L$ and $N$, given by, 
\bea
\label{3b4}
L= N+2 
\eea
The function $\gamma(w)$ of (\ref{3b2}) is then a constant  phase, given by $\gamma =   \bar \omega_0 / \omega _0$. In summary, the differentials $\p_w \cA_\pm$ given in (\ref{3b3}) automatically satisfy $\kappa ^2>0$ in the interior of $\Sigma$ and $\kappa ^2=0$ on $\p \Sigma = \RR$. It remains to enforce the condition $\cG=0$ on $\p \Sigma$. To compute $\cG$, we need $\cA_\pm$ and $\cB$ which we compute in the subsequent subsections.

\subsection{The functions \texorpdfstring{$\cA_\pm$}{A}}\label{sec33}

The functions $\cA_\pm$ may be obtained from the meromorphic differentials $\p_w \cA_\pm$ by integration. Performing this integration is greatly facilitated by expressing the meromorphic differentials of (\ref{3b3}) in terms of their partial fraction decomposition with residues $Z^\ell _\pm$, 
\bea
\label{3c1}
\p_w \cA_\pm (w) = \sum _{\ell=1}^L {  Z^\ell _\pm \over w-p_\ell}
\hskip 0.8in 
\overline{Z^\ell _\pm } = \gamma \, Z^\ell _\mp 
\hskip 0.8in
\sum _{\ell=1}^L Z^\ell _\pm =0
\eea
The residues $Z^\ell _\pm$ may be read off from the product formulas (\ref{3b3}) and are given by, 
\bea
\label{3c2}
Z^\ell _+ & = & \om_0 \lambda _0 \, \prod _{n=1}^N (p_\ell -s_n) \, \prod _{\ell' \not= \ell }^L { 1 \over (p_\ell -p_{\ell'} )}
\no \\  
Z^\ell _- & = & \om_0 \bar \lambda _0 \, \prod _{n=1}^N (p_\ell - \bar s_n) \, \prod _{\ell' \not = \ell}^L { 1 \over (p_\ell -p_{\ell '} )}
\eea
Since $\lambda$ must have at least one zero in the upper half plane we have $N \geq 1$ and the number of poles satisfies $L \geq 3$. We now integrate the differentials to get $\cA_\pm$, as follows,
\bea
\label{3c4}
\cA_\pm (w) = \cA_\pm ^0 + \sum _{\ell=1}^L  Z_\pm ^\ell \ln (w-p_\ell)
\eea
where $\cA_\pm^0$ are two complex integration constants. Throughout, we shall choose the branch cuts of the logarithm so that $\ln (z)$ is real for $z$ real and positive, and use the following continuation formula by circling the branch point $z=0$ in the upper half plane, 
\bea
\label{3f1}
\ln (-z) = \ln (z) + i \pi
\eea
As a result, the functions $\cA_\pm(w)$ have monodromy around each pole $p_\ell$, whose value is given by $i \pi Z^\ell _\pm $. In view of the third equation in (\ref{3c1}), there is no monodromy at $w=\infty$, consistent with treating infinity as a regular point.

\subsection{The functions  \texorpdfstring{$\cB$}{B} and \texorpdfstring{$\cG$}{G}}

Using (\ref{3c1}) and (\ref{3c4}), we integrate the definition for $\cB$ given in (\ref{2a3}), 
\bea
\label{3d1}
\cB (w) =  \cB^0 + \sum _{\ell=1}^L \Big (  \cA_+^0  Z_- ^\ell - \cA_- ^0 Z_+ ^\ell \Big ) \, \ln (w-p_\ell)
+ \sum _{\ell, \ell'=1}^L Z^{[\ell, \ell']} 
\int _ {w_0} ^w dz \, { \ln (z-p_\ell) \over z - p_{\ell'} }
\eea
where $\cB^0$ is an arbitrary  complex integration constant, $w_0$ is an arbitrary reference point, and we have introduced the following convenient abbreviation, 
\bea\label{zdefa}
Z^{[\ell, \ell']} =  Z_+ ^\ell Z_- ^{\ell'} - Z_+ ^{\ell'} Z_- ^\ell
\eea
A change in $w_0$ in (\ref{3d1}) may be compensated by a change in $\cB^0$, so that the parametrization provided above is subject to a redundancy which will be conveniently fixed shortly.
Substituting the expressions for $\cA_\pm$, $\bar \cA_\pm$,  $\cB$, and $\bar \cB$ into the formula for $\cG$ given in (\ref{2a2}), we obtain the following expression, 
\bea
\label{3d2}
\cG (w) & = & \cG^0
+  \sum _{\ell=1}^L \Big ( \bar \cA_+ ^0 Z_+ ^\ell - \bar \cA_-^0 Z_- ^\ell 
+  \cA_+^0  Z_- ^\ell - \cA_- ^0 Z_+ ^\ell \Big ) \Big (  \ln (w-p_\ell ) + \gamma \, \overline{\ln (w-p_\ell) }  \Big )
\no \\ &&
+ \sum _{\ell, \ell'=1}^L Z^{[\ell, \ell']}   \Bigg  \{ 
\gamma \, \ln (w-p_\ell) \, \overline{\ln ( w-p_{\ell'}) } 
\no \\ && \hskip 1in 
+ \int _ {w_0} ^w dz \, { \ln (z-p_\ell) \over z - p_{\ell'} } 
- \gamma ^2 \,  \overline{\int _ {w_0} ^w dz \, { \ln (z -p_\ell) \over z - p_{\ell'} } } \Bigg \}
\eea
where  $\cG^0$ is a real constant given by, 
\bea
\label{3d3}
\cG^0 = |\cA_+^0|^2 - |\cA_-^0|^2 + \cB^0 + \bar \cB^0
\eea
A change in $w_0$ will shift $\cG^0$.  The constants $w_0, \gamma$, $\cA_\pm ^0$ and $\cB^0$, as well as the positions of the zeros $s_n$ and the poles $p_\ell$ remain undetermined at this stage. The integrals are related to dilogarithms, and cannot be evaluated in terms of elementary functions.

\subsection{Necessary conditions for  vanishing \texorpdfstring{$\cG$ on $\p \Sigma$}{G on dSigma}}

We shall now enforce the condition $\cG=0$ on the boundary $\p \Sigma = \RR$, and derive the corresponding conditions on the zeros $s_n$, poles $p_\ell$ and other parameters of the solution. Since the integrals on the last line of (\ref{3d2}) are related to dilogarithms in $w$, the function $\cG$ can vanish  for all $w \in \RR$ only provided the contribution of the dilogarithms vanishes by itself. This requires $\gamma ^2=1$, given the branch cut choice we have made. Since $\gamma $ is then real, the contribution of the second line in (\ref{3d2}) vanishes  for $w\in \RR$ since the combination $Z^{[\ell, \ell']} $ is anti-symmetric in $\ell, \ell'$, while the logarithmic factor becomes symmetric in $\ell, \ell'$ for real $w$. Cancellation of the terms on the first line of (\ref{3d2})  for any $w \in \RR$ further requires,
\bea
\label{3e1}
\gamma =-1 
\eea
The conjugation conditions on the residues then reads as follows for all $\ell=1, \cdots, L$, 
\bea
\label{3e2}
\overline{Z^\ell _\pm } = - Z^\ell _\mp
\eea
and the combinations $Z^{[\ell, \ell']} $ are imaginary-valued. We shall require the following relation between the integration constants, 
\bea
\label{3e2a}
\bar \cA_+ ^0 + \cA_-^0 =0
\eea
Indeed, we may do so without loss of generality since the only supergravity field that involves the combinations $\bar \cA_+ ^0 + \cA_-^0$ is the flux field $\cC$, and the condition (\ref{3e2a})  simply amounts to a (constant) gauge choice for $\cC$. Having imposed (\ref{3e2a}) we see that the above conditions on $Z_\pm ^\ell$ precisely imply the conjugation relations of (\ref{2c7}) advocated on general grounds.

\subsection{Sufficient conditions for vanishing \texorpdfstring{$\cG$ on $\p \Sigma$}{G on dSigma}}

Due to the presence of branch points in $\cA_\pm$ on the real axis, the above conditions are necessary but not sufficient, however, as the monodromy makes the value of $\cG$ jump across the branch points. To impose the further condition that $\cG$ has zero discontinuity on the real axis across a pole $p_\ell$, we proceed as follows. 
Having chosen the branch cuts such that $\ln (z) $ is real for $z$ real and positive, we may order the real poles $p_\ell$ for $\ell=1, \cdots, L$  as follows, 
\bea
\label{3e3}
p_L < p_{L-1} < \cdots < p_2 < p_1 
\eea
We shall choose the arbitrary reference point $w_0$ to be real with $w_0 >p_1$. We then have  $\cG=0$ for $w$ real and $w > p_1$ provided we set $\cG^0 =0$.  

\sm

\begin{figure}
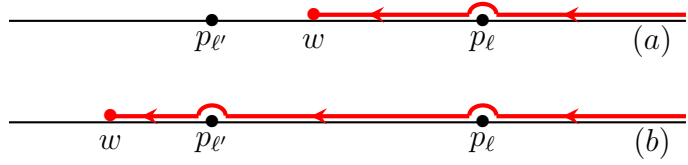

\begin{center}
\tikzpicture[scale=0.9]
\scope[xshift=-5cm,yshift=0cm]
\draw [thick] (0,0) -- (10,0);
\draw [ultra thick, red] (6.8,0.1) arc (180:0:0.2 and 0.15);
\draw [ultra thick, red] (2.8,0.1) arc (180:0:0.2 and 0.15);
\draw (3,0) node{$\bullet$};
\draw (7,0) node{$\bullet$};
\draw [red] (1.5,0.09) node{$\bullet$};
\draw (3,-0.3) node{$p_{\ell'} $};
\draw (7,-0.3) node{$p_\ell $};
\draw (1.5,-0.3) node{$w $};
\draw (9.5,-0.3) node{$(b)$};
\draw [ultra thick, red, directed] (10,0.09) -- (7.2, 0.09);
\draw [ultra thick, red, directed] (6.8,0.09) -- (3.2, 0.09);
\draw [ultra thick, red, directed] (2.8,0.09) -- (1.5, 0.09);
\draw [thick] (0,1.5) -- (10,1.5);
\draw [ultra thick, red] (6.8,1.6) arc (180:0:0.2 and 0.15);
\draw (3,1.5) node{$\bullet$};
\draw (7,1.5) node{$\bullet$};
\draw [red] (4.5,1.59) node{$\bullet$};
\draw (3,1.2) node{$p_{\ell'} $};
\draw (7,1.2) node{$p_\ell $};
\draw (4.5,1.2) node{$w $};
\draw (9.5,1.2) node{$(a)$};
\draw [ultra thick, red, directed] (10,1.59) -- (7.2, 1.59);
\draw [ultra thick, red, directed] (6.8,1.59) -- (4.5, 1.59);
\endscope
\endtikzpicture
\end{center}
\caption{Integration contour (a) when $w \in [p_{\ell'}, p_\ell]$ and (b) when $w \in [-\infty , p_{\ell'}]$.\label{fig2}}
\end{figure}

We make the convergence of the combined integrals in (\ref{3d2}) as $w \to \infty$ explicit by anti-symmetrizing in $\ell, \ell'$, and letting $w _0 \to \infty$. The only dependence of the resulting formula for $\cG$ on the integration constants $\cA_\pm ^0$ is through the combination $2\cA^0 = \cA_+ ^0 - \bar \cA^0_-$. The  resulting expression for $\cG$  is given by, 
\bea
\label{3e5}
\cG (w) & = &
  \sum _{\ell=1}^L \Big ( 2 \cA^0 Z_- ^\ell + 2 \bar \cA ^0  Z_+ ^\ell  \Big ) \Big (  \ln (w-p_\ell ) - \, \overline{\ln (w-p_\ell) }  \Big )
\\ &&
+ \sum _{1 \leq \ell < \ell' \leq L} Z^{[\ell, \ell']}  \Bigg  \{ 
 \ln (w-p_{\ell'}) \, \overline{\ln ( w-p_\ell) } - \ln (w-p_\ell) \, \overline{\ln ( w-p_{\ell'}) }  
\no \\ &&  \quad
+ \int _ \infty ^w dz \, \left ( { \ln (z -p_\ell) \over z - p_{\ell'} } -  { \ln (z -p_{\ell'}) \over z - p_\ell } \right )
-  \overline{\int _ \infty ^w dz \, \left (  { \ln (z -p_\ell) \over z  - p_{\ell'} }  - { \ln (z -p_{\ell'}) \over z - p_\ell }  \right ) }
\Bigg \} 
\no
\eea
To derive the discontinuity of $\cG$ across a pole we evaluate the function $\cG$ as $w$ approaches any point in the open interval $ ( p_{k+1}, p_k )$ from above. To do so, it will be convenient to partition the sums over $\ell, \ell'$ according to their position with respect to $k$. The integrals do not contribute for the partition $k < \ell < \ell'$ since the logarithms are manifestly real there. We evaluate the remaining integrals as a function of the position of $w$ on the real axis  with respect to the points $p_\ell$, and $p_{\ell'}$, using the  contours of Figure \ref{fig2}. The integral  evaluated with the help of contour (a) of Figure \ref{fig2} for $w $ in the open interval $[p_{\ell'}, p_\ell]$ is given by,
\bea
\label{3f3}
\Im  \int _\infty ^w dw' \, \left (  { \ln (z-p_\ell ) \over z - p_{\ell'}  } - { \ln (z - p_{\ell'}  ) \over z - p_{\ell} } \right )
= \pi \, \ln (w-p_{\ell'}) - 2 \pi \ln (p_\ell - p_{\ell'})
 \eea
The integral  evaluated with the help of contour (b) of Figure \ref{fig2} for $w $ in the open interval $[- \infty, p_{\ell'}]$ is given by, 
\bea
\label{3f5}
\Im  \int _\infty ^w dw' \, \left (  { \ln (z-p_\ell ) \over z - p_{\ell'}  } - { \ln (z - p_{\ell'}  ) \over z - p_{\ell} } \right )
= \pi \, \ln (p_{\ell'} - w) -  \pi \ln (p_\ell - w)
 \eea
Substituting the evaluated integrals into the formula for $\cG$, we obtain after the straightforward cancellation of all $w$-dependence, 
\bea
\label{3f6} 
 \sum _{\ell \leq k }    \Big (  \cA^0 Z_- ^\ell +  \bar \cA ^0 Z^\ell _+ \Big ) 
- \sum _{1 \leq \ell \leq k < \ell' \leq L} Z^{[\ell, \ell']}   \ln (p_\ell - p_{\ell'}) =0
\eea
valid for $k = 1, \cdots, L$.  Retaining the above condition for $k=1$ unchanged,  and subtracting from the condition for $k \geq 2$ the condition obtained by letting $k \to k-1$, we obtain a total of $L$ conditions, expressed below for $k=1,\cdots, L$, 
\bea
\label{3f7}
 \cA^0 Z_-^k + \bar \cA^0 Z_+^k 
+ \sum _{{1 \leq \ell \leq L \atop \ell \not= k }} Z^{[\ell, k]}  \ln |p_\ell - p_k| =0
\eea
Both terms are purely imaginary in view of the conjugation property (\ref{3e2}) and the definition (\ref{zdefa}). Only $L-1$ of these conditions are linearly independent since the sum over all $k$ of the above conditions vanishes by anti-symmetry in $\ell, k$ of the pre-factor $Z^{[\ell, k]} $. 
We also note that the final form of the conditions does not depend on a particular ordering of the poles on the real axis.
Subtracting the $k\rightarrow k-1$ condition localizes the calculation around the pole and, 
while convenient for the derivation of the conditions, the assumption on the ordering can be safely relaxed in (\ref{3f7}).

We close this subsection with the explicit expression for $\cG$ which vanishes on the boundary (away from the poles). Using (\ref{3f7}) to eliminate $\cA^0$ in (\ref{3e5}) converts the first line in (\ref{3e5}) to a double sum similar to the remaining terms. Anti-symmetrizing in $\ell$, $\ell^\prime$ yields,
\bea
\label{3f8}
 \cG (w) =
\sum _{\ell \not= \ell^\prime}^L  Z^{[ \ell , \ell^\prime]}  \Bigg  (
\half  \ln \left \{ \frac{w-p_{\ell'}}{(p_\ell-p_{\ell^\prime})^2} \right \} \, 
\overline{\ln \left \{ \frac{ w-p_\ell}{(p_\ell-p_{\ell^\prime})^2} \right \} }
+  \int _ \infty ^w \! dz \,  { \ln (z -p_\ell) \over z - p_{\ell'} }  - \hbox{c.c} \Bigg )
\quad
\eea
where the complex conjugation applies to the entire expression inside the bracket. Note that $\cG$ is real in view of the fact that $Z^{[\ell, \ell']}$ is purely imaginary.

\subsection{\texorpdfstring{$SL(2,\RR)$ acting on $\HH$}{SL(2,R) acting on H}}\label{sl2r}

$SL(2,\RR)$ acting on $\HH$ transforms $w, s_n$ and $p_\ell$ as follows $w \to w' = ( a w + b)/( c w + d)$  with $a,b,c,d \in \RR$ and $ad-bc =1$. The differentials $ dw \, \p_w \cA_\pm$ in  (\ref{3b3}) and the functions $\cA_\pm$ in (\ref{3c4}) are invariant under $SL(2,\RR)$  provided the data transform as follows,
\bea
\omega '_0 \lambda _0 ' & = & \omega _0 \lambda _0 \prod _{n=1}^N (c s_n +d) \prod _{\ell=1}^L (c p_\ell +d)^{-1}
\no \\
(\cA_\pm ^0)' & = & \cA^0_\pm + \sum _{\ell=1}^L Z_\pm ^\ell \ln (c p_\ell+d)
\eea
The residues $Z^\ell_\pm$ in  (\ref{3c1}) are then invariant, as are the  conditions of (\ref{3f7}). The $SL(2,\RR)$-invariance therefore allows us to fix the positions of three of the poles at will.

\subsection{Counting free parameters}\label{sec:counting}

We label the families of solutions by a positive integer $N$ which is the number of zeros $s_n$  of  $\partial_w \cA_+$ in the upper half plane.  The independent parameters of the solutions constructed in section \ref{sec:3} are then given by,
 \bea
\label{3g1}
s_n && 2N  \hbox{  real parameters }
\no \\
p_\ell && N+2  \hbox{  real parameters }
\no \\
\cA^0 && 2  \hbox{  real parameters }
\no \\
\lambda _0, \omega _0  & \hskip 1in & 2 \hbox{  real parameters }
\eea
  Note that, even though $\lambda _0$ and $\omega _0$ are complex, they satisfy $|\lambda _0|=1$ and $\bar \omega _0 = - \omega _0$ which accounts for them representing only one real parameter each. Furthermore, of the two complex constants $\cA_\pm ^0$, only the combination $2\cA^0= \cA_+^0 - \bar \cA_-^0$ enters into the metric factors and the dilaton-axion field, while in the flux potential, the combination $\cA_+^0 + \bar \cA_-^0$ produces merely a  gauge transformation of $\cC$. Therefore, of the constants  $\cA_\pm ^0$ only $\cA^0$ is a genuine physical parameter. Finally, in counting the number of poles $p_\ell$, we have made use of the relation $L=N+2$ of (\ref{3b4}) which relates the number of poles to the number of zeros. The total number of parameters is thus $3N+6=3L$. 

\sm

The parameters of (\ref{3g1}) must satisfy the $L-1=N+1$ regularity relations obtained in (\ref{3f7}), which reduces the number of free parameters by $N+1$. Furthermore, the $SL(2,\RR)$ automorphism group of the upper half plane maps to physically equivalent solutions and therefore further reduces the number of free physical parameters by 3. As a result the number of moduli of the global solutions is given by,
\bea
\label{3h1}
2N+2=2L-2 \quad \hbox{free real parameters}
\eea

\subsection{Asymptotic behavior near the poles}\label{sec:asympt}

The global solutions we have constructed above are regular everywhere in the interior of $\Sigma$, and on the boundary of $\Sigma$ away from the poles $p_\ell$. In this subsection, we shall derive the asymptotic behavior of the supergravity fields near the poles $p_\ell$, and then provide a physical interpretation of this behavior in terms of five-brane webs. The original Ansatz and solutions were formulated in Einstein frame metric, where $SU(1,1)$ symmetry is manifest, and we shall begin by obtaining the asymptotics in this frame. Subsequently to make contact with brane behavior, we shall translate the asymptotics to the string frame.

\subsubsection{Asymptotic behavior near the poles in Einstein frame}

We will use the expressions for the Einstein frame metric factors $f_6^2, f_2^2, \rho^2$ in terms of the function $\cG$, its derivatives, and the functions $ \kappa^2$ and $R$ given in (\ref{2a4}). To do so, we derive first the asymptotic behavior of $\kappa ^2, \cG$ and $R$ near a pole $p_m$, and parametrize the coordinate $w$ near a pole $p_m$ as follows,
\bea
w = p_m + r \, e^{ i \theta}
\eea
where $0 \leq \theta \leq \pi$ and $0 < r \ll |p_m - p_\ell|$ for all $\ell \not= m$. The expansion of $\p_w \cA_\pm$ near $p_m$ in terms of the parametrization by $r, \theta$ is then given by,
\bea
\p_w \cA_\pm (w) = Z^m _\pm \, { e^{-i \theta} \over r} 
+ \sum _{\ell  \not = m} { Z^{\ell} _\pm \over p_m - p_{\ell}}  +\cO(r)
\eea
which implies that the asymptotic expansion of $\kappa ^2$ is given by, 
\bea
\kappa ^2 = \kappa _m^2 \, { \sin \theta \over r} + \cO(r^0)
\hskip 1in 
\kappa _m^2 = 2i  \sum _{\ell  \not = m} { Z^m _- Z^{\ell} _+ - Z^m _+ Z^\ell _-  \over p_m - p_{\ell}} 
\eea
The coefficients $\kappa_m$ are real  in view of the fact that $\kappa ^2 >0$ for $r >0$ and $0 < \theta < \pi$, and may be taken to be positive.

\sm

To calculate the asymptotic form of $\cG$ we use (\ref{3f8}). To exhibit the asymptotics of the integral near the pole $p_m$ we use the fact that $\cG$ vanishes on the real line away from the pole and choose the point $u=p_m+r$ where $\cG(u)=0$ to write $\cG(w) = \cG (w)-\cG(u)$ and thus,
\bea
\cG(w)
&=&
\sum _{\ell < \ell'}^L  Z^{[\ell , \ell^\prime]}  \Bigg  \{
 \ln \frac{w-p_{\ell'}}{(p_\ell-p_{\ell^\prime})^2} \, \overline{\ln \frac{ w-p_\ell}{(p_\ell-p_{\ell^\prime})^2} } 
 -\ln \frac{u-p_{\ell'}}{(p_\ell-p_{\ell^\prime})^2} \, \overline{\ln \frac{ u -p_\ell}{(p_\ell-p_{\ell^\prime})^2} } 
 \no\\ &&
 \hspace*{30mm}
+ \int _ u^w dz \, \left ( { \ln (z -p_\ell) \over z - p_{\ell'} } -  { \ln (z -p_{\ell'}) \over z - p_\ell } \right )
- \hbox{c.c.}\Bigg \} 
\eea
The leading contributions are from the terms in the sum where either $\ell=m$ or $\ell^\prime=m$.
Since the entire contour for the integral is close to the pole, we can then expand the integrand appropriately, and find,
\bea
\cG (w) & = &  2 \kappa ^2_m \,   r \, | \ln r | \, \sin \theta  + \cO(r^2 \ln r) 
\no \\
\partial_w \cG (w) &=&i\kappa_m^2\ln r+\mathcal O(r\ln r)
\no\\
R (w) & = & 1- \sqrt{12} \, | \ln r |^{-\half} \sin \theta + \cO (r)
 \eea
The leading behavior of the metric functions near the pole $p_m$  is given by, 
 \bea\label{3g5}
 f_2^2 & \approx &    {2 \, \kappa _m \over  3^{{3 \over 4}}}    \, r^\half  | \ln r |^{-{1\over 4}} \sin^2 \theta  
  \no\\
 f_6^2 & \approx &  2\cdot 3^{1 \over 4} \,  \kappa _m \, r^\half  |\ln r |^{3\over 4}
 \no\\
 \rho^2 & \approx &   { \kappa _m  \over 2\cdot 3^{{3 \over 4}}}    \,  r^{-{3\over 2}} |\ln r |^{-{1\over 4}}   
 \eea
The dilaton $\phi$ and axion $\chi$ fields are obtained from (\ref{2a6}) with 
\bea
B= { 1+i\tau \over 1-i\tau}
\hskip 1in 
\tau=\chi+ie^{-2\phi}
\eea 
To  leading order, as $r \to 0$, $\phi$ and $\chi$ are given by,
\bea
e^{- 2 \phi} & \approx &  { \sqrt{3}  \, \kappa_m^2  \over |Z_+^m-Z_-^m|^2} \,  r \, |\ln r | ^{-\half}
\no \\
\chi & \approx & i \, { Z_-^m +  Z_+^m \over Z_-^m -  Z_+^m}
\label{3g6}
\eea
Clearly, the dilaton diverges at the pole $p_m$ provided $\kappa _m \not =0$.

\subsubsection{Asymptotic behavior near the poles in string frame}
\label{sec:asy}

Since the dilaton diverges near the poles, there is a qualitative difference in behavior between the metric in the Einstein frame (where duality is manifest) and the metric in the string frame (where the connection with D-branes is transparent).  We shall denote the metric factors in the string frame by $\tilde f_6^2, \tilde f_2^2$ and $\tilde \rho^2$. Their relation with the metric factors $f_6^2, f_2^2, \rho^2$ of the Einstein frame, and their asymptotic behavior near the pole $p_m$ are given by, 
 \bea
 \tilde f_2^2 = e^\phi \, f_2^2 & \approx &    {2\over 3}\,  \left |Z_+^m-Z_-^m \right | \, \sin^2\!\theta
  \no\\
 \tilde f_6^2 = e^\phi \, f_6^2 & \approx &  2 \,  \left |Z_+^m-Z_-^m \right | \cdot | \ln r|
 \no\\
 \tilde \rho^2 = e^\phi \, \rho^2 & \approx &   {1\over 6} \, \left |Z_+^m-Z_-^m \right | \, r^{-2}
\eea
Converting the metric $ds^2$ in the Einstein frame of the Ansatz of  (\ref{2a1}) to the metric $\widetilde d s^2$ in the string frame, and using the relation $dw d\bar w=dr^2+r^2d\theta^2$, the near-pole metric in string frame becomes,
\bea
\widetilde{ds}^2&\approx&
\frac{2}{3} \left |Z_+^m-Z_-^m \right |
\left ( 3 \,   |\ln r| ds^2_{AdS_6}+\frac{dr^2}{r^2}+d\theta^2+\sin^2\theta \,  ds^2_{S^2} \right )
\eea
The last two terms combine into the round metric of  a smooth $S^3$ sphere.
This produces a regular geometry in string frame. The curvature radius of the $AdS_6$ part diverges as the pole is approached, which turns it into a six-dimensional Minkowski space.\footnote{ To see this more explicitly, say we start with the  Poincar\'e patch and $ds^2_{AdS_6}=dy^2+e^{-2ky}\eta_{\mu\nu}dx^\mu dx^\nu$.
Setting $y=\tilde y/\sqrt{|\ln r|}$ and $x^\mu=\tilde x^\mu/\sqrt{|\ln r|}$ yields, up to terms which are subleading in the near-pole limit, $|\ln r|\,ds^2_{AdS_6} \approx d\tilde y^2+\eta_{\mu\nu}d\tilde x^\mu d\tilde x^\nu$,
and we indeed recover flat space. } Since flat space does not have an intrinsic scale, we can absorb overall constants into a rescaling of the coordinates and find the near-pole string-frame metric,
\bea
\label{3g7}
\widetilde{ds}^2&\approx&
ds^2_{\mathds{R}^{1,5}} + \frac{2}{3} \left |Z_+^m-Z_-^m \right |
\left ( \frac{dr^2}{r^2}+ds^2_{\mathrm{S}^3} \right )
\eea
It is important to note that while our supergravity solutions are singular at the poles $p_\ell$, we shall argue in the discussion below that these singularities have a clear and compelling physical interpretation in terms $(p,q)$ five-branes. This is to be contrasted with previously constructed Type IIB solutions  \cite{Apruzzi:2014qva,Kim:2015hya,Kim:2016rhs}.

\subsubsection{Asymptotic behavior of the flux form near the poles}

Finally, the shift in the flux potential $\cC$, as the pole at $p_m$ is being circumnavigated in the upper half plane  from the right to the left, namely from  $\theta=0$ to $\theta=\pi$, is determined by examining (\ref{2a7}). The variables $\partial_w \cA_\pm$,  $\kappa^2$, and $\cG$ do not shift. The shift in $\partial_w\cG$ cancels, and with $\cA_\pm\rightarrow \cA_\pm+i\pi Z_\pm^m$ we find the shift in $\cC$,
\bea
\label{3g8}
\Delta \cC  =  {4 \pi \over 3} \, Z_+^m
\eea
This expression is exact.  Evaluating the asymptotic behavior of the flux potential itself we find,
\bea
\label{3g9}
 d C_{(2)}&=&d\cC \wedge e^{78}=\frac{8}{3 }\;Z_+^m \, \sin^2\!\theta \, d\theta\wedge e^{78}
 +\mathcal \cO \Big( |\ln r|^{-3/2} dr /r \Big)
\eea
As $r\rightarrow 0$, $dr/r$ is regular and the leading neglected term vanishes. Finally, the factor $\sin^2\!\theta d\theta\wedge e^{78}$ gives the volume form of the S$^3$ sphere in (\ref{3g7}). Integrating (\ref{3g9}) from $\theta=0$ to $\theta=\pi$ reproduces (\ref{3g8}).

\subsection{Identification of the poles with \texorpdfstring{$(p,q)$}{(p,q)} 5-branes}
\label{sec:poles-branes}

With the asymptotic behavior of the solution and the shift in the flux potential around the poles in hand, 
we can now give a physical  interpretation of the poles. To this end we compare the near-pole behavior to the well-known supergravity solution for a $(p,q)$ 5-brane as given in \cite{Lu:1998vh}. With the string-frame metric (\ref{3g7}), axion and dilaton (\ref{3g6}) and complex three-form field strength (\ref{3g9}), we find an exact correspondence with the geometry near a $(p,q)$ 5-brane, with matching axion,  dilaton, and 3-form field strength. More precisely, matching the 3-form field strength given above with the results presented in \cite{Lu:1998vh} and denoting $p=q_1Q$ and $q=q_2 Q$ yields the identification,
\bea
p&=& + \frac{8 }{3} \,  \Re(Z_+^m)
\no\\
q&=&-\frac{8 }{3} \,  \Im(Z_+^m)
\label{3g0}
\eea
The near-pole behavior of our global solutions then matches the near-brane behavior of the $(p,q)$ 5-brane solution not only in scaling and  functional form of the various fields and involved geometry, but also in the overall coefficients.\footnote{For an explicit match, the radial coordinate $\rho$ of \cite{Lu:1998vh} should be identified with $r$ used here as $\rho=r|\ln r|^{-1/2}$. The leading-order expansion of (22), (23) (24) of \cite{Lu:1998vh} near the 5-brane  then matches the expansions of the corresponding quantities near the poles derived in sec.~\ref{sec:asympt}. The complex 3-form field strength $dC_{(2)}$ is related to the real $H^{(1)}$, $H^{(2)}$ of \cite{Lu:1998vh} by $dC_{(2)}=H^{(1)}-iH^{(2)}$. }
This demonstrates that the poles in our solutions correspond to external 5-branes as they typically 
appear in the brane web diagrams. As we have already advocated  briefly  in \cite{DHoker:2016ysh}, this evidence strongly suggests that the solutions correspond to fully localized intersections of $(p,q)$ 5-branes.

\newpage

\section{Example solutions and relation to 5-brane webs}
\setcounter{equation}{0}
\label{sec:examples}

In this section we provide explicit realizations and corresponding numerical plots for the solutions on the upper half plane in the case where the number of poles is three and four. These numerical results are used to illustrate the validity of the  general regularity arguments made in the previous sections and to  provide explicit physically acceptable supergravity solutions. In addition, we confirm the claim, briefly advocated already in \cite{DHoker:2016ysh} and discussed in greater detail in the present paper, that the general $L$-pole solutions for the upper half plane can be identified precisely with fully localized intersections of $(p,q)$ 5-branes. We show that the corresponding features discussed in the previous sections indeed admit a natural interpretation as properties of the intersection. In \ref{sec:5web} we shall extend this discussion and speculate on how the solutions relate to $(p,q)$ 5-brane webs and the corresponding SCFTs.

\subsection{Solutions with three poles}

\begin{figure}
  \centering
  \begin{tabular}{lll}
  \includegraphics[width=59mm]{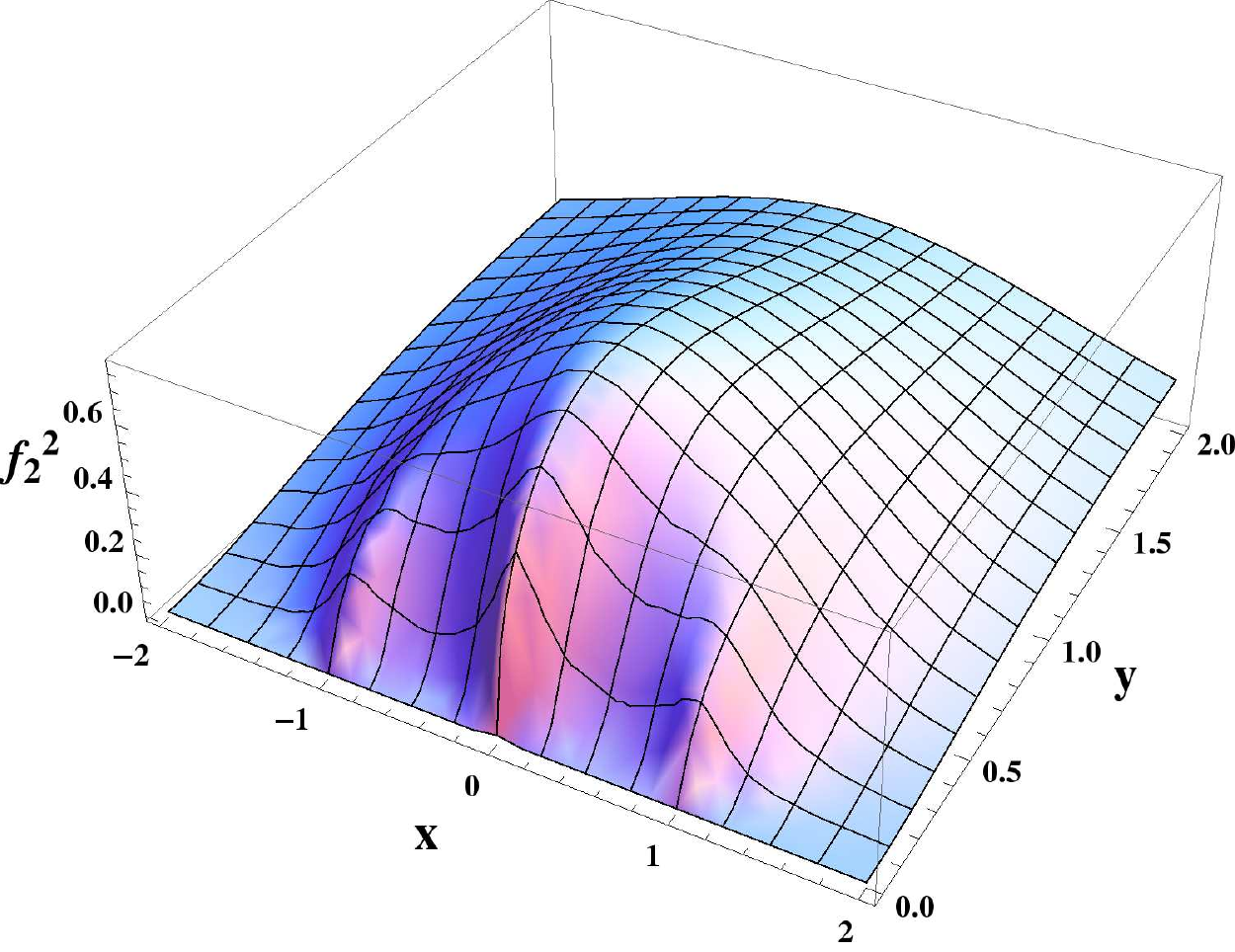}&\hskip 1in
    \includegraphics[width=59mm]{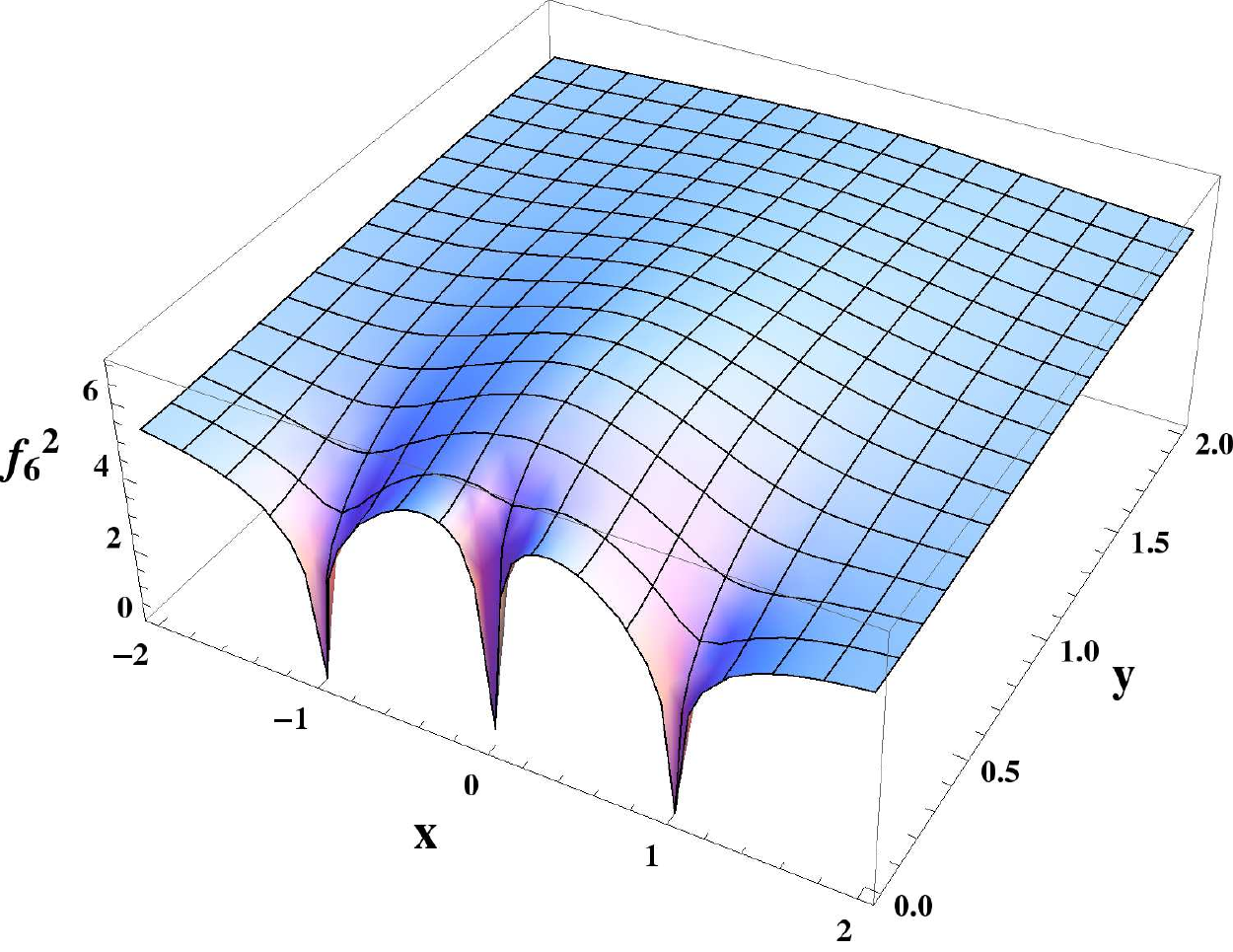}\\
    \includegraphics[width=59mm]{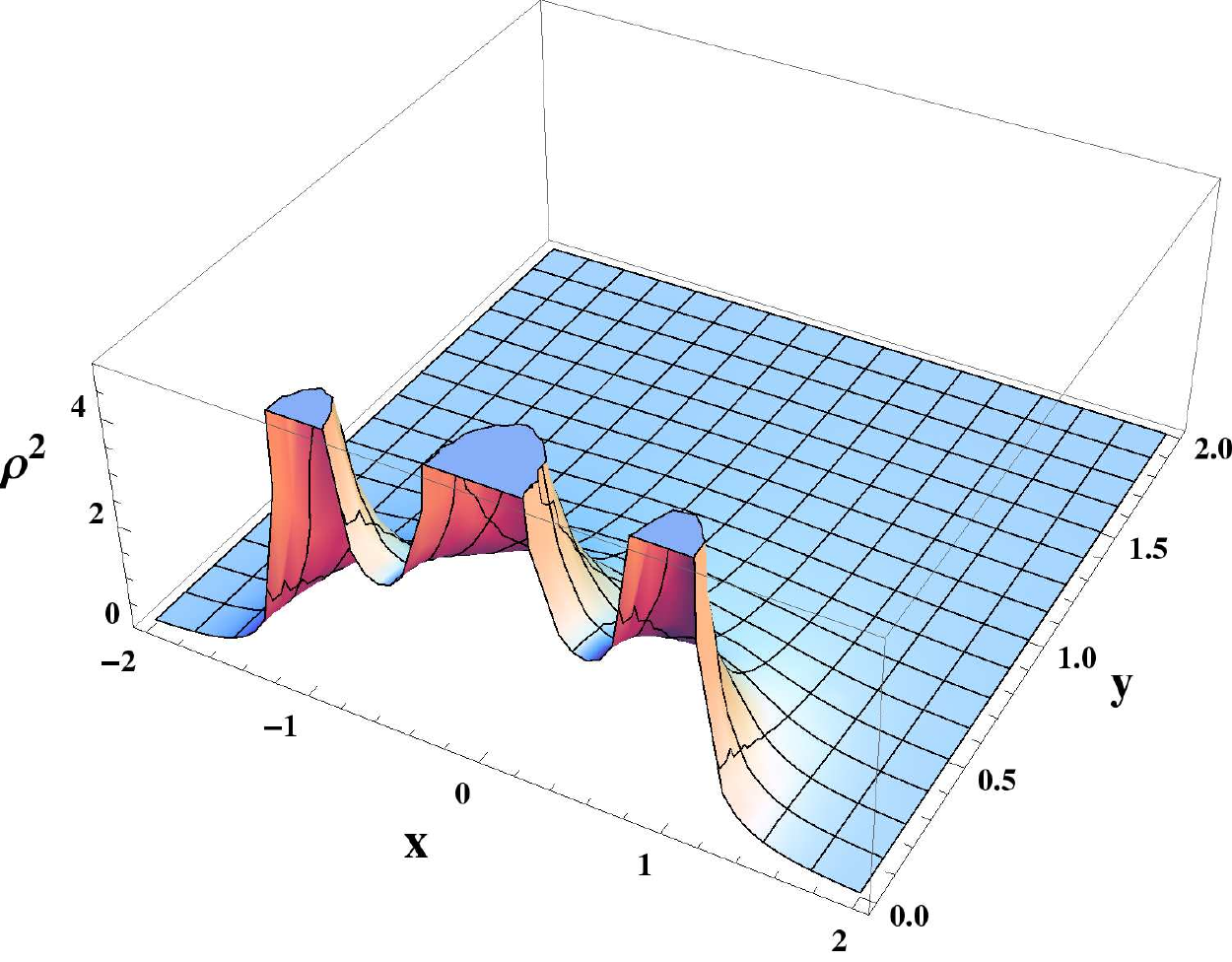}\\
    \includegraphics[width=59mm]{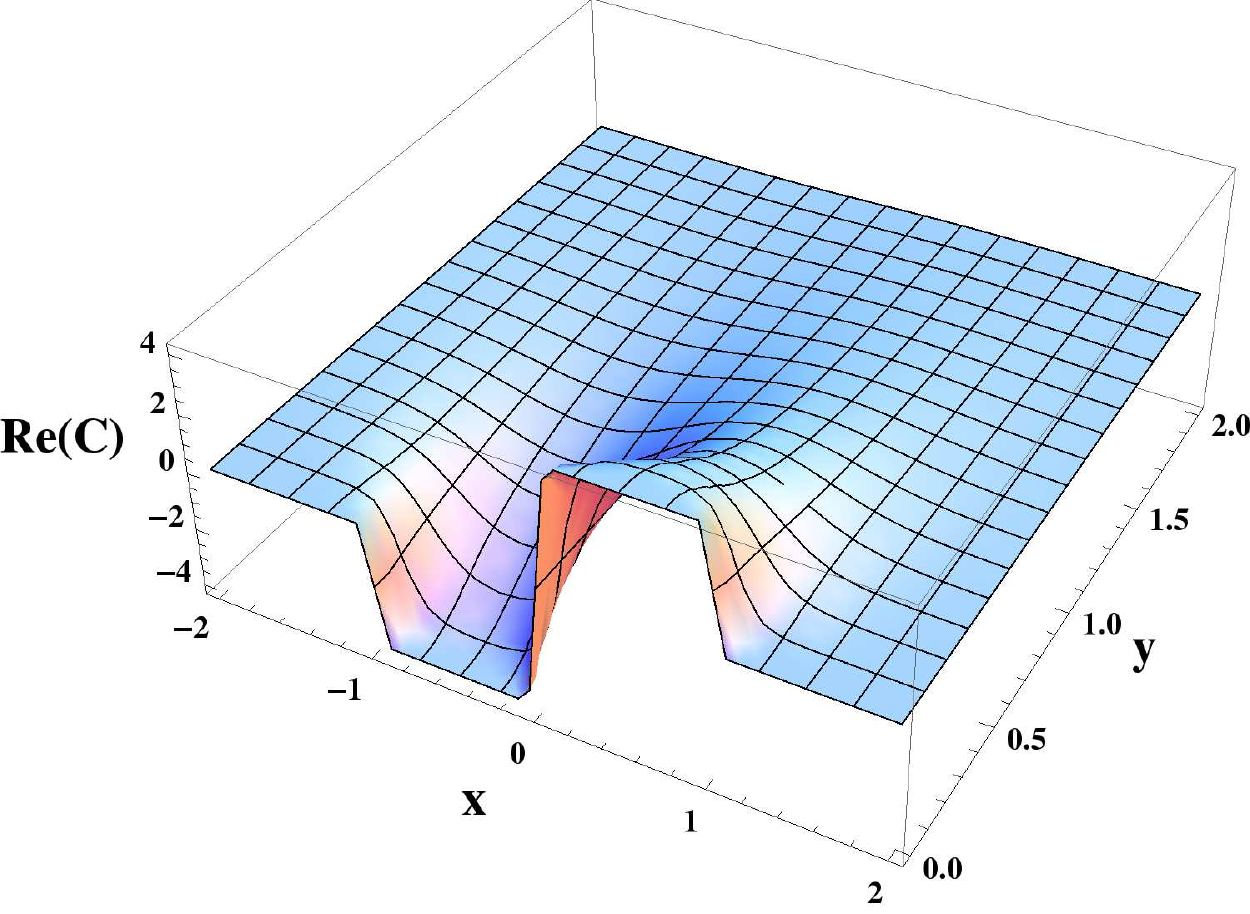}&\hskip 1in
    \includegraphics[width=59mm]{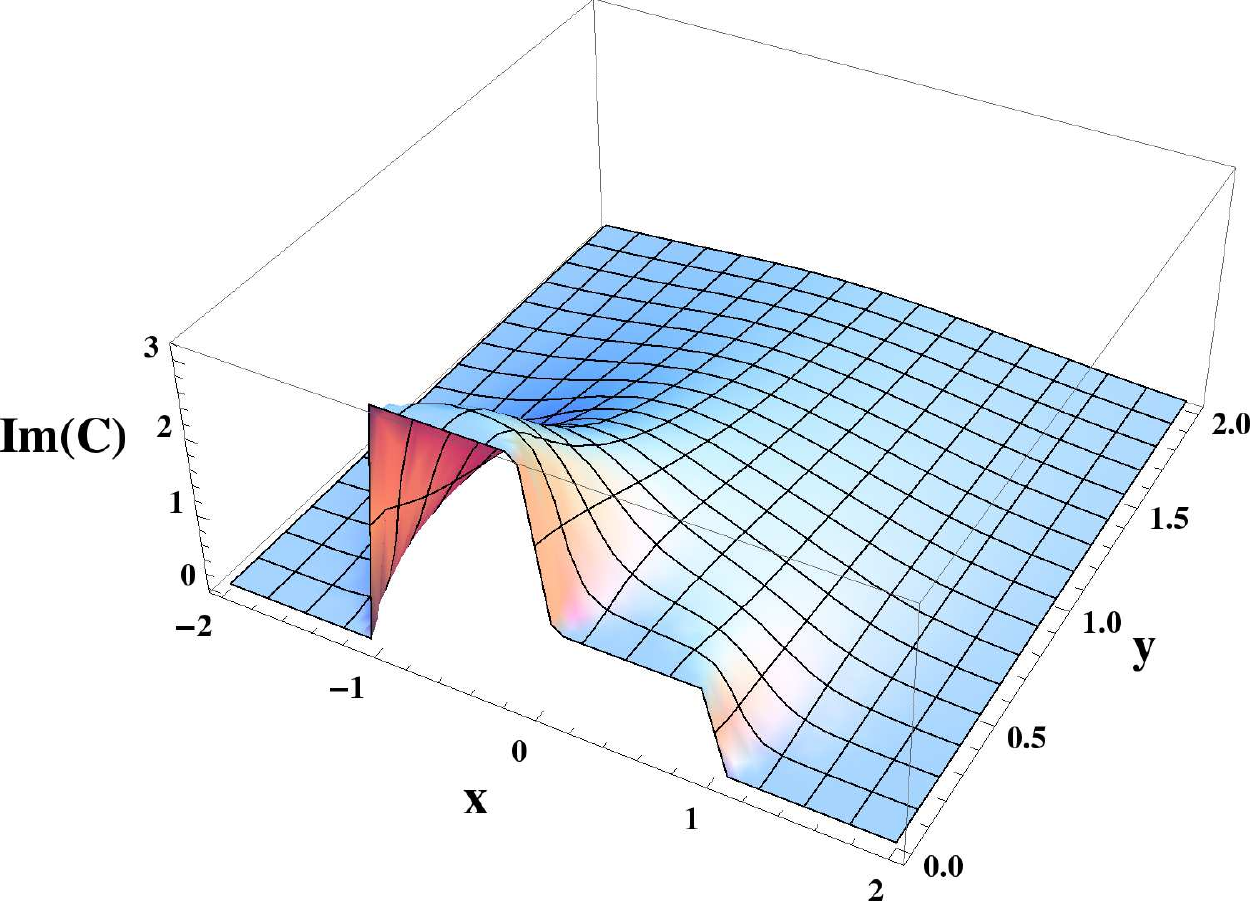}\\
    \includegraphics[width=59mm]{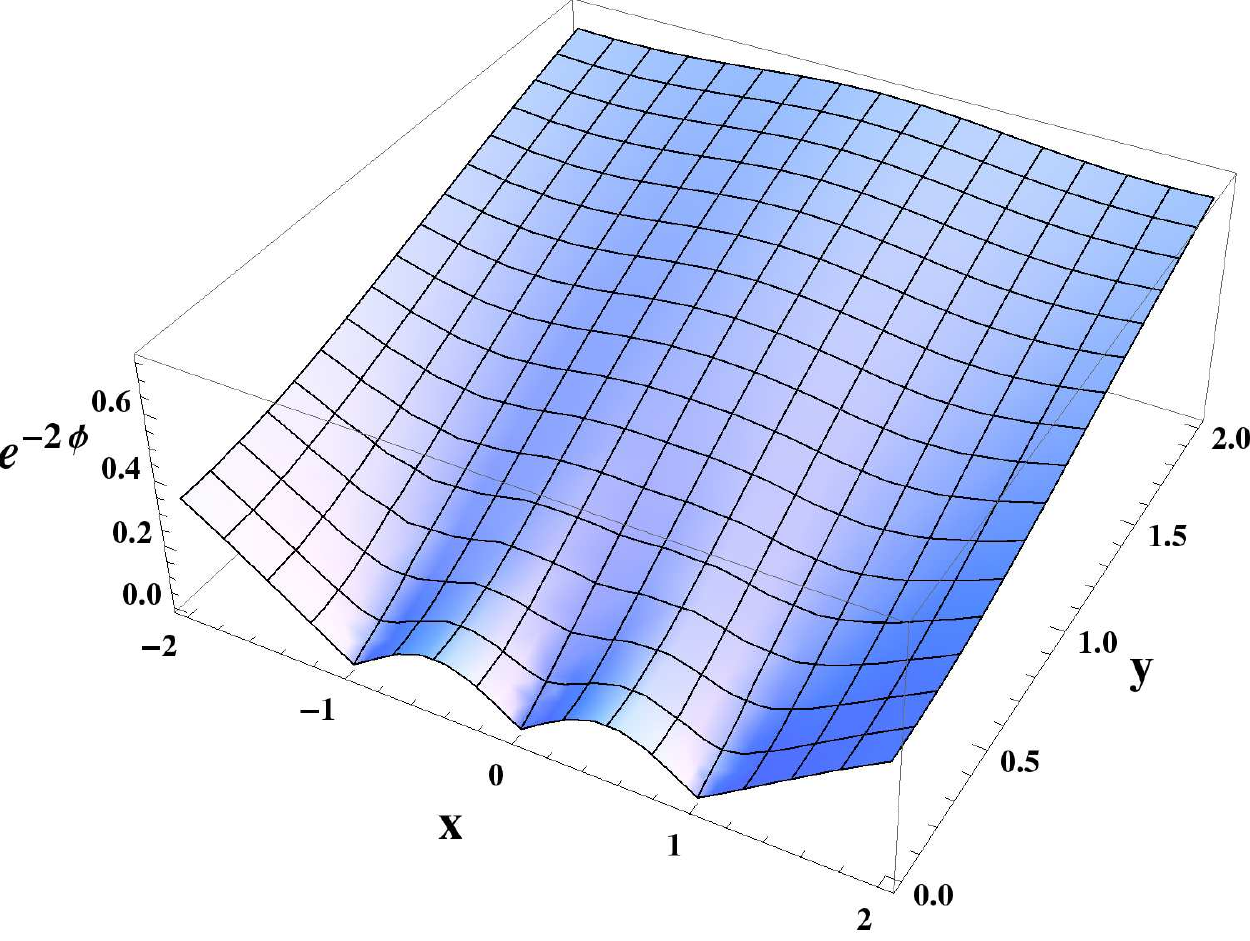}&\hskip 1in
    \includegraphics[width=59mm]{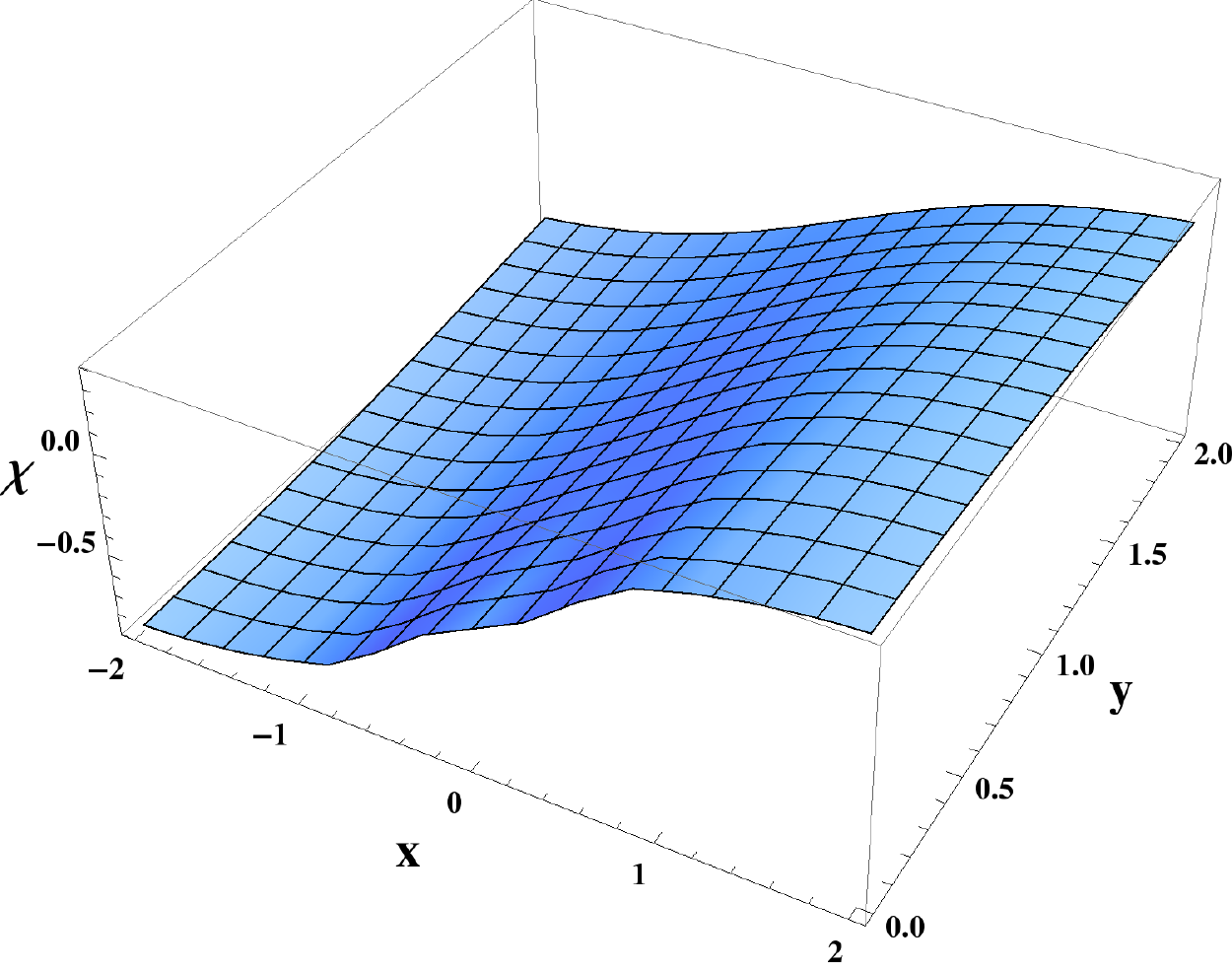}
  \end{tabular}
  \caption{
The three-pole solution: the metric factors $f_2^2,f_6^2$ and $\rho^2$,  the real and imaginary parts of the two-form potential $\cC$  and axion and dilaton  corresponding to the parameters given in (\ref{3j2}),  as functions of $w=x+iy$.
 }
  \label{fig3a}
\end{figure}

In this subsection, we discuss the family of minimal solutions with $N=1$ zeros and $L=3$  poles in the differentials $\p_w \cA_\pm$. It follows from the counting given in (\ref{3h1}) that this family of  solutions has four real free parameters.  Using the $SL(2,\RR)$ automorphism group of the upper half plane, we may choose the poles to reside at $0, \pm 1$. As argued in the previous section,  the single zero $s$ must lie strictly in the upper half plane,
\bea
\label{3j0}
p_1=1 \hskip 0.5in p_2=0 \hskip 0.5in p_3=-1 \hskip 1in \Im (s)>0
\eea
The four real parameters are the complex zero $s$ and the overall complex normalization $\omega_0\lambda_0$. The residues are calculated from (\ref{3c2}), and we obtain, 
\bea
Z_+ ^1 & = & \half \omega _0 \lambda _0 \, (+1-s) 
\no \\
Z_+ ^2 & = & \omega _0 \lambda _0 \, s 
\no \\
Z_+ ^3 & = & \half \omega _0 \lambda _0 \, (-1-s) 
\label{3j1}
\eea
the residues $Z_- ^\ell$ being given by $Z^\ell _- = - \overline{Z^\ell _+}$ for $\ell=1,2,3$.  We then have, 
\bea
Z^{[1,2]}=Z^{[2,3]} =Z^{[3,1]}=\frac{1}{2}|\omega_0|^2|\lambda_0|^2(s-\bar s)
\eea
and the relations (\ref{3f7}) are solved by $\cA^0=\omega_0\lambda_0 s \ln 2$. The metric functions, dilaton/axion fields and flux potential all involve polylogarithms and do not admit a simple form. Therefore, we shall directly resort to a numerical evaluation of the supergravity fields of the solution. In Figure~\ref{fig3a} we illustrate  the behavior of the metric functions in the Einstein frame,  as well as the axion, dilaton and the two-form potential ${\cal C}$ for the following choice of the four free parameters, 
\bea
\label{3j2}
s=\frac{1}{2}+2i
\hskip 0.5in
\lambda_0=1
\hskip 0.5in
\omega_0=i
\eea
For stable numerical evaluation we note that, since $\cG$ assumes a maximum in the interior of $\Sigma$
and $\partial_w \cG=0$ at that point, the expression for $\rho^2$ in (\ref{2a4}) has to be treated with care.
The expression is regular since $R\rightarrow 0$ when $\partial_w\cG\rightarrow 0$, and to make the regularity 
of $\rho^2$ explicit we rewrite it in a manifestly regular form. This leads to the following exact alternative 
expressions for the metric factors,
\bea
f_6^2=c_6 \sqrt{6\cG} \left ( \frac{1+R}{1-R} \right ) ^{1/2}
\hskip 0.4in
f_2^2=\frac{c_6}{9}\sqrt{6\cG} \left ( \frac{1-R}{1+R} \right ) ^{3/2}
\hskip 0.4in
\rho^2={ c_6\kappa^2 \over \sqrt{6\cG} } \left (\frac{1+R}{1-R} \right ) ^{1/2}
\eea
They were derived directly from (\ref{2a4}) and we have therefore included the appropriate powers of $c_6$.
The plots show that the exponentiated dilaton is positive, as required, and goes to zero at the poles, as derived in sec.~\ref{sec:asympt}. The metric factors are real and positive, as desired,  and behave at the poles precisely as discussed for the Einstein frame in sec.~\ref{sec:asympt}. At the boundary of $\Sigma$, $f_2$ vanishes  as required to have a smooth ten-dimensional geometry without boundary, and $f_6$ stays finite except for at the poles. The two-form potential $\cC$ is piecewise constant at the boundary, and jumps at the poles in accordance with (\ref{3g8}). Moreover, it approaches the same value to the left and to the right of all poles, reflecting charge conservation.

\subsection{Solutions with four poles}
\label{sec:4pole-plots}

Solutions with four poles have six real free parameters. We use $SL(2,\RR)$ invariance to fix the positions of  three of the four poles on the real line. We may choose the remaining free parameters  to be the two complex zeros and one complex overall normalization, with $\cA^0$ and the position of the fourth pole determined from (\ref{3f7}). A more physically motivated choice of parameters is as follows. Via the identification in sec.~\ref{sec:poles-branes}, the poles correspond to four stacks of semi-infinite external 5-branes, and the residues are related to their $(p,q)$ charges.  Alternatively, we may choose three complex residues at three of the poles as the free parameters, the residue at the fourth pole then being determined by the condition that the sum of the four residues vanishes in (\ref{3c1}).  We will look at the special family of solutions where the residues are related as follows,
\bea
\label{3h1a}
Z_+^1=-Z_+^3 \hskip 1in Z_+^2=-Z_+^4
\eea
That is, the stacks of semi-infinite external 5-branes have pairwise opposite charges.
Due to the relation $\sum_\ell Z_+^\ell=0$ the two relations above are equivalent to one another.
Using $SL(2,\RR)$ to fix the position of three poles as follows, 
\bea
p_1=1 \hskip 0.5in
p_2=\frac{2}{3} \hskip 0.5in
p_3=\frac{1}{2}
\eea
the position of the remaining pole, $p_4$ remains a free parameter. The conditions (\ref{3h1a}) and (\ref{3f7}) are solved by, 
\bea
p_4=0
\hskip 0.5in
s_2=\frac{3s_1-2}{5s_1-3}
\hskip 0.5in
\cA^0=\frac{3\lambda_0\omega_0}{5s_1-3}(s_1^2 \ln432-s_1\ln576+\ln 4)
\eea
Since the relation between $s_1$ and $s_2$ is an $SL(2,\RR)$ transformation, both $s_1,s_2$ will be in the upper half plane provided either one is. The resulting charges are,
\bea
Z_+^1=\frac{6\lambda_0\omega_0}{3-5s_1}(1-s_1)(1-2s_1)
\hskip 0.7in
Z_+^2=\frac{3\lambda_0\omega_0}{3-5s_1}s_1(2-3s_1)
\eea
with $Z_+^3$ and $Z_+^4$ given by (\ref{3h1a}).
\begin{figure}
  \centering
  \begin{tabular}{lll}
  \includegraphics[width=59mm]{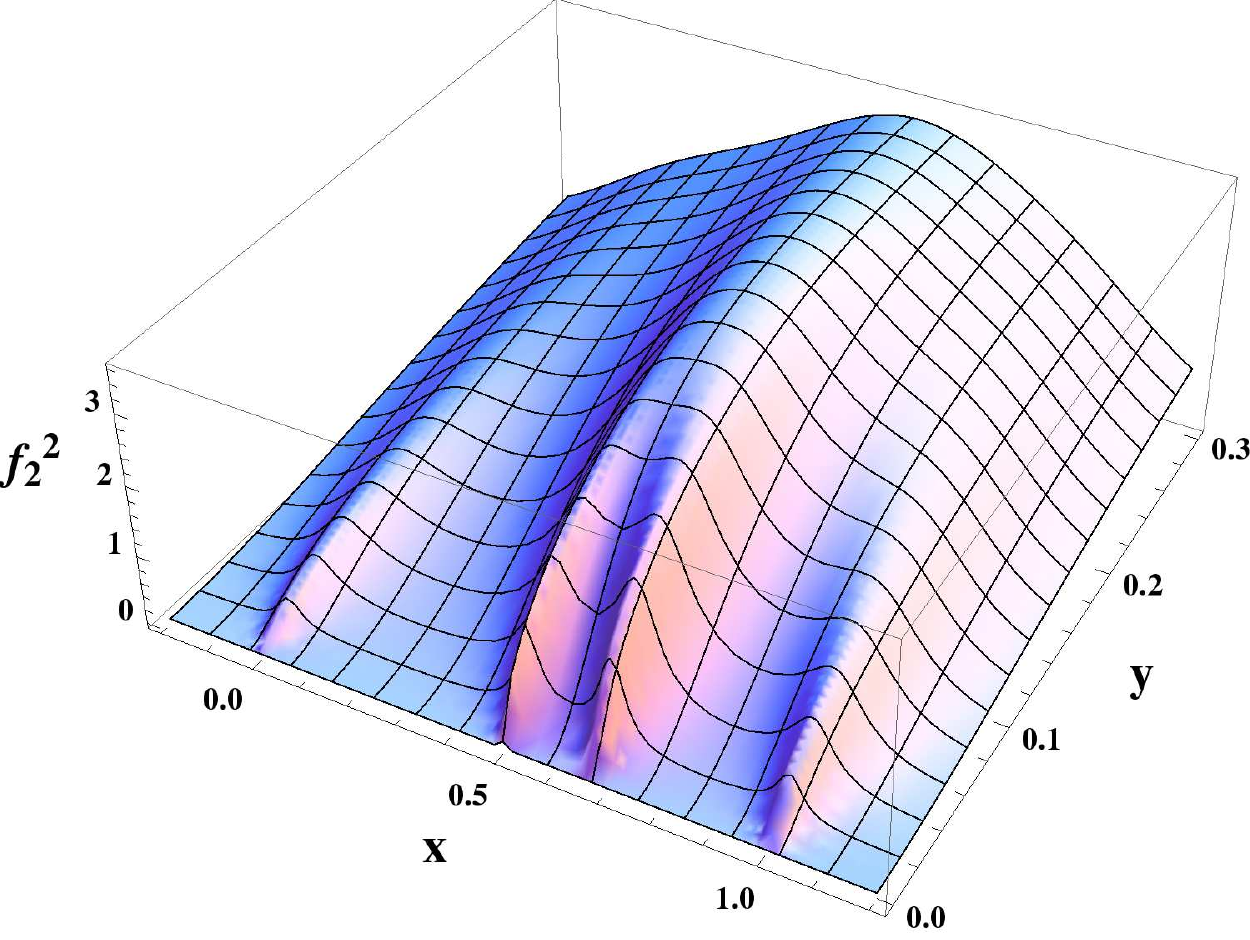}&\hskip 1in
    \includegraphics[width=59mm]{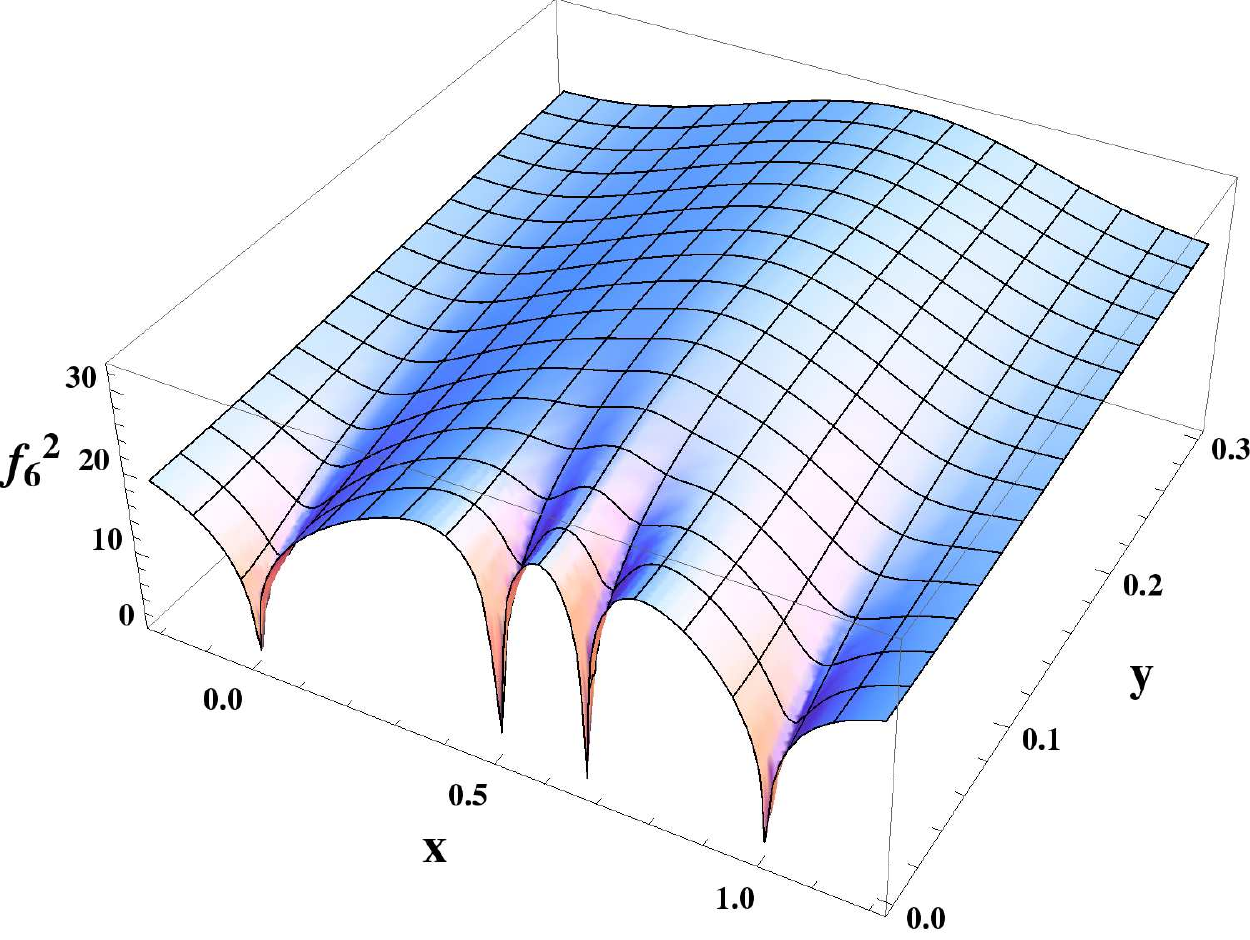}\\
    \includegraphics[width=59mm]{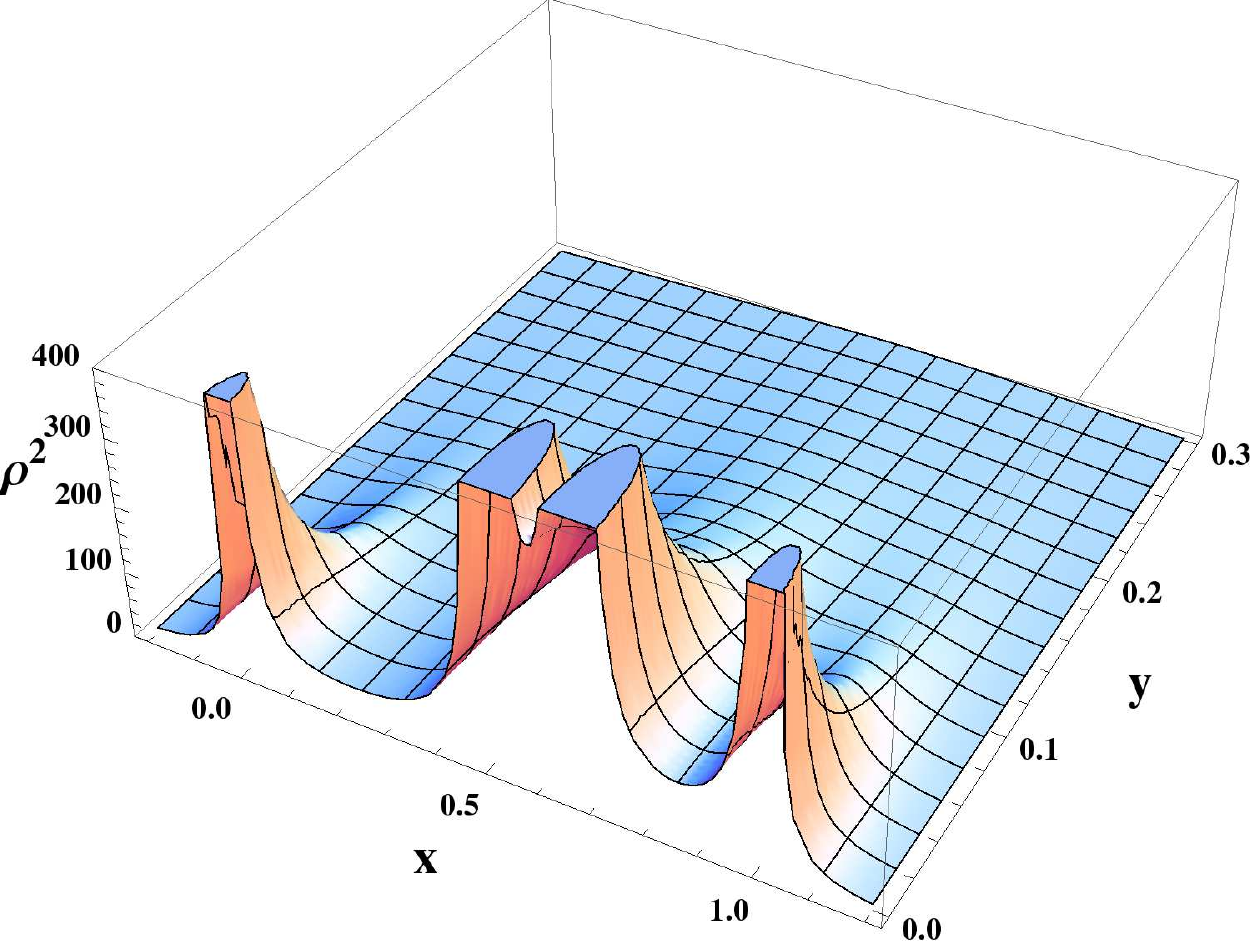}\\
    \includegraphics[width=59mm]{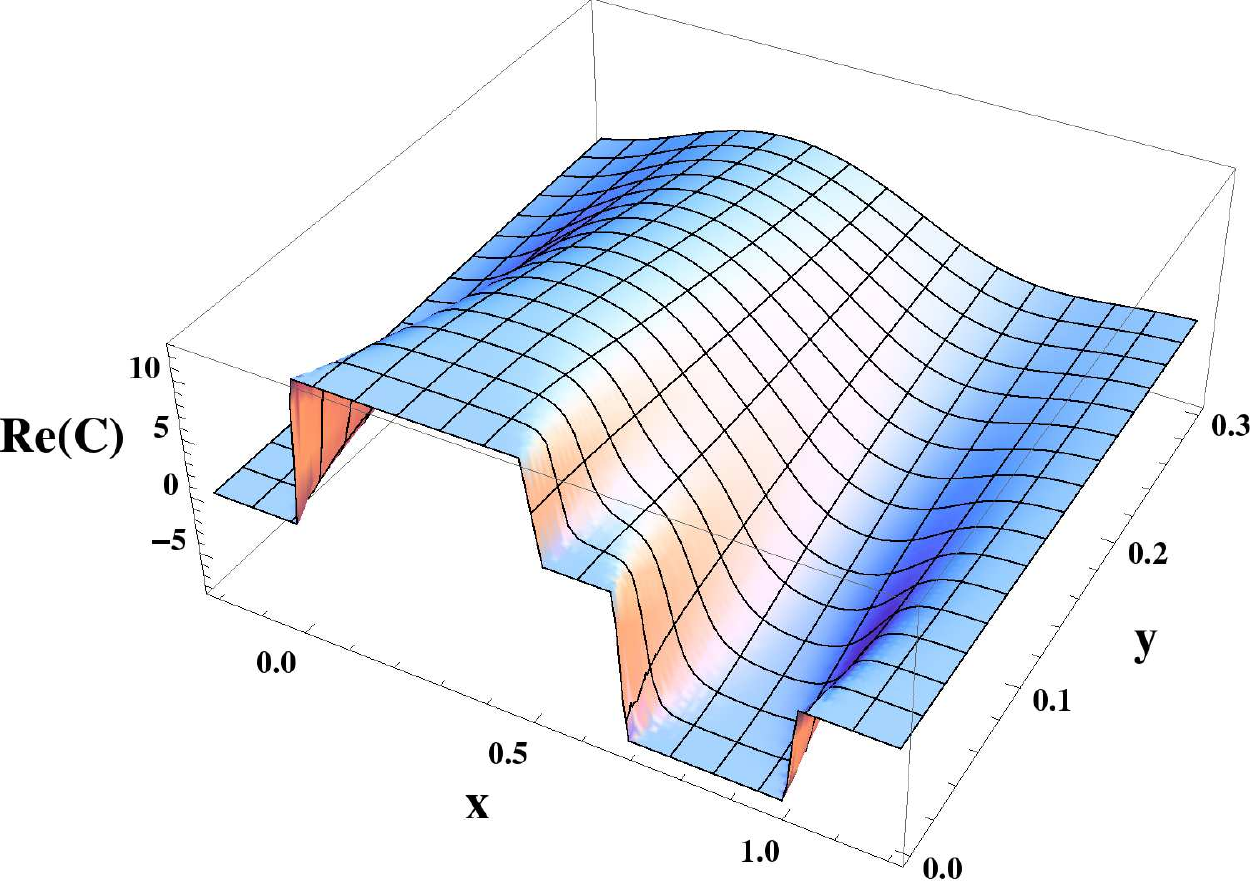}&\hskip 1in
    \includegraphics[width=59mm]{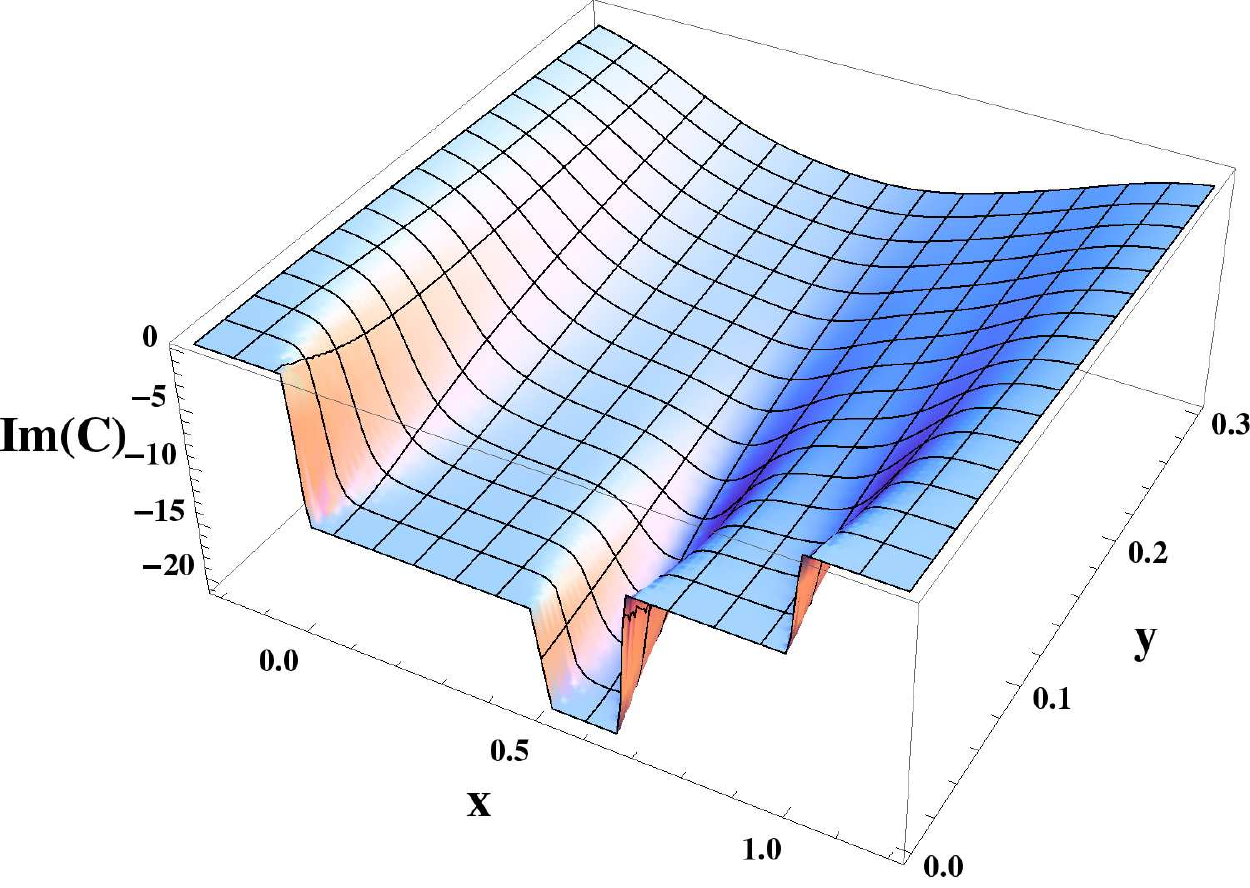}\\
    \includegraphics[width=59mm]{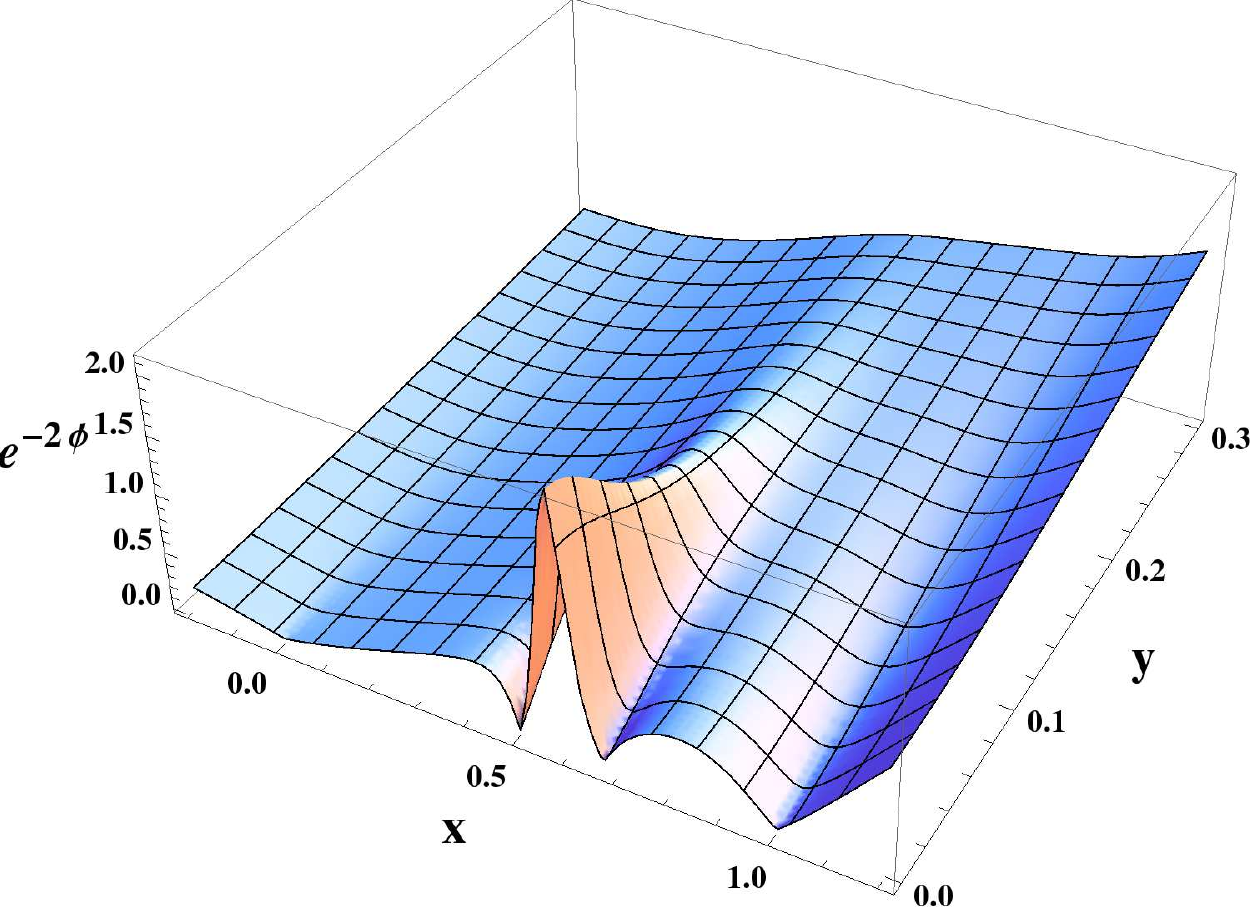}&\hskip 1in
    \includegraphics[width=59mm]{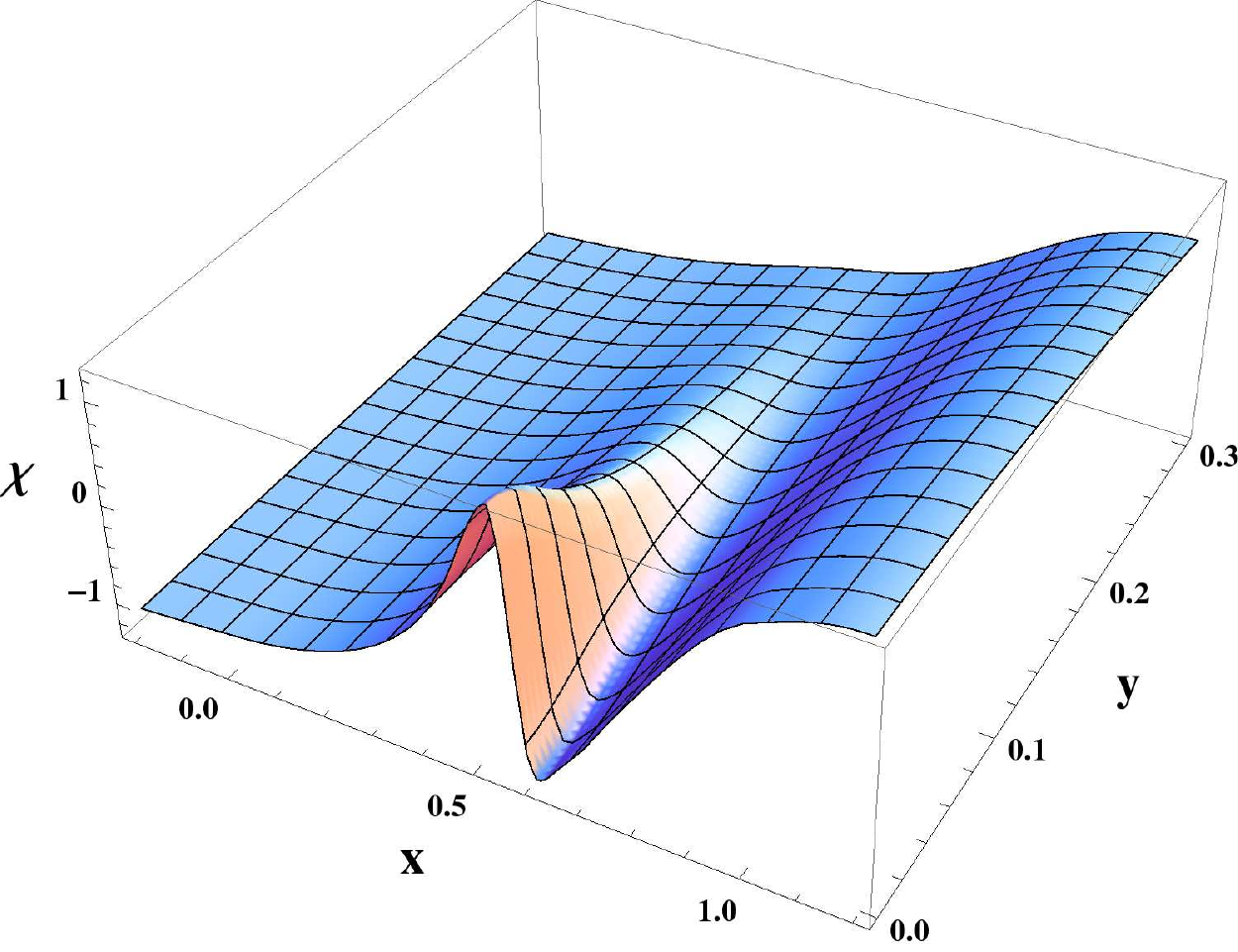}
  \end{tabular}
  \caption{The four-pole solution: the metric factors $f_2^2,f_6^2$, and $\rho^2$,  the real and imaginary parts of the two-form potential $\cC$  and of the axion and dilaton  corresponding to the parameters given in (\ref{3h6}),
  with $N=2$, $M=3$.  All supergravity functions are regular throughout $\Sigma$ except for the poles,
  where their behavior is precisely as discussed in sec.~\ref{sec:asympt}.
  }
  \label{fig3b}
\end{figure}
We will take a closer look at the particular solutions obtained by choosing
\bea\label{3h6}
s_1=\frac{3M-i N+i\sqrt{N^2-M^2}}{4M-3i N}
\hskip 0.5in
\omega_0\lambda_0=\frac{1-i}{6}(4M-3i N)
\eea
For $M, N>0$ the imaginary part is positive for both zeros $s_1$ and $s_2$,
corresponding to the location of the charges in the electrostatics analogy.
They are interchanged upon changing the branch of the square root
and become coincident for $M=N$.
The residues, fixing the charges of the external 5-branes via (\ref{3g0}), are
\bea
-Z_+^1=Z_+^3=(1+i)N
\hskip 0.5in
Z_+^2=-Z_+^4=(1-i)M
\eea
Plots illustrating the solution are shown in Figure~\ref{fig3b}.
The qualitative features are the same as for the 3-pole solution:
The metric factors are positive and the fields satisfy the desired regularity conditions throughout,
except for the poles where they behave as derived in sec.~\ref{sec:asympt}.

\subsection{Relation to 5-brane webs}\label{sec:5web}

We have argued already briefly in \cite{DHoker:2016ysh} that our supergravity solutions can be identified with fully localized intersections of $(p,q)$ 5-branes, and the detailed derivations presented in this paper provide the justification for the arguments used in \cite{DHoker:2016ysh}:
By construction, our supergravity solutions have the correct superconformal $F(4)$ symmetry, and by the results of sec.~\ref{sec:counting} the parameter count precisely matches the parameter count of $(p,q)$ five-brane webs in the conformal limit. Moreover, as the discussion of sec.~\ref{sec:poles-branes} shows the solutions also have the correct external states and the parameters can be directly translated to those specifying a 5-brane intersection.

\sm

Many other features of the solutions admit a natural interpretation in the context of 5-brane intersections as well.
The minimal number of poles being three, as discussed in sec.~\ref{sec33}, corresponds to the fact that three external branes are needed to produce a codimension-1 intersection. Moreover, there are no solutions with either only D5 charge or only NS5 charge. From (\ref{zdefa}) we have $Z^{[\ell,k]}=2i\Im(\bar Z_+^k Z_+^\ell)$. Therefore, if the $Z_+^\ell$ are either all real or all imaginary, $Z^{[\ell,k]}$ vanishes for all $\ell,k$, which implies that $\cG=0$ everywhere and the solution degenerates. Regular solutions therefore necessarily involve D5 and NS5 charge, and this corresponds to the fact that D5 and NS5 charges are needed to realize a codimension-1 intersection with 5-branes.

\sm

The identification of our solutions with 5-brane intersections certainly suggests a holographic relation to the 5d SCFTs obtained by taking the conformal limit of 5-brane webs describing 5d supersymmetric gauge theories \cite{Aharony:1997ju,Aharony:1997bh}.
Holographic relations usually involve some form of a large-$N$ limit, and we indeed note that the charges of the external 5-branes in our solutions are assumed to be large for the supergravity description to be valid. There is therefore no constraint
from charge quantization, similarly to the familiar case of e.g.\ $\cN=4$ SYM and its $AdS_5\times S^5$ dual.
Generically, the brane intersections described by our solutions therefore involve both, large D5 charge and large NS5 charge.
Brane webs related by the $SL(2,\ZZ)$ duality of type IIB string theory describe the same field theory, and in type IIB supergravity this is enhanced to an $SL(2,\RR)$ symmetry.\footnote{
Not to be confused with the $SL(2,\RR)$ automorphisms of the upper half plane (c.f.\ sec.~\ref{sl2r}).
Solutions related by automorphisms of the upper half plane are actually equivalent solutions,
while solutions related by $SU(1,1)\sim SL(2,\RR)$ duality transformations are generally not.}
We thus expect solutions related by $SU(1,1)\sim SL(2,\RR)$ to describe field theories with the
same ``large-$N$'' limit.

\sm

\begin{figure}
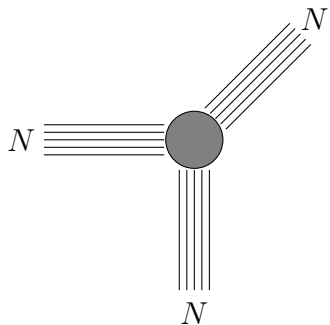

\begin{center}
\tikzpicture
\draw (-2,0.2) -- (-0.4,0.2);
\draw (-2,0.1) -- (-0.4,0.1);
\draw (-2,0.0) -- (-0.4,0.0);
\draw (-2,-0.1) -- (-0.4,-0.1);
\draw (-2,-0.2) -- (-0.4,-0.2);

\draw (-0.2,-0.4) -- (-0.2,-2);
\draw (-0.1,-0.4) -- (-0.1,-2);
\draw (-0.0,-0.4) -- (-0.0,-2);
\draw (0.1,-0.4) -- (0.1,-2);
\draw (0.2,-0.4) -- (0.2,-2);

\draw (0.28,0.28) -- (1.41,1.41);
\draw (0.35,0.21) -- (1.48,1.34);
\draw (0.42,0.14) -- (1.55,1.27);
\draw (0.21,0.35) -- (1.34,1.48);
\draw (0.14,0.42) -- (1.27,1.55);

\draw[fill=gray] (0,0) circle (0.38);

\node at (-2.3,0) {$N$};
\node at (0,-2.3) {$N$};
\node at (1.6,1.6) {$N$};
\endtikzpicture
\end{center}
\caption{
The $N$-junction, a $(p,q)$ 5-brane intersection with three external 5-brane stacks,
obtained by taking $N$ copies of the basic junction of a D5 and an NS5 brane.
}
\label{tn-web}
\end{figure}

We will now use the explicit solutions with three and four poles for a more explicit discussion and to illustrate further points.
The solutions with three poles correspond to 5-brane intersections with three external 
$(p,q)$ 5-branes, with charges given by (\ref{3g0}).
They have a total number of four parameters, corresponding to the choice of charges subject to charge conservation.
For the field theory interpretation, $SU(1,1)\sim SL(2,\RR)$ further reduces the number of parameters by $3$,
which leaves only one free parameter.
Indeed, a generic 3-fold intersection of $(p,q)$ 5-branes can be mapped to the form 
shown in Figure~\ref{tn-web} by an $SU(1,1)$ transformation.
This ``$N$-junction'' is obtained by combining $N$ copies of 
the basic 5-brane junction with external charges $(1,0)$, $(0,1)$ and $(-1,-1)$ (in all in-going convention)
and allows for a large-$N$ limit.
A choice of parameters for the supergravity solution to realize this form of the web via (\ref{3j0}), (\ref{3j1}) is
\bea
s=\frac{1+2i}{5}
\hskip 0.5in
\lambda_0\omega_0=(2+i)N
\eea
Brane webs of the form shown in Figure~\ref{tn-web} have been identified in \cite{Benini:2009gi}
as five-dimensional uplifts of the four-dimensional $T[A_{N-1}]$ theories, obtained by wrapping $N$ M5-branes on a sphere with 3 punctures.
For the field theory interpretation it is crucial whether and how the external 5-branes defining the intersection end on 7-branes.
In \cite{Aharony:1997ju,Aharony:1997bh} the external 5-branes were taken to be semi-infinite, 
but one can also terminate them on 7-branes such that they are of finite extent in the plane in which the brane web is drawn \cite{DeWolfe:1999hj}.
Taking groups of 5-branes to end on the same 7-brane yields a different field theory than having each 5-brane semi-infinite or terminate on its own 7-brane, and 
for the intersection of Figure~\ref{tn-web} various options were discussed in \cite{Benini:2009gi}.

\sm

In our solutions we only see the external 5-brane geometries with no indication for the presence of 7-branes,
and at least at the level of disk solutions there are also no obvious moduli corresponding to the choice which 5-branes terminate on which 7-brane.
This suggests that the solutions correspond to the original form of the brane webs and intersections, where the external 5-branes are 
indeed semi-infinite \cite{Aharony:1997ju,Aharony:1997bh}.
However, one could also argue that our solutions only cover the near-intersection limit, where the external 7-branes may not be directly accessible and more indirect methods will be required to precisely pin down the dual field theory. 
We will leave a more detailed investigation for the future and leave this question open for the remaining discussion.

\sm

The lift of the isolated 4d $T[A_{N-1}]$ theories to 5d via intersections of the form in Figure~\ref{tn-web}
offers a chance to find deformations that permit a Lagrangian description as gauge theories in the IR,
and such deformations have been constructed in \cite{Bergman:2014kza}.
For the case where each external 5-brane ends on a separate 7-brane, it results in a web describing the quiver 
\bea
N-SU(N-1)\times\dots\times SU(2)-2
\eea
That is, a gauge theory with product gauge group $SU(N-1)\times\dots\times SU(2)$,
with one hypermultiplet in the bi-fundamental representation for
each pair of adjacent gauge group factors
and in addition $N$ massless hypermultiplets in the fundamental representation of $SU(N-1)$
along with $2$ massless hypermultiplets in the fundamental of $SU(2)$.
The 5-brane intersection only has one large parameter controlling both D5 and NS5 charge, and we see that in this case it also translates to two large parameters in the field theory defining the UV fixed point: the length of the quiver and the rank of the largest gauge group factor.

\begin{figure}
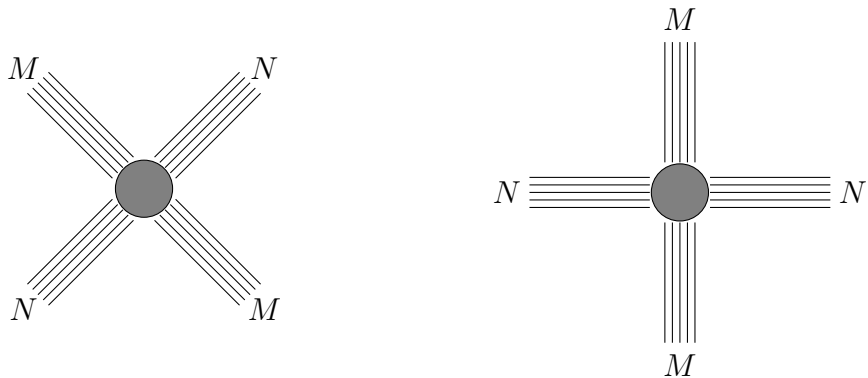

\begin{center}
\tikzpicture

\draw (0.28,0.28) -- (1.41,1.41);
\draw (0.35,0.21) -- (1.48,1.34);
\draw (0.42,0.14) -- (1.55,1.27);
\draw (0.21,0.35) -- (1.34,1.48);
\draw (0.14,0.42) -- (1.27,1.55);

\draw[fill=gray] (0,0) circle (0.38);

\draw (0.28,-0.28) -- (1.41,-1.41);
\draw (0.35,-0.21) -- (1.48,-1.34);
\draw (0.42,-0.14) -- (1.55,-1.27);
\draw (0.21,-0.35) -- (1.34,-1.48);
\draw (0.14,-0.42) -- (1.27,-1.55);

\draw (-0.28,-0.28) -- (-1.41,-1.41);
\draw (-0.35,-0.21) -- (-1.48,-1.34);
\draw (-0.42,-0.14) -- (-1.55,-1.27);
\draw (-0.21,-0.35) -- (-1.34,-1.48);
\draw (-0.14,-0.42) -- (-1.27,-1.55);

\draw (-0.28,0.28) -- (-1.41,1.41);
\draw (-0.35,0.21) -- (-1.48,1.34);
\draw (-0.42,0.14) -- (-1.55,1.27);
\draw (-0.21,0.35) -- (-1.34,1.48);
\draw (-0.14,0.42) -- (-1.27,1.55);

\node at (-1.6,-1.6) {$N$};
\node at (-1.6,1.6) {$M$};
\node at (1.6,-1.6) {$M$};
\node at (1.6,1.6) {$N$};

\node[white] at (0,-2.3) {$\tilde M$};

\endtikzpicture
\hskip 1in
\tikzpicture
\draw (-2,0.2) -- (-0.4,0.2);
\draw (-2,0.1) -- (-0.4,0.1);
\draw (-2,0.0) -- (-0.4,0.0);
\draw (-2,-0.1) -- (-0.4,-0.1);
\draw (-2,-0.2) -- (-0.4,-0.2);

\draw (-0.2,-0.4) -- (-0.2,-2);
\draw (-0.1,-0.4) -- (-0.1,-2);
\draw (-0.0,-0.4) -- (-0.0,-2);
\draw (0.1,-0.4) -- (0.1,-2);
\draw (0.2,-0.4) -- (0.2,-2);

\draw (2,0.2) -- (0.4,0.2);
\draw (2,0.1) -- (0.4,0.1);
\draw (2,0.0) -- (0.4,0.0);
\draw (2,-0.1) -- (0.4,-0.1);
\draw (2,-0.2) -- (0.4,-0.2);

\draw (-0.2,0.4) -- (-0.2,2);
\draw (-0.1,0.4) -- (-0.1,2);
\draw (-0.0,0.4) -- (-0.0,2);
\draw (0.1,0.4) -- (0.1,2);
\draw (0.2,0.4) -- (0.2,2);

\draw[fill=gray] (0,0) circle (0.38);

\node at (-2.3,0) {$N$};
\node at (0,-2.3) {$M$};
\node at (2.3,0) {$N$};
\node at (0,2.3) {$M$};
\endtikzpicture
\end{center}
\caption{
The left hand side shows the intersection realized in sec.~\ref{sec:4pole-plots}.
The intersection on the right hand side is related to the one on the left by an SL(2,$\mathds{R}$) transformation
combined with a rescaling of the charges.
}
\label{4web}
\end{figure}

\sm

Moving on to 4 external branes, the 5-brane intersection with the external charges of sec.~\ref{sec:4pole-plots} is shown on the left hand side in Figure~\ref{4web}.
Once again the field theory interpretation depends on whether and how the external 5-branes end on 7-branes.
Taking $N=M$ and all 5-branes within each external 5-brane stack to end on the same 7-brane would yield a 5-brane construction for the USp($N$) theory \cite{Bergman:2012kr}, while the original form of the webs with no 7-branes again leads to long quivers:
The configuration is SL(2,$\mathds{R}$) dual (up to a rescaling of the charges $M$ and $N$) to the intersection shown on the right hand side in Figure~\ref{4web}.
Without the introduction of 7-branes this intersection has been discussed already in \cite{Aharony:1997bh},
and a deformation of the fixed-point SCFT to a gauge theory is described by the quiver
\bea
N - \underbrace{SU(N)\times\dots\times SU(N)}_{SU(N)^{M-1}} - N
\eea
This example explicitly exhibits the presence of two independently large parameters, 
the rank of the gauge group and the length of the quiver,
corresponding to the D5 charge and NS5 charge in the brane intersection.
Moreover, we note that the number of matter fields is directly linked to the rank of the gauge group factors,
and the large-$N$ limit is a form of a Veneziano limit rather than a `t Hooft limit.

\newpage
  
\section{The annulus}
\setcounter{equation}{0}
\label{sec:annulus}

In this section, we shall investigate the existence of supergravity solutions when $\Sigma$ has the topology of an annulus, or equivalently of a finite cylinder. This is the next-simplest topology after the case of the disk, since the annulus has genus zero but two boundary components. We shall provide an explicit parametrization of the differentials $\p_w \cA_\pm$ under the assumption that these differentials are single-valued in $\Sigma$ and no axion monodromy is allowed. 
 We will explicitly construct the functions $\cA_\pm$ as well and derive the general physical regularity conditions.

\sm

 We have used a combination of analytical and numerical methods to explore whether the physical regularity conditions on the supergravity fields can all be satisfied simultaneously for the annulus. These explorations are not exhaustive, but the outcome so far has consistently been negative, and no example of a physically regular solution has been found to date. We have no analytical proof that such solutions cannot exist, but the regularity conditions are structurally different when there are multiple boundary components and the negative results so far may well be due to the fact that such solutions do generally not exist.
We will leave a more systematic analysis for the future and turn to Riemann surfaces with non-trivial topology and a single boundary component in the next section.

\subsection{Parametrization of the annulus}

Compared to the situation of the disk, the annulus has two new features. First, the annulus has two disconnected boundary components instead of one for the disk; second the annulus has a non-trivial fundamental group $\pi_1 (\Sigma)=\ZZ$.

\sm

\begin{figure}
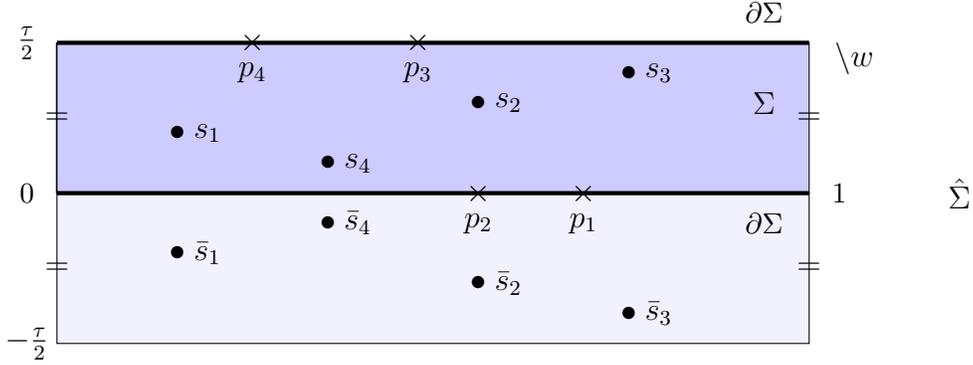

\begin{center}
\tikzpicture[scale=2]
\scope[xshift=-5cm,yshift=-0.4cm];
\draw [thick] (0,0) -- (0,1);
\draw [fill=blue!20!white] (5,0) rectangle (0,1);
\draw [fill=blue!5!white] (5,-1) rectangle (0,0);
\draw [ultra thick] (0,0) -- (5, 0);
\draw [ultra thick] (0,1) -- (5, 1);
\draw (5.2,0) node{\small $1$};
\draw (-0.2,0) node{\small $0$};
\draw (-0.2,1) node{${\tau \over 2} $};
\draw (-0.2,-1) node{$-{\tau \over 2} $};
\draw (5.3,0.9) node{$\backslash w$};
\draw (4.7,0.6) node{$\Sigma$};
\draw (6,0) node{$\hat \Sigma$};
\draw (4.7,1.2) node{\small $\p \Sigma$};
\draw (4.7,-0.2) node{\small $\p \Sigma$};
\draw (1,0.4) node{$s_1$};
\draw (1,-0.4) node{$\bar s_1$};
\draw (0.8,0.4) node{$\bullet$};
\draw (0.8,-0.4) node{$\bullet$};
\draw (3,0.6) node{$s_2$};
\draw (3,-0.6) node{$\bar s_2$};
\draw (2.8,0.6) node{$\bullet$};
\draw (2.8,-0.6) node{$\bullet$};
\draw (4,0.8) node{$s_3$};
\draw (4,-0.8) node{$\bar s_3$};
\draw (3.8,0.8) node{$\bullet$};
\draw (3.8,-0.8) node{$\bullet$};
\draw (2,0.2) node{$s_4$};
\draw (2,-0.2) node{$\bar s_4$};
\draw (1.8,0.2) node{$\bullet$};
\draw (1.8,-0.2) node{$\bullet$};
\draw (3.5,-0.2) node{$p_1$};
\draw (3.5,0) node{$\times$};
\draw (2.8,-0.2) node{$p_2$};
\draw (2.8,0) node{$\times$};
\draw (2.4,0.8) node{$p_3$};
\draw (2.4,1) node{$\times$};
\draw (1.3,0.8) node{$p_4$};
\draw (1.3,1) node{$\times$};
\draw (0,0.5) node{$=$};
\draw (5,0.5) node{$=$};
\draw (0,-0.5) node{$=$};
\draw (5,-0.5) node{$=$};
\endscope
\endtikzpicture
\caption{The annulus $\Sigma$ is represented by the dark blue rectangle with periodic identification $w \equiv w+1$; its boundary $\p \Sigma$  has two components denoted by thick black lines; the extension to the double surface $\hat \Sigma$ is indicated by the light blue rectangle; four zeros $s_n$ of $\lambda$ in the interior of $\Sigma$, and their complex conjugates $\bar s_n$, are indicated by black dots, while four poles $p_n$ distributed amongst the two boundary components are indicated by crosses.}
\label{fig7}
\end{center}
\end{figure}

The function theory on the annulus, which we need  to construct the differentials $\p_w \cA_\pm$ and the associated functions $\cA_\pm$, is conveniently obtained in terms of the function theory on the double surface. The annulus $\Sigma$ may be represented in the complex plane by a rectangle with two opposing edges periodically identified.  The {\sl double surface} $\hat \Sigma$ of the annulus $\Sigma$ is a torus whose periods may be chosen to be 1 and $\tau$ and whose modulus is purely imaginary $\tau = i \tau_2$ with $\tau_2 \in \RR^+$. The surface $\Sigma$ and its boundary  $\p \Sigma$  may then be represented by,
\bea
\label{4a1}
\Sigma & = & \left \{ w \in \CC, ~ 0 \leq \Re(w) \leq 1, ~ 0 \leq \Im (w) \leq { \tau_2 \over 2} \right \}
\no \\
\p \Sigma & = & \Big \{ w \in \Sigma, ~ \Im (w)=0  \Big \}  
\cup \left \{ w \in \Sigma, ~ \Im (w)={\tau_2\over 2} \right \}
\eea
both periodically identified under $w \equiv w+1$, as represented in Figure \ref{fig7}. We choose $\hat \Sigma$ by symmetry across the real axis, such that $0 \leq \Re(w) \leq 1$ and $|\Im (w) | \leq  \tau_2 /2$. Complex conjugation $\mI$ is the anti-conformal involution which maps between the components of $\hat \Sigma$ in the upper and lower half planes. We may view $\Sigma$ as the quotient $\Sigma = \hat \Sigma / \mI$, and the boundary $\p \Sigma$ as the fixed set under $\mI$.

\subsection{The function  \texorpdfstring{$\lambda$}{lambda}}

To investigate the existence of physically regular supergravity solutions, we closely follow the general strategy outlined in subsection \ref{sec:23}, and begin with the construction of the holomorphic function $\lambda$. 
To construct  $\lambda(w|\tau )$ we use the scalar Green function $G(w,z|\tau)=G(z,w|\tau)$ on $\Sigma$ which vanishes whenever $w$ is on $\p\Sigma$. To construct $G$, we make use of the scalar Green function $G_0$ on the double surface $\hat \Sigma =  \CC/ ( \ZZ + \ZZ \tau)$,
\bea
\label{4b1}
G_0(w|\tau) = - \ln \left | { \vartheta _1 (w|\tau) \over \tet _1 '  (0 |\tau) } \right |^2  - {\pi \over 2 \tau_2} (w-\bar w)^2
\eea
where $\tet_1$ is the Jacobi $\tet$-function. By construction, $G_0$ is a real and doubly periodic function,
\bea
\label{4b2}
G_0(\bar w |\tau)  & = & G_0(w|\tau)
\no \\
G_0(w + m + n \tau |\tau) & = & G_0(w|\tau) \hskip 1in m,n \in \ZZ
\eea
in view of the standard translation properties of the Jacobi $\tet$-function,
\bea
\label{4b3}
\tet_1(w+1 |\tau) & = & - \tet_1 (w|\tau)
\no \\
\tet_1(w+\tau |\tau) & = & - \tet_1 (w|\tau) \, \exp \left (-  i \pi \tau  - 2 \pi i w \right )
\eea
The corresponding Green function $G$ on the annulus is given by, 
\bea
\label{4b5}
G(w, s|\tau) & = &  G_0(w-s|\tau) - G_0 (w-\bar s |\tau)
\no \\
& =  &  - \ln \left | { \tet _1 (w-s |\tau) \over \tet _1 (w-\bar s | \tau) } \right |^2
+ { 2 \pi \over \tau_2} (w-\bar w) (s-\bar s)
\eea
The properties of reality and double periodicity of (\ref{4b2}) ensure $G(w,s|\tau)=0$ whenever $w \in \RR + \ZZ { \tau \over 2} $, so that $G$ vanishes on both boundary components of $\Sigma$.

\sm

The function $\lambda (w|\tau )$ for the annulus with modulus $\tau \in i \RR^+$ may be inferred from the electrostatic potential obtained  by adding the contributions from an array of positive unit charges placed at points $s_n \in \Sigma$  with $n=1,\cdots, N$, and satisfying $s_n \not \in \RR + \ZZ {\tau \over 2}$.  Therefore, the general electrostatic potential used in (\ref{2c2}) takes the from, 
\bea
\label{4b4}
- \ln |\lambda (w)|^2 = \sum _{n=1}^N G(w, s_n|\tau)
\eea
To extract the holomorphic function $\lambda$, we use the fact that $G$ is real and harmonic in $w \in \Sigma$ away from $s$ to split (\ref{4b4}) into a sum of holomorphic and anti-holomorphic parts in $w$,  
\bea
\label{4b6} 
\lambda (w)  = \lambda _0^2 \, \prod  _{n=1}^N   
  { \vartheta _1 (w-s_n |\tau) \over \vartheta _1 (w- \bar s_n |\tau) } 
  \times \exp \left \{ 
- { 2 \pi \over \tau_2} w \sum _{n=1}^N (s_n - \bar s_n)  \right \}
\eea
leaving a constant phase factor $\lambda _0$ undetermined. This expression is automatically single-valued under $w \to w+\tau$, and is single-valued under $w \to w+1$  provided we impose the following condition on the points $s_n$, 
\bea
\label{4b5a}
\sum _{n=1}^N(s _n - \bar s_n ) & \in & \ZZ \tau
\eea
We recognize this relation as the divisor condition for meromorphic functions on the torus, applied to the special case where the zeros and poles come in complex conjugate pairs.
The condition implies that we must have $N \geq 2$, since it is clearly unattainable with $N=1 $
in view of the requirements that at least one of the zeros $s_n $ should lie in the interior of $\Sigma$.

\subsection{The differentials \texorpdfstring{$\p_w \cA_\pm$}{dA}}

To construct the meromorphic differentials $\p_w \cA_\pm$, we begin by splitting $\lambda$ into two complex conjugate functions $\lambda _\pm$, 
\bea
\label{4c1}
\lambda (w|\tau) = { \lambda _+ (w|\tau) \over \lambda _- (w |\tau)}
\hskip 1in 
\overline{\lambda _\pm (\bar w | \tau )} = \lambda _\mp (w |\tau )
\eea
The splitting is not unique, as a common factor of a real function to $\p_w \cA_\pm$ cancels out in $\lambda$. The irreducible solution is given as follows, 
\bea
\label{4c2}
\lambda _+ (w|\tau) & = & \lambda _0 \prod _{n=1}^N  \vartheta _1 (w-s_n |\tau) 
\exp \left \{ - { 2 \pi \over \tau_2} w \sum _{n=1}^N s_n  \right \}
\no \\
\lambda _- (w|\tau) & = & \bar \lambda _0 \prod _{n=1}^N  \vartheta _1 (w-\bar s_n |\tau) 
\exp \left \{ - { 2 \pi \over \tau_2} w \sum _{n=1}^N \bar s_n  \right \}
\eea
Neither  function $\lambda _\pm$  is single-valued on $\hat \Sigma$. The differentials $\p_w \cA_\pm$ are given in terms of  a meromorphic function $\f$ on $\hat \Sigma$ by (\ref{2c5}) so that, 
\bea
\p_w \cA_\pm (w|\tau) = \lambda _\pm (w |\tau) \, \f (w |\tau)
\eea
Applying the general conditions on the meromorphic 1-forms $\p_w \cA_\pm$ of subsection \ref{sec:232} to the case of the annulus allows us to narrow the choices of $\f$ as follows. 

\sm

Since $\p_w \cA_-$ cannot vanish in the interior of $\Sigma$ in view of the condition $\kappa ^2 >0$, the function $\f$ cannot have zeros in the interior of $\Sigma$. Furthermore, $\p_w \cA_\pm$ cannot have poles in the interior of $\Sigma$  since otherwise the functions $\cA_\pm$ would have monodromy in the interior of $\Sigma$. Therefore,  all zeros and poles of $\f$ must be on the boundary $\p \Sigma$. Zeros on the boundary may be viewed as corresponding to a degenerate situation of the general case where all zeros $a_n$ are in the interior, and we shall therefore assume that no zeros occur on the boundary, so that $\f$ has no zeros at all. 
Since  meromorphic differentials on the torus $\hat \Sigma$ have equal numbers of zeros and poles, $\f$ must have precisely $N$ poles on $\hat \Sigma$.   As a result, the zeros of $\p_w \cA_+$ in the interior of $\Sigma$ are precisely the $N$ zeros  $s_n$ of $\lambda_+$. 

\sm

Finally, we shall adopt the condition (\ref{2c6}) which for the torus reads, 
\bea
\label {4c3}
\overline{ \p_{\bar w} \cA_\pm (\bar w |\tau) } = - \p_w \cA_\mp (w |\tau) 
\eea 
With the above assumptions, we parametrize $\f$ by its $N$ poles at points $p_n$ with $n=1, \cdots, N$ distributed amongst the two boundary components of $\p \Sigma$ given in (\ref{4a1}). Combining the above conditions, we find the following expressions for the differentials $\p_w \cA_\pm$, 
\bea
\label{4c4}
\p_w \cA_+ (w |\tau) 
& = & 
\omega _0 \, \lambda _0 \, \prod _{n=1}^N { \tet _1 (w-s_n |\tau)  \over  \tet _1 (w-p_n |\tau )}
\times \exp \left \{ - { 2 \pi \over \tau_2} w \Lambda_+  \right \} 
\no \\
\p_w \cA_- (w |\tau)
& = & 
\omega _0 \, \bar \lambda _0 \,  \prod _{n=1}^N{ \tet _1 (w-\bar s_n |\tau)  \over \tet _1 (w-p_n |\tau )}
\times \exp \left \{ - { 2 \pi \over \tau_2} w \Lambda_- \right \} 
\eea
where $\omega _0$ is a complex constant while $\Lambda_\pm$ are given by,
\bea
\Lambda_+  = \sum _{n=1}^N (s_n  -  p_n)
\hskip 1in 
\Lambda _-  = \sum _{n=1}^N (\bar s_n  -  p_n )
\eea
By construction, and with the help of (\ref{4b3}), the differentials $\p_w \cA_\pm$ are invariant under $ w \to w+\tau$, but their invariance under $w \to w+1$ requires the extra conditions,
\bea
\label{4c5}
\Lambda_+, \, \Lambda_- \, \in ~ \ZZ \, \tau
\eea
These conditions on $\Lambda _\pm$ combined imply the divisor relation (\ref{4b5a}) derived earlier. 

\sm

It remains to enforce the conjugation condition of (\ref{4c3}). In the special case where all the poles $p_n$ lie on the real boundary component,  the condition amounts to requiring $\bar \omega _0 = - \omega _0$. In the general case when  poles are allowed to lie on both boundary components, enforcing (\ref{4c3}) is more delicate, and we have instead the general relation,
\bea
\label{4c8}
\omega _0  =  \omega _0' \, \exp  \sum _{n=1}^ N { \pi \over \tau_2} ( -p_n^2 - p_n  \Big )
\eea
where $\omega _0'$ is constant and is subject to the relation $\bar \omega _0' = - \omega _0'$. Satisfying the combination of the conditions (\ref{4c5}) and (\ref{4c8}) thus provides single-valued meromorphic differentials $\p_w \cA_\pm$ on the annulus which satisfy the conjugation condition (\ref{4c3}).

\subsection{The functions \texorpdfstring{$\cA_\pm$}{A}}

In this subsection, we shall integrate the differentials $\p_w \cA_\pm$ to obtain the functions $\cA_\pm$. The product representation (\ref{4c4}) obtained in the preceding subsection is inconvenient to carry out this integration. Instead, we shall derive here an equivalent representation as a sum over meromorphic Abelian differentials which may be easily integrated.

\sm

Since the meromorphic differentials $\p_w \cA_\pm $ of (\ref{4c4})  are single-valued  on the double surface $\hat \Sigma$,  the sum of their residues $Z^\ell_\pm$ at the poles $p_\ell$ must vanish, and the differentials  can be expressed as a sum over meromorphic Abelian differentials $ \p_w \ln \tet _1 (w-p_\ell |\tau)  $, which are single-valued under $w \to w+1$, and transform  under $w \to w +\tau$ by a constant shift,
\bea
\label{4d2}
\p_w \cA_+ (w|\tau) & = & \alpha_+ + \sum _{\ell =1}^N Z_+ ^\ell \, \p_w \ln \tet_1 (w-p_\ell |\tau)
\hskip 1in 
\sum _{\ell=1}^N Z^\ell _\pm =0
\no \\
\p_w \cA_- (w|\tau) & = & \alpha_- + \sum _{\ell =1}^N Z_- ^\ell \, \p_w \ln \tet_1 (w-\bar p_\ell |\tau)
\eea
The residues $Z^\ell_\pm $ may be obtained from (\ref{4c4}) by matching poles and are given by,
\bea
\label{4d3}
Z_+ ^\ell & = & {\omega _0' \lambda _0 \over \tet _1 '(0 |\tau)} \, {  \prod _{n=1}^N  \vartheta _1 (p_\ell -s_n |\tau)  \over \prod _{ n\not = \ell } \tet _1 (p_\ell -p_n |\tau )}
\exp \left \{ - { 2 \pi \over \tau_2} \, p_\ell \, \Lambda_+ - { \pi \over \tau_2} \sum _{n=1}^N  (p_n^2 + p_n) \right \} 
\no \\
Z_- ^\ell & = & {\omega _0' \bar \lambda _0 \over \tet _1 '(0 |\tau)} \, {  \prod _{n=1}^N  \vartheta _1 (\bar p_\ell - \bar s_n |\tau)  \over \prod _{n \not = \ell }  \tet _1 (\bar p_\ell - \bar p_n |\tau )}
\exp \left \{ - { 2 \pi \over \tau_2} \, \bar p_\ell \, \bar \Lambda_+  - { \pi \over \tau_2} \sum _{n=1}^N  (\bar p_n^2 + \bar p_n) \right \} 
\eea
The presence of the constants $\alpha _\pm$ is due to the fact that the torus $\hat \Sigma$ has a one-dimensional space of holomorphic Abelian differentials which, in local complex coordinates $w$, is generated by the differential $dw$. A meromorphic one-form such as $\p_w \cA_\pm (w)$ will generically have a non-trivial component along $dw$.  The values of $\alpha _\pm$ may be obtained  by using (\ref{4c3}) and the vanishing of  $\p_w \cA_+$ at any of the zeros  $s_*  \in \{ s_1 , \cdots, s_N\}$, 
\bea
\label{4d4}
 \alpha _+ = -  \sum _{\ell=1}^N Z_+ ^\ell \, \p_{s_*} \ln \tet_1 (s_*-p_\ell | \tau) 
\hskip 1in 
\alpha _- = - \, \bar \alpha _+
\eea
The integrals of $\p_w \cA_\pm$ are now easily computed,  and we have,
\bea
\label{4d5}
\cA_+ (w|\tau) & = &   \cA^0 _+  + \alpha _+ w +  \sum _{\ell =1}^L Z_+ ^\ell \, \ln \tet _1  (w - p_\ell  |\tau)
\no \\
\cA_- (w|\tau) & = &   \cA^0 _-  + \alpha _- w +  \sum _{\ell =1}^L Z_- ^\ell \, \ln \tet _1  (w - \bar p_\ell  |\tau)
\eea
where $\cA^0_\pm$ are complex integration constants. 

\sm

To secure useful conjugation properties, care is needed in the choice of branch cuts for the logarithm.  Given a choice of branch cut in $\cA_+$, the proper branch cut in $\cA_-$ is actually more properly written as, 
\bea
\label{4d6}
\cA_- (w|\tau) =   \cA^0 _-  + \alpha _- w +  \sum _{\ell =1}^L Z_- ^\ell \, \overline{\ln \tet _1  (\bar w - p_\ell  |\tau)}
\eea
where the branch cuts chosen in $\cA_\pm$ are now the same. With this choice, we may use a constant gauge transformation of the flux field $\cC$ to set  $\cA_-^0+\bar\cA_+^0$ to zero, just as we did in the case of the upper half plane, upon which we obtain the following conjugation relation,
\bea
\label{4d7}
\overline{ \cA_\pm (\bar w |\tau) } = - \cA_\mp (w |\tau)
\eea
 In the next subsection, we obtain the conditions under which $\cG=0$ on the boundary.

\subsection{Conditions for \texorpdfstring{$\cG=0$}{G=0} on the boundary} 

The general arguments of subsection \ref{sec:232} along with the conjugation conditions (\ref{4c3}) and (\ref{4d7}) lead us to conclude that $\cG$ is constant along any line segment free of poles $p_\ell$ on the boundary $\p \Sigma$. To enforce the boundary condition $\cG=0$ on the entire boundary, it will therefore suffice to enforce $\cG=0$ on any one line segment free of poles on each boundary, and then require that the monodromy across every pole on that boundary component vanishes.

\sm

\begin{figure}
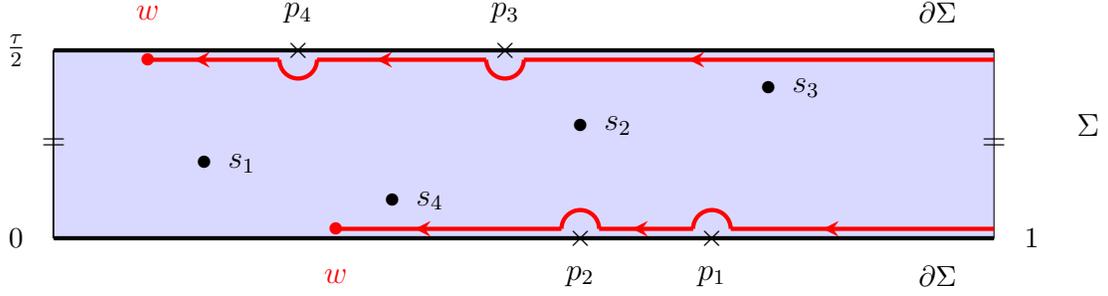

\begin{center}
\tikzpicture[scale=2.5]
\scope[xshift=-5cm,yshift=-0.4cm];
\draw [thick] (0,0) -- (0,1);
\draw [fill=blue!15!white] (5,0) rectangle (0,1);
\draw [ultra thick] (0,0) -- (5, 0);
\draw [ultra thick] (0,1) -- (5, 1);
\draw (5.2,0) node{\small $1$};
\draw (-0.2,0) node{\small $0$};
\draw (-0.2,1) node{${\tau \over 2} $};
\draw [ultra thick, red] (3.4,0.05) arc (180:0:0.1 and 0.1);
\draw [ultra thick, red] (2.7,0.05) arc (180:0:0.1 and 0.1);
\draw [ultra thick, red, directed] (5,0.05) -- (3.6, 0.05);
\draw [ultra thick, red, directed] (3.4,0.05) -- (2.9, 0.05);
\draw [ultra thick, red, directed] (2.7,0.05) -- (1.5, 0.05);
\draw [ultra thick, red] (2.3,0.95) arc (-180:0:0.1 and 0.1);
\draw [ultra thick, red] (1.2,0.95) arc (-180:0:0.1 and 0.1);
\draw [ultra thick, red, directed] (5,0.95) -- (2.5, 0.95);
\draw [ultra thick, red, directed] (2.3,0.95) -- (1.4, 0.95);
\draw [ultra thick, red, directed] (1.2,0.95) -- (0.5, 0.95);
\draw (5.5,0.6) node{$\Sigma$};
\draw (4.7,1.2) node{\small $\p \Sigma$};
\draw (4.7,-0.2) node{\small $\p \Sigma$};
\draw (1,0.4) node{$s_1$};
\draw (0.8,0.4) node{$\bullet$};
\draw (3,0.6) node{$s_2$};
\draw (2.8,0.6) node{$\bullet$};
\draw (4,0.8) node{$s_3$};
\draw (3.8,0.8) node{$\bullet$};
\draw (2,0.2) node{$s_4$};
\draw (1.8,0.2) node{$\bullet$};
\draw (3.5,-0.2) node{$p_1$};
\draw (3.5,0) node{$\times$};
\draw (2.8,-0.2) node{$p_2$};
\draw (2.8,0) node{$\times$};
\draw (2.4,1.2) node{$p_3$};
\draw (2.4,1) node{$\times$};
\draw (1.3,1.2) node{$p_4$};
\draw (1.3,1) node{$\times$};
\draw (0,0.5) node{$=$};
\draw (5,0.5) node{$=$};
\draw [red] (1.5, 0.05) node{$\bullet$};
\draw [red] (1.5, -0.2) node{$w$};
\draw [red] (0.5, 0.95) node{$\bullet$};
\draw [red] (0.5, 1.2) node{$w$};
\endscope
\endtikzpicture
\caption{The annulus $\Sigma$ is represented as in Figure \ref{fig7}. The contours of integration along the upper and lower boundary components of $\p \Sigma$ are indicated with thick red lines.\label{fig5}}
\end{center}
\end{figure}

We begin by requiring the vanishing of the monodromy of $\cG$  across an arbitrary pole $p_k$ along  the contour illustrated in Figure \ref{fig5}. To keep track of the behavior of the functions $\cA_\pm$ near their branch cuts, we introduce $\ep_k$ defined by,  
\bea
 \ep_k= r e^{i\delta_k\phi}
 \hskip 0.6in
  0<r,\phi \ll 1
\eea
where $\delta_k=1$ for poles on the real line and $\delta_k=-1$ for poles on the second boundary component.
The resulting $\ep_k$ has a small positive real part and an imaginary part reflecting the contours in Figure \ref{fig5}. To find the jump conditions we start out from, 
\bea
\Delta_k \cG & = & 
 |\cA_+ (p_k- \bar\ep_k|\tau)|^2 - |\cA_+ (p_k+ \ep_k|\tau)|^2
 \no \\ &&
 - |\cA_- (p_k- \bar\ep_k|\tau)|^2 + |\cA_- (p_k+ \ep_k|\tau)|^2
 +\Delta_k\cB+\Delta_k\bar\cB
 \eea
where
\bea
\Delta_k \cB &=& \int _{C_k} dz \Big ( \cA_+ (z|\tau) \, \p_z \cA_- (z |\tau) - \cA_- (z|\tau) \, \p_z \cA_+ (z |\tau) \Big )
\eea
with $C_k$ the half circle part of the contour around the pole $p_k$.
For the change in $\cA_\pm$ we find 
\bea
 \cA_\pm(p_k-\bar\ep_k|\tau) =A_\pm(p_k+\ep_k|\tau)+i\pi \delta_k Z_\pm^k
\eea
which leads to
\bea
  \Delta_k \cG &=& 
  i\pi \delta_k Z_+^k\left(\overline{\cA_+(p_k+\ep_k|\tau)}-\cA_-(p_k+\ep_k|\tau)\right)
  \nonumber\\&&
  +i\pi \delta_k Z_-^k\left(\cA_+(p_k+\ep_k|\tau)-\overline{\cA_-(p_k+\ep_k|\tau)}\right)
   +\Delta_k\cB+\Delta_k\bar\cB
\eea
Working this out more explicitly yields
\bea
\Delta_k\cG&=&
 i\pi\delta_k\Big[
 Z_-^k\left(2\cA_+^0+\alpha_+(p_k+\overline{p_k})\right)
 -Z_+^k\left(2\cA_-^0+\alpha_-(p_k+\overline{p_k})\right)
 \nonumber\\
 &&\hphantom{i\pi\delta_k\Big[}
 +\sum_{\ell=1}^L Z_+^\ell Z_-^k\left(\ln\vartheta_1(p_k+\ep_k-p_\ell)+\overline{\ln\vartheta_1(p_k+\ep_k-\overline{p_\ell})}\right)
 \\ \nonumber
 &&\hphantom{i\pi\delta_k\Big[}
 -\sum_{\ell=1}^L Z_+^kZ_-^\ell\left(\overline{\ln\vartheta_1(p_k+\ep_k-p_\ell)}+\ln\vartheta_1(p_k+\ep_k-\overline{p_\ell})\right)\Big]
 +\Delta_k\cB+\Delta_k\bar\cB
\eea
Using $\overline{p_\ell}=p_\ell-(1-\delta_\ell)\tau/2$ and the quasi-periodicity of the $\vartheta_1$ functions, 
the log terms in the round brackets can be evaluated more explicitly.
The integrals in $\Delta_k\cB+\Delta_k\bar\cB$ for the jump are restricted to a half circle around the pole, which means the integrand can be expanded for $|w-p_k|\ll 1$.
Upon explicit evaluation we find that the contribution from $\Delta_k\cB+\Delta_k\bar\cB$ precisely matches the change in $|\cA_+(w|\tau)|^2-|\cA_-(w|\tau)|^2$ and merely produces an overall factor of $2$.
Altogether, we find
\bea
\label{eqn:sumDelta-cG}
 \frac{\Delta_k\cG}{2\pi i \delta_k}&=&
  Z_-^k\left(2\cA_+^0+\alpha_+(p_k+\overline{p_k})\right)
  -Z_+^k\left(2\cA_-^0+\alpha_-(p_k+\overline{p_k})\right)
 +\sum_{\ell\neq k} Z^{[\ell ,k]}\ln|\vartheta_1(p_k-p_\ell)|^2
 \nonumber\\&&\ 
 -\sum_{\ell=1}^L i\pi\frac{p_\ell-\overline{p_\ell}}{\tau}\left[
  (1+p_\ell+\overline{p_\ell}-2p_k)Z_+^kZ_-^\ell+(1+p_\ell+\overline{p_\ell}-2\overline{p_k})Z_+^\ell Z_-^k
 \right]
\eea
The sum then evaluates to
\bea
\label{eq:DeltaG-sum}
 \sum_{k=1}^L\frac{\Delta_k\cG}{2\pi i \delta_k}&=&\sum_{k=1}^L(p_k+\overline{p_k})\left(Z_-^k\alpha_+-Z_+^k\alpha_-\right)
 \nonumber\\ &&
 +\frac{2\pi i}{\tau}\sum_{k=1}^L\sum_{\ell=1}^L(p_\ell-\overline{p_\ell})\left(p_k Z_+^k Z_-^\ell+\overline{p_k}Z_+^\ell Z_-^k\right)
\eea
The sum manifestly vanishes if $\alpha_+=0$ and all poles are on one boundary component. In this case, the conditions (\ref{eqn:sumDelta-cG}) simplify considerably, and we have,
\bea
\label{5f30}
Z_-^k \cA_+^0  - Z_+^k \cA_-^0
 + \sum_{\ell\neq k} Z^{[\ell ,k]}\ln|\vartheta_1(p_k-p_\ell)|
=0
\eea
which is analogous to the condition for the disk in (\ref{3f7}).

\subsection{Minimal number of poles}

For the case of the disk, a minimum of three poles in $\p_w \cA_\pm$ was required in order to have a single zero  in the upper half plane, lest  the solution be trivial. For the case of the annulus, the number of zeros of $\p_w \cA_\pm$ must equal the number of its poles instead. We shall now analyze the minimal number of zeros and poles required for the annulus case.

\sm

When all the poles $p_\ell$ of $\p_w \cA_\pm$ are on one boundary component the requirement $\Lambda_+=\sum_{n=1}^N(s_n-p_n)\in \ZZ \tau$  implies  $\Lambda_-\in \ZZ \tau$, and hence both conditions in eq.~(\ref{4c5}). Periodicity of the functions $\cA_\pm$ as $w\rightarrow w+1$ requires $\alpha_\pm=0$. This is one complex condition due to $\alpha_-=-\bar\alpha_+$, and is equivalent to the condition, 
\bea
\label{eq:alpha-plus}
 \alpha_+=
 -\sum_{\ell=1}^{N}Z_+^\ell\partial_s\ln\vartheta_1(s-p_\ell|\tau) = 0
\eea
where $s$ is any one of the zeros of $\p_w \cA_\pm$ (cf.~(\ref{4d4})).

\sm

The condition $\Lambda_+\in \ZZ \tau$ has no solutions with two poles on the same boundary, since this would require two zeros in the interior of $\Sigma$, whose imaginary parts add up to a non-zero multiple of $\tau$ by conditions (\ref{4c5}). The only possibility would be to have their imaginary parts both equal to $\tau_2/2$, but this implies in turn that both zeros are on the boundary of $\Sigma$ leading to an unphysical solution.

\sm

The condition $\Lambda_+\in \ZZ \tau$ can be solved with two poles and two zeros inside the annulus, provided we have one pole on each boundary component. Without loss of generality, we use translation invariance to set $p_1 =0$ and  $p_2\in \RR+\frac{\tau}{2}$. Assuming furthermore that $\alpha_+=0$, the condition (\ref{eq:DeltaG-sum}) with $\sum_k Z_\pm^k=0$ then implies $\Re(p_2)=0$ and thus
$p_2=\tau/2$. The condition $\alpha_+=0$, via (\ref{eq:alpha-plus}) and with $\sum_k Z_\pm^k=0$, then implies,
\bea
\partial_s\left[\ln\vartheta_1(s |\tau )-\ln\vartheta_1\left(s-\frac{\tau}{2} |\tau \right)\right] =0
\eea
This equation only has two unacceptable solutions below the real line, and no solutions for $s$ inside the annulus.  We thus find that there are no solutions with two poles, and that at least three poles are a necessary condition for a physically regular solution to exist.

\subsection{Investigating solutions with at least three poles}
\label{sec:57}

Starting with 3 poles it becomes considerably more involved to disentangle the collection of conditions developed earlier. While we cannot offer an exhaustive analysis here, we shall present the conclusions we draw from a partial numerical analysis we have carried out.  

\sm

With $N$ poles for $N\geq 3$, the divisor conditions (\ref{4c5}) allow for all poles to be distributed over both boundary components including the case where all $N$ poles are on one boundary component. Our numerical analysis of the conditions $\alpha _\pm=0$, which are required to make $\cA_\pm$ single-valued under $w \to w+1$,  in the case of $N=3, 4, 5$ poles appears to exclude systematically the cases where all poles of $\p_w \cA_\pm$ are {\sl not} on the same boundary. Indeed, when poles occur on both boundary components, we find that the condition $\alpha _\pm=0$, for a given distribution of poles on the boundary and for $N-2$ zeros inside $\Sigma$,  always forces one of the remaining zeros of $\p_w \cA_+$ to be outside of $\Sigma$. So far, we have no analytical proof of this claim.

\sm

With $N \geq 3$ poles, and all $N$ poles on a single boundary component, it is possible to satisfy the conditions $\alpha _\pm=0$ and have all $N$ zeros of $\p_w \cA_+$ in the interior of $\Sigma$. A simple family of such solutions may be obtained by choosing the distribution of poles and zeros to be invariant under $\ZZ_N$ translations, and  given by,
\bea
p_n = { n \over N} \hskip 1in s_n = { n \over N} + {\tau \over N} \hskip 1in n=1,\cdots, N
\eea
For given $N$ the zeros are well in the interior of $\Sigma$. Therefore, by continuity, we know that there will exist an open set in the space of all solutions which contains the above solution as  a point, and has maximal dimension, given by the real coordinates  of poles $N$, the complex coordinates of $N$ zeros, minus the complex condition $\alpha_+=0$, and minus the real divisor condition, adding up to $3N-3$. 

\sm

Having established that there are at least the partial solutions exhibited above, we more generally explored configurations with
all poles on a single boundary, and showed numerically that the conditions for continuity of the function $\cG$ across the poles, namely the conditions $\Delta_n \cG=0$ for $n=1,\dots,N$, may be solved as well. The solutions to all these conditions combined now provide configurations for which all zeros of $\p_w \cA_+$ are in the interior of $\Sigma$, the condition $\alpha _+=0$ is satisfied, and $\cG$ is constant on each boundary component. 

\sm

There now remain two further conditions to be implemented, namely that $\cG=0$ on both boundary components. There is a natural free parameter, namely the integration constant $\cB^0$ of the composite holomorphic function $\cB$ which may be used to set $\cG=0$ on one boundary component. The value of $\cG$ on the other boundary component is then determined by the parameters of the solution. Our numerical analysis for the case of $N=3,4,5$ poles appears to show that the condition $\cG=0$ can not be satisfied simultaneously on both boundary components. Rather, the difference between the values of $\cG$ on the boundary with no poles and on the boundary with all poles appears to consistently be positive. Again, at this time, we have no analytical proof of this claim, but hope to return to this question in the future. 

\sm

If the  indications from the numerical analyses carried out so far are correct, then the annulus may not support physically regular solutions in the absence of axion monodromy. More generally, we may speculate that the obstruction to the existence of solutions resides in the presence of more than one boundary component, in the case of the annulus as well as for surfaces of arbitrary genus.

\newpage

\section{Riemann surfaces of arbitrary topology}
\setcounter{equation}{0}
\label{sec:generalRiemann}

In this last section, we shall construct suitable differentials $\p_w \cA_\pm$ for general Riemann surfaces, and spell out the conditions required for the corresponding supergravity solutions to be physically regular. The final equations are so complicated, however, that so far we have not succeeded in constructing acceptable solutions, even numerically. 

\sm

Consider an orientable Riemann surface $\Sigma$ of genus $g$ with $\nu \geq 1$ boundary components which topologically are circles. Functions and differential forms on $\Sigma$ may be constructed in terms of functions and differential forms on the double surface $\hat \Sigma$, equipped with an anti-conformal involution which we shall denote by $\mI$ (for a standard reference, see  \cite{Fay}).  The boundary $\p \Sigma$ of $\Sigma$ is fixed, point by point,  under $\mI$ so that $\mI (\p \Sigma) = \p \Sigma$. We may view $\Sigma$ as the quotient $\Sigma = \hat \Sigma / \mI$. The genus $\hat g$ of  $\hat \Sigma$ is related to the genus $g$ and the number of boundary components $\nu$ of the original surface $\Sigma$ by the standard relation,
\bea
\label{6a0}
\hat g = 2g + \nu -1
\eea
This construction is depicted in Figure \ref{fig9} for the case $g=2$ and $\nu=3$. The conventions for the basis of homology cycles indicated in the figure will be given in section \ref{sec:62} below.

\subsection{Generalizing the electrostatics analogy}

The electrostatics analogy, used to construct the function $\lambda$ for the cases of the upper half plane and the annulus, may be generalized without complications to the case of an oriented Riemann surface of arbitrary genus $g$, and an arbitrary number $\nu$ of boundary components the  topology of each of which is that of a circle. In the electrostatics analogy, we seek first to construct an electrostatic potential $ - \ln |\lambda|^2$ which vanishes on the boundary $\p\Sigma$ and is strictly positive everywhere in the interior of $\Sigma$. To achieve such an electrostatic potential, we ground the system to zero potential on every  component of $\p \Sigma$ and place an arbitrary arrangement of positive charges in the interior of $\Sigma$. By the min-max principle for harmonic functions, used already earlier in subsection \ref{sec:23}, the potential $ - \ln |\lambda|^2$ is then guaranteed to be positive everywhere in the interior of $\Sigma$. 

\sm

Mathematically, we start from the original Riemann surface $\Sigma$ of genus $g$ with $\nu$ boundary components and construct its double  $\hat \Sigma$ endowed  with a conformal involution $\mI$. To obtain a potential  which is strictly positive in the interior of $\Sigma$, and which vanishes on the boundary $\p \Sigma$, we place an arrangement of an arbitrary number $N$ of positive electric charges $q_n$ at arbitrary points $s_n$ in the interior of $\Sigma$,  with $n=1,\cdots, N$, and place mirror charges $- q_n$ at the mirror locations $\mI (s_n)$. Since by construction the potential $ - \ln |\lambda|^2$ is odd under the involution $\mI$, it is guaranteed to vanish on the boundary $\p \Sigma$. Since all charges on $\Sigma$ are positive, and the potential  vanishes on $\p \Sigma$, we use again the min-max principle for harmonic functions to argue that $ - \ln |\lambda|^2$ must then be strictly positive in the interior of $\Sigma$, just as we had already done in the case of the upper half plane and the 
annulus. Clearly,  such solutions will always exist. To obtain explicit formulas for  $\lambda$, we shall provide the necessary mathematical set-up in the next subsection.

\subsection{Double surface \texorpdfstring{$\hat \Sigma$}{Sigma-hat} and involution \texorpdfstring{$\mI$}{I}}
\label{sec:62}

\begin{figure}
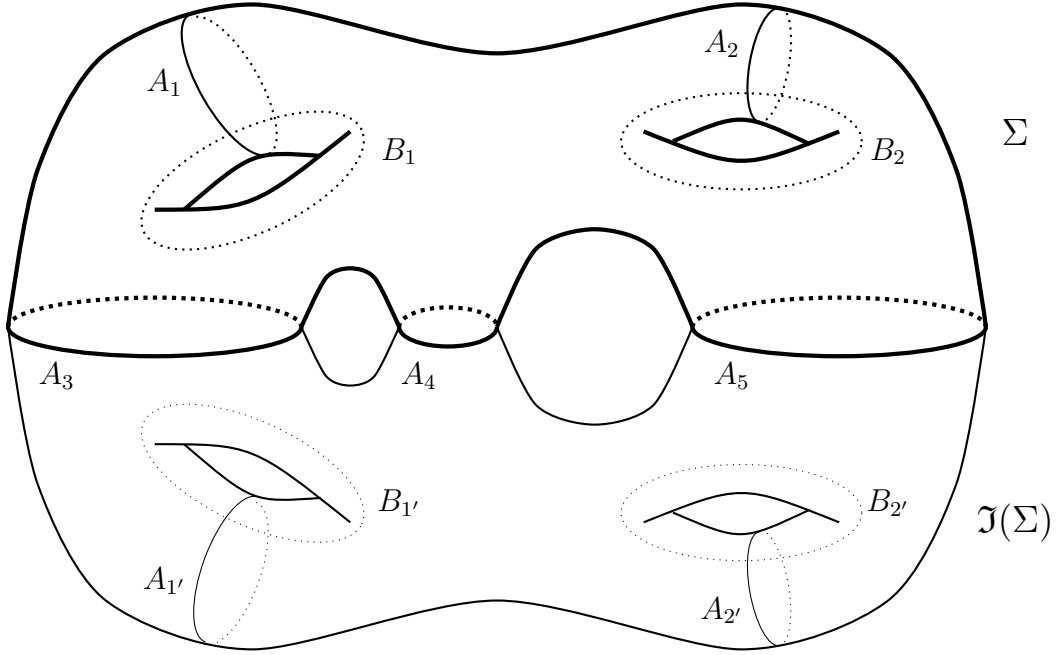

\begin{center}
\tikzpicture[scale=1.3]
\scope[xshift=-5cm,yshift=0cm]
\draw [ultra thick, dotted] (0,0) arc (180:0:1.5 and 0.3);
\draw [ultra thick] (0,0) arc (-180:0:1.5 and 0.3);
\draw [ultra thick, dotted] (4,0) arc (180:0:0.5 and 0.2);
\draw [ultra thick] (4,0) arc (-180:0:0.5 and 0.2);
\draw [ultra thick, dotted] (7,0) arc (180:0:1.5 and 0.3);
\draw [ultra thick] (7,0) arc (-180:0:1.5 and 0.3);
\draw[ultra thick] plot [smooth] coordinates {(0,0) (0.3,1.6) (1,2.8) (2.5, 3.3)  (5,2.8) (7.5, 3.3) (9,2.8) (9.7,1.6) (10,0)};
\draw[ultra thick] plot [smooth] coordinates {(3,0) (3.25, 0.5) (3.5,0.6) (3.75, 0.5) (4,0)};
\draw[ultra thick] plot [smooth] coordinates {(5,0) (5.4, 0.8) (6,1) (6.6, 0.8) (7,0)};
\draw[ultra thick] (1.5,1.2) .. controls (2.5, 1.2) .. (3.5,2);
\draw[ultra thick] (1.8,1.2) .. controls (2.5, 1.8) .. (3.2,1.75);
\draw[ultra thick] (6.5,2) .. controls (7.5, 1.6) .. (8.5,2);
\draw[ultra thick] (6.8,1.9) .. controls (7.5, 2.2) .. (8.2,1.87);
\draw[thick, dotted,  rotate=25] (2.9,0.3)  ellipse (35pt and 15pt);
\draw[thick, dotted,  rotate=00] (7.5,1.9)  ellipse (35pt and 14pt);
\draw[thick, dotted,  rotate=30] (3.2,0.2)  arc (-90:90:0.3 and 0.8);
\draw[thick,  rotate=30] (3.2,1.84)  arc (90:270:0.3 and 0.82);
\draw[thick, dotted,  rotate=-10] (7.2,3.4)  arc (-90:90:0.2 and 0.6);
\draw[thick,  rotate=-10] (7.2,4.58)  arc (90:270:0.2 and 0.59);
\draw[thick] plot [smooth] coordinates {(0,0) (0.3,-1.6) (1,-2.8) (2.5, -3.3)  (5,-2.8) (7.5, -3.3) (9,-2.8) (9.7,-1.6) (10,0)};
\draw[thick] plot [smooth] coordinates {(3,0) (3.25, -0.5) (3.5, -0.6) (3.75, -0.5) (4,0)};
\draw[thick] plot [smooth] coordinates {(5,0) (5.4, -0.8) (6,-1) (6.6, -0.8) (7,0)};
\draw[thick] (1.5,-1.2) .. controls (2.5, -1.2) .. (3.5,-2);
\draw[thick] (1.8,-1.2) .. controls (2.5, -1.8) .. (3.2,-1.75);
\draw[thick] (6.5,-2) .. controls (7.5, -1.6) .. (8.5,-2);
\draw[thick] (6.8,-1.9) .. controls (7.5, -2.2) .. (8.2,-1.87);
\draw[ dotted,  rotate=-25] (2.9,-0.3)  ellipse (35pt and 15pt);
\draw[ dotted,  rotate=00] (7.5,-1.9)  ellipse (35pt and 14pt);
\draw[ dotted,  rotate=-20] (3,-2.36)  arc (-90:90:0.3 and 0.77);
\draw[ rotate=-20] (3,-0.76)  arc (90:270:0.3 and 0.79);
\draw[ dotted,  rotate=10] (7.2,-4.58)  arc (-90:90:0.2 and 0.6);
\draw[ rotate=10] (7.2,-3.4)  arc (90:270:0.2 and 0.59);
\draw (1.6,2.5) node{$A_1$};
\draw (7.3,2.9) node{$A_2$};
\draw (4,1.8) node{$B_1$};
\draw (9,1.8) node{$B_2$};
\draw (1.6,-2.6) node{$A_{1'}$};
\draw (7.3,-2.9) node{$A_{2'}$};
\draw (4,-1.8) node{$B_{1'}$};
\draw (9,-1.8) node{$B_{2'}$};
\draw (0.5,-0.5) node{$A_3$};
\draw (4.2,-0.5) node{$A_4$};
\draw (7.4,-0.5) node{$A_5$};
\draw (10.3,2) node{{\large $\Sigma$}};
\draw (10.3,-2) node{{\large $\mI (\Sigma)$}};
\endscope
\endtikzpicture
\end{center}
\caption{A surface $\Sigma$ of genus $g=2$ and $\nu=3$ boundary components $A_3, A_4, A_5$, along with its double surface without boundary  $\hat \Sigma = \Sigma \cup \mI (\Sigma)$  of genus $\hat g = 6$.\label{fig9}}
\end{figure}

We shall sort the homology cycles on the double surface $\hat \Sigma$, and their dual holomorphic 1-forms, according to their parity  under  $\mI$. Our labeling of the cycles generalizes to arbitrary $g$ and $\nu$  the labeling indicated in Figure~\ref{fig9} for  $g=2$ and $\nu=3$, and we have \cite{Fay}, 
\bea
\label{6a1}
\hbox{$2g$ cycles belonging to $\Sigma$}  \hskip 0.35in & \hskip 0.6in & A_I, \, B_I \hskip 0.48in I=1, \ldots, g
\\
\hbox{$2g$ cycles conjugate under $\mI$} \hskip 0.11in && A_{I'} , \, B_{I'} \hskip 0.4in I'=1', \ldots, g
\no \\
\hbox{$\nu-1$ boundary cycles for $\p \Sigma$} \hskip 0in &&  A_i  \hskip 0.8in i=g+1, \ldots, g+\nu-1
\no \\
\hbox{$\nu-1$ conjugate cycles for $\p \Sigma$} &&  B_i  
\no
\eea
There is one further boundary component, $A_{g+\nu}$ which is homologically trivial on $\hat \Sigma$. 
Throughout, the indices $I$,  $I'$, $i$ will run over the ranges given in (\ref{6a1}) and we shall conveniently use the composite index  $\hat I = ( I, I',i)$. We define the involution matrix $\cS$ by, 
\bea
\label{6a2}
\cS = \left ( \matrix{ 0 & I_g & 0 \cr I_g & 0 & 0 \cr 0 & 0 & I_{\nu-1} \cr } \right )
\eea
where $I_g$ and $I_{\nu-1}$ are the identity matrices respectively in dimensions $g$ and $\nu-1$.
The cycles may be arranged to have definite parity under $\mI$,  which we choose as follows, 
\bea
\label{6a3}
\mI (A_{\hat I}) = + \sum _{\hat J} \cS_{\hat I \hat J} A_{\hat J} \hskip 1in 
\mI (B_{\hat I}) = - \sum _{\hat J} \cS_{\hat I \hat J} B_{\hat J}
\eea
The normalization of the holomorphic 1-forms on the $A$-cycles and their corresponding behavior under the pull-back $\mI^*$  of the involution $\mI$ to 1-forms, are given as follows, 
\bea
\label{6a4}
\oint _{A_{\hat I}} \om _{\hat J} = \delta _{\hat I \hat J} 
\hskip 1in
\mI ^* (\om _{\hat I}) = \sum _{\hat J} \cS_{\hat I \hat J} \, \bar \om _{\hat J}
\eea
where $\bar \omega _{\hat J}$ is the complex conjugate of $\omega _{\hat J}$.
The period matrix $\Omega _{\hat I \hat J}$ of $\hat \Sigma$ is defined by, 
\bea
\label{6a5}
\oint _{B_{\hat I}} \om_{\hat J} = \Omega _{\hat I \hat J}
\eea
By the Riemann bilinear relations, $\Omega_{\hI \hJ}$  is symmetric in $\hat I, \hat J$ and has positive definite  imaginary part. As a result of the action of $\mI$ on the cycles in (\ref{6a3}), and on the holomorphic 1-forms in (\ref{6a4}),  $\Omega _{\hat I \hat J}$ satisfies further relations. To derive them, we make use of the formulas, 
\bea
\label{6a5a}
\int _{\mI (z) } ^ {\mI (w)} \mI ^* \omega _{\hat I} = \int ^z _w \om _{\hat I}
\hskip 1in 
\overline{\int ^z _w \om _{\hat I} } = \int ^z _w \bar \om _{\hat I}
\eea
Combining these results, we proceed by the following manipulations,
\bea
\label{6a6}
\Omega _{\hat I \hat J } = \oint _{B_{\hat I} } \om_{\hat J} 
= - \sum _{\hat K, \hat L} \cS_{\hat I \hat K} \cS _{\hat J \hat L}  \oint  _{\mI (B_{\hat K} )} \mI ^* (\bar \om _{\hat L}) 
= - \sum _{\hat K, \hat L} \cS_{\hat I \hat K} \cS _{\hat J \hat L}  \oint  _{B_{\hat K} } \bar \om _{\hat L}
\eea
The result is the conjugation condition on $\Omega$, 
\bea
\label{6a7}
\bar \Omega = - \cS \, \Omega \, \cS 
\eea
Decomposing $\Omega = X+i Y$ into real matrices $X,Y$ decomposes   (\ref{6a7})  into 
$\cS X \cS = - X$ and $\cS Y \cS = Y$. The conjugation condition guarantees that the genus $\hat g$ Riemann surface with period matrix $\Omega$ indeed has an involution $\mI$. It is equivalent to a reality condition on $i \Omega$ up to conjugation by $\cS$, and reduces  the number of free parameters of $\Omega$ to half.

\subsection{The function \texorpdfstring{$\lambda$}{lambda}}

To construct $\lambda$ we follow the strategy of section \ref{sec:23}. We construct a general electrostatic potential  which is odd under the involution $\mI$ and which  is strictly positive in the interior of $\Sigma$. 
To do so we place an arbitrary arrangement of positive charges in $\Sigma$, and add their image charges  in $\mI (\Sigma)$.  The  scalar Green function $G_0$ on $\hat \Sigma$ may be chosen as follows, 
\bea
G_0(w,z |\Omega ) = - \ln | E(w,z) |^2 + 2 \pi  \sum _{\hat I,  \hat J} \left ( \Im \int _w^z \om _{\hat I} \right ) Y^{-1} _{\hat I \hat J}  \left ( \Im \int _w ^z \om _{\hat J} \right )
\eea
where $E(w,z)$ is the prime form on $\hat \Sigma$. The prime form is a holomorphic form of weight $(-\half, 0)$ in $w$ and $z$, which satisfies $E(z,w) = - E(w,z)$,  and whose asymptotics is given by $E(w,z) = w-z + \cO((w-z)^3)$ for $w$ near $z$. For its definition in terms of $\tet$-functions, and further properties, see for example \cite{Fay, D'Hoker:1988ta}. Monodromy transformations act as follows,
\bea
\label{EAB}
E(w+A_{\hat I}, z) & = & - E(w,z)
\no \\
E(w+B_{\hat I}, z) & = & - E(w,z) \exp \left \{ - i \pi \Omega _{\hat I \hat I} - 2 \pi i \int ^w _z \om _{\hat I} \right \} 
\eea
Under the involution $\mI$, the prime form behaves as follows,
\bea
\mI ^* E(w,z) = E\Big (\mI (w), \mI (z) \Big ) = \overline{ E(w,z)}
\eea
While the scalar Green function is not uniquely defined, since it would correspond to the electrostatic potential of a single electric charge, the Green function for a pair of opposite unit charges placed at conjugate points $z$ and $\mI (z)$ is well-defined, and given by, 
\bea
G(w,z|\Omega ) = G_0(w,z|\Omega ) - G_0 (w, \mI (z) |\Omega )
\eea
Its expression in terms of the prime form and Abelian integrals simplifies as follows,
\bea
\label{6c5}
G(w,z|\Omega ) = - \ln \left | { E(w,z)  \over E(w, \mI (z))  } \right |^2
-2  \pi \sum _{\hat I,  \hat J}  \left ( \Im \! \int _{\mI (w) } ^w \om _{\hat I} \right )
Y^{-1} _{\hat I \hat J} \left ( \Im \! \int  _{\mI (z) }^z \om _{\hat J} \right )
\eea
In order to produce a function $\lambda$ without branch cuts the positive charges should be chosen to be all equal to unity, giving the following general solution for the electrostatic potential,
\bea
 \ln |\lambda (w|\Omega)  |^2  =  \sum _{n=1}^N 
 \ln \left | { E(w,s_n)  \over E(w, \mI (s_n))  } \right |^2 
+2  \pi \sum _{n=1}^N  \sum _{\hat I,  \hat J} \left ( \Im \! \int _{\mI (w) } ^w \om _{\hat I} \right ) 
Y^{-1} _{\hat I \hat J} \left ( \Im \! \int  _{\mI (s_n) }^{s_n} \om _{\hat J} \right )
\quad
\eea
To split the holomorphic dependence in $w$ from the anti-holomorphic dependence we begin by splitting the line integral using (\ref{6a5a}),
\bea
\label{6c8}
\Im \int _{\mI (w) } ^w \om _\hI = { 1 \over 2i} \int ^w _{\mI (w)} \om _\hI 
- { 1 \over 2i} \overline{\int ^w _{\mI (w)} \om _\hI}
= { 1 \over 2i} \left ( \delta _{\hI \hJ} + S_{\hI \hJ} \right ) \int ^w _{\mI (w)} \om _\hJ
\eea
Splitting off the holomorphic dependence in $w$ now clearly amounts to splitting off the dependence in $w$ from  the dependence in $\mI (w)$. Using furthermore the fact that $\cS^2=I$ and that $\cS$ commutes with $Y$, and exponentiating the holomorphic part to obtain $\lambda$, we find,  
\bea
\label{6c9}
\lambda (w |\Omega) = \lambda _0^2  \prod _{n=1} ^N { E(w,s_n) \over E(w, \mI (s_n))} 
\exp \left \{ -  2 \pi i \sum _{\hat I, \hat J} \left ( \int _{w_0 } ^w \om _{\hat I} \right ) Y^{-1} _{\hI \hJ} \left ( \Im \! \int  _{\mI (s_n) }^{s_n} \om _\hJ \right ) \right \}
\eea
While $|\lambda (w|\Omega)|$ was single-valued on $\hat \Sigma$, the splitting into $\lambda (w|\Omega)$ and its complex conjugate generally introduces monodromy by phase factors. To obtain a single-valued $\lambda$ will necessitate divisor relations on the zeros $s_n$, which we now obtain.

\sm

The monodromy properties of the function $\lambda (w |\Omega)$  as $w$ is moved around the homology  cycles  $A_{\hat I}$ and $ B_{\hat I}$ on $\hat \Sigma$ are readily evaluated using (\ref{EAB}).   We state here, without detailed derivation, the conditions for the absence of  monodromies around all homology cycles,
\bea
\label{6c10}
 \sum _{n=1}^ N \int ^{s_n } _{\mI (s_n)} \om _\hI = m_\hI + \sum _{\hJ} \Omega _{\hI \hJ} n_\hJ
\eea
where the column matrices $m,n$ have integer entries, restricted in the following way,
\bea
 \matrix{ (I+\cS) m & = & 0 \cr
(I-\cS) n & = & 0 \cr } 
\hskip 1in
m = \left ( \matrix{ m_I \cr - m_I \cr 0 \cr } \right )
\hskip 0.5in 
n =  \left ( \matrix{ n_J \cr  n_J \cr n_j \cr } \right ) 
\eea
where the integers $m_I, n_I$ for $I=1,\cdots, g$ and $n_j$ for $j=g+1,\cdots, g+\nu-1$ are arbitrary. 
These conditions state that the zeros $\{s_1 , \cdots , s_N \}$ and poles $ \{ \bar s_1, \cdots , \bar s_N \}$ indeed satisfy the customary divisor condition of a meromorphic function $\lambda$ on a Riemann surface $\hat \Sigma$  which possesses an involution $\mI$ (see for example \cite{Fay, D'Hoker:1988ta}).

\subsection{The differentials \texorpdfstring{$\p_w \cA_\pm$}{dA}}

To construct the differentials $\p_w \cA_\pm$ we split the meromorphic function $\lambda$ on $\hat \Sigma$ into the ratio of two conjugate multiple-valued  holomorphic differentials on $\hat \Sigma$ as follows,
\bea
\label{6d1}
\lambda (w|\Omega) = { \lambda _+ (w |\Omega) \over \lambda _- (w |\Omega)}
\hskip 1in
\overline{ \lambda _\pm ( \mI(w) |\Omega ) } = \lambda _\mp (w |\Omega)
\eea
The differential form $\lambda _+$ has its zeros at $s_n$, while $\lambda _-$ has its zeros at $\mI (s_n)$. The splitting of the prime form factors in (\ref{6c9}) is manifest. To split  the exponential factor, we make use of (\ref{6c8}) but with $s_n$ replacing $w$.  The splitting of $\lambda$ is now manifest, and we find, 
\bea
\label{6d2}
\lambda _+ (w |\Omega) & = & 
\lambda _0  \prod _{n=1} ^N  E(w,s_n) \, 
\exp \left \{ -  2 \pi i \sum _{\hat I, \hat J} \left ( \int _{w_0 } ^w \om _{\hat I} \right )
Y^{-1} _{\hI \hJ}  \, \Im \int  _{w_0 }^{s_n} \om _\hJ  \right \}
\\
\lambda _- (w |\Omega) & = & 
\bar \lambda _0  \prod _{n=1} ^N  E(w,\mI (s_n)) \,
\exp \left \{ -  2 \pi i\sum _{\hat I, \hat J} \left ( \int _{w_0 } ^w \om _{\hat I} \right )
Y^{-1} _{\hI \hJ}  \, \Im \int  _{w_0 }^{\mI (s_n)} \om _\hJ  \right \}
\no
\eea
where the point $w_0$  lies on the boundary of $\Sigma$, so that we have $\mI (w_0) = w_0$. 
As constructed above, the  differential forms $\lambda _\pm$ have  weight $(-N/2,0)$ in $w$. 

\sm

Next, we construct the holomorphic differentials $\p_w \cA_\pm$ to be well-defined and single-valued on $\hat \Sigma$.  The basic equations  for $\p_w \cA_\pm$ are as follows, 
\bea
\p_w \cA_\pm (w |\Omega ) = \lambda _\pm (w|\Omega) \, \f  (w |\Omega ) 
\eea
The differential form $\f$ must have weight $(1+N/2,0)$ and  has neither zeros nor poles in the interior of $\Sigma$, so that its poles and zeros are on $\p \Sigma$. As in the cases of the upper half plane and the annulus, we shall view any real zeros as degenerations of pairs of conjugate zeros  $(s_n, \mI(s_n))$ to the boundary. We shall denote the poles by $p_\ell$ with $\ell =1 , \cdots, L$. To have well-defined and single-valued meromorphic differentials $\p_w \cA_\pm$ on $\hat \Sigma$, the number of their poles and zeros are related as follows,
\bea
\label{6d3}
N = L + 2 \hat g -2
\eea
For $\hat g\geq 2$, 
one may view the positions of the $L$ poles  and of $N-\hat g +1 $ of the zeros as arbitrary, while the remaining $\hat g -1$ zeros are determined by the divisor condition for forms of weight $(1,0)$. 
Taking these considerations into account,  we obtain the following  expressions for $\p_w \cA_\pm$,
\bea
\label{6d4}
\p_w \cA_+ (w|\Omega) & = & 
\lambda _0 \omega _0 \,  \sigma (w)^2 \,   \, {  \prod _{n=1} ^N  E(w,s_n) \over \prod _{\ell =1}^L E(w,p_\ell) } 
\exp \Bigg \{ - 2 \pi i \sum _{\hat I, \hat J}  \left ( \int _{w_0} ^w \om _\hI \right ) Y^{-1}_{\hI \hJ} \, \Im \Lambda ^+ _{\hat J} \Bigg \}
\\
\p_w \cA_- ( w |\Omega) & = &
\bar \lambda _0 \omega _0 \, \sigma (w)^2 \, 
 { \prod _{n=1} ^N  E(w,\mI (s_n))   \over \prod _{\ell =1}^L E(w,p_\ell)} 
\exp \Bigg \{ - 2 \pi i \sum _{\hat I, \hat J} \left ( \int _{w_0} ^w \om _\hI \right ) Y^{-1}_{\hI \hJ} \, \Im \Lambda ^-_{\hat J} \Bigg \}
\qquad 
\no
\eea
where $\Lambda ^\pm_\hJ$ are given by,
\bea
\label{6d5}
  \Lambda _\hJ^+    & = & \sum _{n=1}^N \int  _{w_0 }^{s_n} \om _{\hat J} - \sum _{\ell=1}^L \int ^{p_\ell} _{w_0} \omega _\hJ
  \no \\
   \Lambda _\hJ^-    & = & \sum _{n=1}^N \int  _{w_0 }^{\mI(s_n)} \om _{\hat J} - \sum _{\ell=1}^L \int ^{p_\ell} _{w_0} \omega _\hJ
 \eea
The holomorphic  form $\sigma (w)$ has weight $(\hat g /2,0)$ and has neither zeros nor poles. Its role in (\ref{6d4})  is to guarantee that the forms $\p_w \cA_\pm$ have the  weight $(1,0)$ in $w$, and are single-valued. The monodromy of the form $\sigma(w)$ around $A$-cycles on $\hat \Sigma$ is trivial; its monodromy around $B$-cycles and its expression in terms of the prime form are given below, 
\bea
\label{6d6}
\sigma (w+B_\hI) & = & \sigma (w) \, \exp \left \{ \pi i (\hat g -1) \Omega _{\hI \hI} + 2 \pi i (\hat g -1) \int _{w_0} ^w \omega _\hI  - 2 \pi i \Delta _\hI (w_0) \right \} 
\no \\
\Delta _\hI (w_0) & = & \half - \half \Omega _{\hI \hI} + \sum _{\hJ \not = \hI} \oint _{\hJ} \omega _\hJ (z) \int _{w_0} ^z \omega _\hI
\no \\
\sigma (w) & = & \exp \Bigg \{ - \sum _{\hJ=1}^{\hat g} \oint _{A_\hJ} \omega _\hJ (z) \ln E(z,w) \Bigg \} 
\eea
Here, $\Delta _\hI (w_0)$ is the Riemann vector on the surface $\hat \Sigma$. Choosing the base point $w_0$ to be invariant under the involution $\mI$,  the Riemann vector is real in the Jacobian $\CC^{\hat g} /(\ZZ^{\hat g} + \Omega \ZZ^{\hat g})$.
The conjugation property of $\sigma$ under conjugation follows that of the prime form and we have,
\bea
\mI^* (\sigma (w)) = \sigma (\mI(w)) = \overline{\sigma (w)}
\eea
Using it, we readily establish the conjugation relations for the differentials,
\bea
\p_w \cA_\pm (w) = - \overline{ \p_{\mI(w)} \cA_\mp(\mI(w))} = - \p_w \overline{\cA_\mp (\mI (w)) }
\eea
The transformation laws given above for the prime form, the exponential in (\ref{6d4}), and $\sigma$ allow us to compute the monodromies of $\p_w \cA_+$ around  $A_\hI$ and $B_\hI$ cycles. Requiring the monodromies to vanish around all cycles imposes the following conditions on $\Lambda _\hI^\pm$, 
\bea
\label{6d7}
\Lambda _\hI ^\pm  - 2 \Delta _\hI (w_0)  =  m_\hI^\pm + \sum _{\hJ} \Omega _{\hI \hJ} n_\hJ^\pm
\eea
where $m^\pm$ and $n^\pm $ are column matrices whose entries are integers which satisfy,
\bea
\label{6d8}
 \matrix{ (I+\cS) m^\pm  & = & 0 \cr
(I-\cS) n^\pm  & = & 0 \cr } 
\hskip 1in
m^\pm  = \left ( \matrix{ m_I^\pm  \cr - m_I^\pm  \cr 0 \cr } \right )
\hskip 0.5in 
n^\pm =  \left ( \matrix{ n_J^\pm  \cr  n_J^\pm  \cr n_j^\pm  \cr } \right ) 
\eea
where the integers $m_I^\pm , n_I^\pm $ for $I=1,\cdots, g$ and $n_j^\pm $ for $j=g+1,\cdots, g+\nu-1$ are arbitrary. 
These conditions imply the relations for single-valuedness of $\lambda$ of (\ref{6c10}) with,
\bea
\label{6d9}
m= m^+ - m^- \hskip 1in n= n^+ - n^-
\eea
The total number  in (\ref{6d7}) is $ 2 \hat g$ real conditions.

\subsection{The functions \texorpdfstring{$\cA_\pm$}{A}}

To integrate $\p_w \cA_\pm$ we decompose these differentials onto the following Abelian differentials of the third kind, 
\bea
\label{6e1}
 \p_w \ln { E(w,z) \over E(w,w_0)}
\eea
with simple poles at $w=z$ and $w=w_0$ with unit residues of opposite signs. In view of the cancellation of the sum of the residues $Z^\ell _\pm$ at the poles $p_\ell$, all dependence on $w_0$ cancels out, and we have the following decomposition, 
\bea
\label{6e2}
\p_w \cA_\pm (w |\Omega) = \sum _{\hI=1}^{\hat g} \alpha _\pm^\hI \omega _\hI (w) 
+ \sum _{k =1}^L Z^k _\pm \, \p_w \ln E(w,p_k)
\eea
where $\alpha ^\hI_\pm$ are the coefficients of the holomorphic one-forms $\omega _\hI$ on $\hat \Sigma$.
The residues satisfy,
\bea
\label{6e2a}
\overline{Z^\ell _\pm} = - Z^\ell _\mp
\hskip 1in 
\sum _{\ell=1}^L Z_\pm ^\ell=0
\eea
and their values may be read off from the product representations in (\ref{6d4}), 
\bea
\label{6e3}
Z^k _+  & = & 
 \lambda _0 \omega _0  \,  \sigma (p_k)^2 \,   \, {  \prod _{n=1} ^N  E(p_k ,s_n) \over \prod _{\ell \not= k}^L E(p_k,p_\ell) } 
\exp \Bigg \{ - 2 \pi i \sum _{\hat I, \hat J}  \int _{w_0} ^{p_k} \om _\hI \, Y^{-1}_{\hI \hJ} \, \Im \Lambda ^+ _{\hat J} \Bigg \}
\no \\
Z^k _- & = &
\bar \lambda _0 \omega _0 \, \sigma (p_k)^2 \, 
 { \prod _{n=1} ^N  E(p_k,\mI (s_n))   \over \prod _{\ell \not= k}^L E(p_k,p_\ell)} 
\exp \Bigg \{ - 2 \pi i \sum _{\hat I, \hat J}  \int _{w_0} ^{p_k} \om _\hI \, Y^{-1}_{\hI \hJ} \, \Im \Lambda ^-_{\hat J} \Bigg \}
\qquad 
\eea
The coefficients $\alpha _\pm^\hI $ may be obtained by evaluating (\ref{6e2}) at $\hat g$ of the zeros, denoted by  $s_n$, of $\p_w \cA_+$ and $\hat g$ of the zeros $\mI(s_n)$ of $\p_w \cA_-$, 
\bea
\label{6e4}
0 & = & \sum _{\hI=1}^{\hat g} \alpha _+^\hI \omega _\hI (s_n) 
+ \sum _{k =1}^L Z^k _+ \, \p_{s} \ln E(s,p_k) \Big |_{s=s_n}
\no \\
0 & = & \sum _{\hI=1}^{\hat g} \alpha _-^\hI \omega _\hI (\mI(s_n)) 
+ \sum _{k =1}^L Z^k _- \, \p_{\tilde s} \ln E(\tilde s,p_k) \Big |_{\tilde s = \mI(s_n)}
\eea
and then solving each $\hat g \times \hat g$ linear system in turn.
The functions $\cA_\pm$ are obtained by integrating (\ref{6e2}), 
\bea
\cA_\pm (w|\Omega) = \cA^0_\pm + \sum _{\hI=1}^{\hat g} \alpha _\pm^\hI \int _{w_0} ^w \omega _\hI 
+ \sum _{k =1}^L Z^k _\pm \, \ln E(w,p_k)
\eea
 The functions $\cA_\pm$ do not necessarily have to be  single-valued on $\hat \Sigma$. Indeed, monodromy of $\cA_\pm$  by constant shifts, as given in (\ref{Atrans}) subject to the condition (\ref{moncon}),   is allowed as it gives rise to single-valued supergravity fields. We shall now investigate the conditions this kind of allowed monodromy imposes on the parameters of the holomorphic functions $\cA_\pm$. First, by comparing the complex conjugate of the second line in (\ref{6e4}) with the first line of (\ref{6e4}), we obtain the relation,
\bea
\overline{\alpha _- ^\hI} = - \sum _{\hJ} \cS_{\hI \hJ} \alpha _+ ^\hJ
\eea
or simply $\bar \alpha _- = - \cS \alpha _+$ in matrix notation. When the index $\hI=i$ corresponds to a boundary cycle, this relation generalizes the one obtained for the annulus in (\ref{4d4}). Second, the monodromies around cycles $A_\hI$ and $B_\hI$  are given by,
\bea
\cA_\pm (w+A_\hI) - \cA_\pm (w) & = & \alpha _\pm ^\hI
\no \\
\cA_\pm (w+B_\hI) - \cA_\pm (w) & = & \beta _\pm ^\hI= \sum _\hJ \Omega _{\hI \hJ} \alpha _\pm ^\hJ
+ 2 \pi i \sum _{k=1}^L Z_\pm ^k \int ^{p_k} _{w_0} \om _\hI
\eea
There are no conditions on the monodromy around $B_i$ cycles, as such cycles extend beyond the surface $\Sigma$. 
Thus, the conditions for allowed monodromy around these cycles are given by,
\bea
\overline{\alpha ^\hI _-} & = & \alpha _+ ^\hI
\no \\
\overline{\beta ^\hJ _-} & = & \beta _+ ^\hJ \hskip 1in \hJ = J, J'=1, \cdots, h
\eea
or simply $\bar \alpha _- = \alpha _+$ and $(I-\cS) (\bar \beta _- - \beta _+)=0$ in matrix notation.
Combining these relations, we obtain the following constraints on the parameters of the solution, expressed here in matrix notation,
\bea
\bar \alpha _- = \alpha _+ & \hskip 1in & (I+\cS) \alpha _+ = (I+\cS) \alpha _-=0
\no \\
\bar \beta _- = \beta _+ && (I-\cS) \beta _+ = (I-\cS) \beta _- =0
\eea
Translated in components, the first set of equations become,
\bea
\alpha _\pm ^i =0 \hskip 1in \alpha _\pm ^I + \alpha _\pm ^{I'}=0
\eea
while the second set of equations gives $\alpha_\pm^I$ in terms of the residues $Z^k_\pm$ and the poles $p_k$. We see that, by contrast with the case of the annulus, for higher genus the parameters $\alpha _\pm$ need not be forced to vanish by the monodromy conditions alone.

\subsection{The conditions required for \texorpdfstring{$\cG=0$}{G=0} on the boundary}

The general arguments of subsection \ref{sec:232} guarantee that $\cG$ is piecewise constant on any line segment of the boundary which is free of poles.  The condition for continuity of the function $\cG$ across a pole  $p_\ell$ is given by the same arguments as in the case of the upper half plane and the annulus. Here, we shall specialize to the case where we have a single boundary component and hence $\nu =1$, since the expressions considerably simplify in this case. We shall also assume that the condition for single-valuedness around $A_\hI$ cycles has been imposed, so that $\alpha _\pm ^\hI=0$. The conditions are local around $p_\ell$, and are given as follows, 
\bea
\label{6f1}
\cA_+^0  Z^k _-  - \cA_- ^0  Z^k _+ 
+\sum _{\ell \not= k} \Big ( Z_+^\ell Z_- ^k - Z_+ ^k Z_- ^\ell \Big )  \ln \Big |E ( p_k , p_\ell|\Omega) \Big | =0
\eea
When only a single boundary component is present, the vanishing of $\cG$ on $\p \Sigma$ may be achieved by solving for the integration constant $\cB^0$ of the holomorphic function $\cB$. When $\p \Sigma$ has $\nu \geq 2$ boundary components, there are $\nu-1$ further real conditions required to guarantee the vanishing of $\cG$ on each boundary component.

This completes the construction of the ansatz for general Riemann surfaces and the derivation of the regularity conditions.
We leave a detailed investigation of the existence of solutions with non-trivial topology for the future and close by noting that, based on the numerical investigations for the annulus described in subsection \ref{sec:57}, we may speculate that no solutions will exist if $\Sigma$ has multiple boundary components.

\newpage

\section{Discussion}
\setcounter{equation}{0}
\label{sec:6}

In this paper we  have constructed  a large family of Type IIB supergravity solutions which are candidates for  holographic duals to five-dimensional superconformal field theories. The space-time of these solutions consists of the manifold   $AdS_6\times S^2$ warped over a two-dimensional Riemann surface $\Sigma$ with boundary.  The supergravity solutions preserve sixteen supersymmetries and give a holographic realization of the $F(4)$ superconformal algebra. Unlike corresponding solutions in dimensions 3, 4 and 6, there is no asymptotic enhancement of the supersymmetry, in accord with the fact that there exist no five-dimensional superconformal theories with  thirty two supersymmetries.

\sm

The solutions become singular at isolated points on the boundary of the Riemann surface, where the poles in the differentials $\p_w \cA_\pm$ which characterize the solutions are located. The singularities have a clear physical interpretation, as the supergravity  fields  near the poles take the form of the singular ``near  horizon'' limit for a $(p,q)$ five-brane.
This gives a natural identification of the poles as remnants of the external five-branes which are used to construct brane web realizations of five-dimensional field theories.
The $SO(2,5)$ symmetry of our solutions implies that we are describing the conformal fixed points where the web collapses to a five-brane intersection with all branes intersecting at a single point. This offers a clear path to identifying the dual SCFTs, as we have discussed for example solutions with three and four poles corresponding to intersections with three or four external branes.

\sm

For the large class of explicit solutions presented in this work we have assumed the topology of the Riemann surface to be that of a disk. For higher topologies, we have presented the general set-up for the construction of physically regular supergravity solutions, but we have not shown that such solutions exist. For the case of the annulus, our numerical investigations point in the direction that physically regular solutions without axial monodromy may not exist, though we have no analytical proof. From the analysis and numerical results for the annulus, it is natural to speculate that physically regular solutions will exist only when there is precisely one boundary component. Whether solutions with higher genus exist, and what their physical interpretation in terms of five-brane webs may be, is at this time an open question, to which we hope to return in the future.

\sm

Many aspects of the supergravity solutions that we have constructed and the proposed dual superconformal field theories to these solutions remain to be explored as well. We end the paper with a discussion of some  of the open questions and avenues for future research. 

\sm

One might wonder whether the presence of the singularities at the poles of the differentials $\p_w \cA_\pm$, and the fact that the supergravity solution near the singularities approaches the near horizon limit of a five-brane, can invalidate the  interpretation of the solution as a dual of a five-dimensional theory. It is a natural next step to study field theory observables holographically in order to address this question.  

\sm

In the AdS/CFT correspondence the spectrum of operators and   correlation functions are the most direct observables which can be obtained by considering the KK-spectrum of linearized perturbations and Witten diagrams, respectively. However, due to the fact that the $AdS_6\times S^2$ space is warped over the Riemann surface $\Sigma$  this analysis is considerably more complicated than in more familiar cases such as the $AdS_5\times S^5$ solution of Type IIB.  It would  be interesting to investigate the operator spectrum  for our warped $AdS$ solutions, as little is known about the operator spectrum of the five dimensional SCFTs.

\sm

Other CFT observables such as the entanglement entropy can be calculated more straightforwardly using the Ryu-Takyanagi prescription \cite{Ryu:2006bv,Ryu:2006ef}. Such a calculation for a spherical region has been performed for the singular Type IIA $AdS_6$ supergravity solutions in \cite{Jafferis:2012iv}, and we are currently investigating the holographic entanglement entropy  for the solutions presented in this paper \cite{Marasinoutoappear}. Preliminary results suggest that the presence of the poles does not obstruct the holographic computations of either the entanglement entropies or of the free energy directly.

\sm

Another interesting prospect for the future is to further generalize our solutions. S-duality acts naturally  on the holomorphic functions which parameterize them, but the solutions presented in this paper all exhibit vanishing monodromy under the $SU(1,1)$ S-duality group. This means that as one moves along a closed loop on  the Riemann surface $\Sigma$ all supergravity fields come back to the same value. It  is well known, however,  that seven-branes induce  nontrivial monodromies on some supergravity fields. For example, a D7-brane induces a shift of the axion along a closed contour enclosing the D7-brane in the two transverse directions. We note that seven-branes can naturally be incorporated into five-brane webs \cite{DeWolfe:1999hj} and they are indeed needed for the most general construction of five dimensional field theories and their moduli spaces. Consequently, it  is an interesting question whether it is possible to incorporate seven-branes and their monodromies into the  framework of our solutions.  
We plan to address this question in future work.

\section*{Acknowledgements}

We are happy to thank Andreas Karch for collaboration on an earlier paper that led to the present work, and to Oren Bergman and Diego Rodriguez-Gomez for extensive and helpful discussions on five-brane webs.
We also acknowledge the Aspen Center for Physics, which is supported by National Science Foundation grant PHY-1066293, for hospitality during the workshop ``Superconformal Field Theories in $d\geq 4$'' and thank the organizers and participants for the enjoyable and inspiring conference.
The work of all three authors  is supported in part by the National Science Foundation under grant   PHY-16-19926.  

\newpage

\appendix

\section{Absence of  poles for \texorpdfstring{$\partial_w \cA_\pm$}{dA}  in the interior of \texorpdfstring{$\Sigma$}{Sigma} }
\label{appenda}
\setcounter{equation}{0}

In the main part of the paper we have assumed that all poles of $\partial_w \cA_\pm$ are located on  the boundary of $\Sigma$. In this appendix we shall rule out the presence of poles in $\partial_w \cA_\pm$ in the interior of $\Sigma$. The arguments are local. For concreteness, we shall present them here for the explicit solutions on the upper half plane, but their locality guarantees that their validity extends to surfaces of arbitrary topology. 

\sm

The starting point for the construction of the physically regular supergravity solutions  is the function $\lambda$ which, for the upper half plane, is given by,
\bea
\lambda(w)=\prod_{n=1}^N{w-s_n\over w-\bar s_n}
\eea
To construct the differentials $\partial_w \cA_\pm$ from $\lambda$, we had originally assumed that all zeros of $\lambda$  become zeros of $\partial_w \cA_+$, while all poles of $\lambda$ become zeros of $\partial_w \cA_-$. But this is not the most general possibility available for given $\lambda$. Indeed, some of the zeros of $\lambda$ can actually become poles of $\partial_w \cA_-$ and the accompanying complex conjugate poles of $\lambda$  can become poles of $\partial_w \cA_+$. We illustrate this by singling out the pair of zeros and poles of $\lambda$, $(s_1, \bar s_1)$, 
\bea
\lambda(w) = \left({w-s_1 \over w-\bar s_1}\right) \prod_{n=2}^N {w-s_n\over w-\bar s_n}
\hskip 1in {\rm Im}(s_n)>0
\eea
The locality of the argument will guarantee that it applies equally well when more than one pair of zeros and poles are reversed. We now construct $\p_w \cA_\pm$   accordingly,
\bea
\partial_w \cA_+&=& {i\over w-\bar s_1} {\prod_{n=2}^N (w-s_n) \over\prod_{\ell=1}^L (w-p_\ell)}
\nonumber \\
\partial_w \cA_-&=& {i\over w- s_1} {\prod_{n=2}^N (w-\bar s_n) \over\prod_{\ell =1}^L (w-p_\ell)}
\eea
Note that we have $L=N$ poles to ensure regularity at infinity.
We automatically have $\kappa^2 > 0$ in the interior of $\Sigma$, since this condition was already guaranteed by the form of $\lambda$ alone. Recall that the pre-factor of $i$ is present (previously denoted $\omega_0$) to enforce the conjugation property $ \overline{\partial_{\bar w} \cA_+(\bar w)} =-\partial_w \cA_-$,
which is essential for the piece-wise vanishing of $\cG$ along the real line. Decomposing $\p_w \cA_\pm$ into partial fractions, we have,
\bea
\partial_w \cA_+&=&  {i\alpha\over w-\bar s_1} +\sum_{\ell =1}^L {Z_+^\ell \over w-p_\ell} 
\nonumber \\
\partial_w \cA_-&=&  {i\bar\alpha\over w- s_1} +\sum_{\ell=1}^L {Z_-^\ell\over w-p_\ell} 
\eea
and integrating,
\bea
 \cA_+&=&  \cA_+^0+ i\alpha\ln( w-\bar s_1)+\sum_{\ell=1}^L Z_+^\ell \ln(w-p_\ell) 
 \nonumber \\
 \cA_-&=&  \cA_-^0+ i\bar\alpha\ln( w-s_1)+\sum_{\ell=1}^L Z_-^\ell \ln(w-p_\ell) 
\eea
Next, $\cG$  is computed by  first  evaluating its derivative,
\bea
\partial_w \cG &=& \left( \bar \cA_+-\cA_-\right) \partial_w \cA_++ \left(\cA_+-\bar \cA_-\right)\partial_w \cA_-
\eea
The dominant behavior near  the pole $w = s_1$ in the upper half plane is carried by $\cA_-$ which diverges logarithmically there, while $\cA_+$  is regular. Hence near $w = s_1$ we have,
\bea
\partial_w \cG\sim - \bar \cA_-\partial_w \cA_-
\eea
This implies that the leading behavior of $\cG$ is given by
\bea
\cG\sim -\left| \cA_-\right|^2 \sim -|\alpha|^2 \left( \ln |w-s_1|\right)^2
\eea
Hence, near $w = s_1$, the function $\cG$  is negative and divergent. Hence $\cG$ fails to be positive everywhere in the interior of $\Sigma$ and the solution  fails to be regular.

\newpage

\bibliographystyle{JHEP.bst}
\bibliography{ads6}
\end{document}